\def\mearth{{\rm\,M_\oplus}}
\def\rearth{{\rm\,R_\oplus}}

\documentclass[preprint2]{pp7}

\input{pp7.h}

\begin{document}

\title{\textbf{\LARGE Planet Formation Theory in the Era of ALMA and Kepler:\\from Pebbles to Exoplanets }}

\author{ {\textbf {Joanna Dr\c{a}\.{z}kowska}}$^{1,2}$, {\textbf {Bertram Bitsch}}$^3$, {\textbf {Michiel Lambrechts}}$^{4,5}$, {\textbf {Gijs D. Mulders}}$^{6,7}$, {\textbf {Daniel Harsono}}$^{8,9}$, {\textbf {Allona Vazan}}$^{10}$, {\textbf {Beibei Liu}}$^{11}$, {\textbf {Chris W. Ormel}}$^{12}$, {\textbf {Katherine Kretke}}$^{13}$ \\ and {\textbf {Alessandro Morbidelli}}$^{14}$}
\vspace{2mm}
\affil{$^1$\small\it {University Observatory, Faculty of Physics, Ludwig-Maximilians-Universit\"at M\"unchen, Scheinerstr.~1, 81679 Munich, Germany}}
\affil{$^2$\small\it {Max Planck Institute for Solar System Research, Justus-von-Liebig-Weg 3, 37077 G\"ottingen, Germany}}
\affil{$^3$\small\it Max-Planck-Institut f\"ur Astronomie, K\"onigstuhl 17, 69117 Heidelberg, Germany}
\affil{$^4$\small\it {Lund Observatory, Department of Astronomy and Theoretical Physics, Lund University, Box 43, 22100 Lund, Sweden}}
\affil{$^5$\small\it {Center for Star and Planet Formation, GLOBE Institute, University of Copenhagen, \O ster Voldgade 5-7, 1350 Copenhagen, Denmark}}
\affil{$^6$\small\it {Facultad de Ingenier\'{i}a y Ciencias, Universidad Adolfo Ib\'{a}\~{n}ez, Av.\ Diagonal las Torres 2640, Pe\~{n}alol\'{e}n, Santiago, Chile}}
\affil{$^7$\small\it{Millennium Institute for Astrophysics, Chile}}
\affil{$^8$\small\it {Institute of Astronomy, Department of Physics, National Tsing Hua University, Hsinchu, Taiwan}}
\affil{$^9$\small\it {Academia Sinica Institute of Astronomy and Astrophysics, No.1, Sec. 4., Roosevelt Road, Taipei 10617, Taiwan}}
\affil{$^{10}$\small\it Astrophysics Research Center of the Open University (ARCO), Department of Natural Sciences, The Open University of Israel, 4353701 Raanana, Israel}
\affil{$^{11}$\small\it {Zhejiang Institute of Modern Physics, Department of Physics \& Zhejiang University-Purple Mountain Observatory Joint Research Center for Astronomy, Zhejiang University, 38 Zheda Road, Hangzhou 310027, China} }
\affil{$^{12}$\small\it Department of Astronomy, Tsinghua University, Beijing 100084, China}
\affil{$^{13}$\small\it Southwest Research Institute, 1050 Walnut St. Suite 300, Boulder, CO, USA}
\affil{$^{14}$\small\it Laboratoire Lagrange, UMR 7293, Universit\'e de Nice Sophia-Antipolis, CNRS, Observatoire de la C\^ote d’Azur, Boulevard de l’Observatoire, F-06304 Nice Cedex 4, France}

\begin{abstract}
\baselineskip = 11pt
\leftskip = 1.5cm 
\rightskip = 1.5cm
\parindent=1pc
{\small Abstract

{Our understanding of the planet formation has been rapidly evolving in recent years. The classical planet formation theory, developed when the only known planetary system was our own Solar System, has been revised to account for the observed diversity of the exoplanetary systems. At the same time, the increasing observational capabilities of the young stars and their surrounding disks bring new constraints on the planet formation process. In this chapter, we summarize the new information derived from the exoplanets population and the circumstellar disks observations. We describe the new developments in planet formation theory, from dust evolution to the growth of planetary cores by accretion of planetesimals, pebbles, and gas. We review the state-of-the-art models for the formation of diverse planetary systems, including the population synthesis approach which is necessary to compare theoretical model outcomes to the exoplanet population. We emphasize that the planet formation process may not be spatially uniform in the disk and there are preferential locations for the formation of planetesimals and planets. Outside of these locations, a significant fraction of solids is not growing past the pebble-sizes. The reservoir of pebbles plays an important role in the growth of planetary cores in the pebble accretion process. The timescale of the emergence of massive planetary cores is an important aspect of the present models and it is likely that the cores within one disk form at different times. In addition, there is growing evidence that the first planetary cores start forming early, during the circumstellar disk buildup process.}
 \\~\\~\\~}
 %leave this in to get the correct vertical space after the abstract
\end{abstract}  

%\maketitle

%%%%%%%%%%%%%%%%%%%%%%%%%%%%%%%%%%%%%%%%%%%%%%%%%%%%%%%%%%%%%%%%%%%%%
\section{\textbf{INTRODUCTION}}

Many exoplanets and exoplanet systems have been discovered in the past years, demonstrating that the demographics of planetary systems varies widely from what we observe in our own Solar System. At the same time, thanks to the progress in observational capabilities of the planet-forming disks surrounding young stars, we got new limits on the expected mass budgets and timescales of planet formation. These findings challenged our understanding of planetary origins.

Circumstellar disks initially consist mainly of hydrogen-helium gas. The condensible material referred to as solids constitutes only about 1\% of the disk mass. Initially, all the solids are in form of tiny dust grains, which are tightly coupled to the gas but with time they may grow to so-called pebbles. We must stress that the astrophysical definition of pebble does not rely on its size but on its aerodynamic properties, as we discuss in \S\ref{sect:theory}. Classical models of planet formation have mostly neglected dust evolution and started with the assumption that the gravitationally bound building blocks of planets, the planetesimals, formed quickly \citep{safronov1972, Goldreich1973}. In recent years, the observations of millimeter-sized solids in disks and the development of the pebble accretion paradigm have revived the interest in models of dust growth and its connection to planetesimal and planet formation. 

Still, much of the theory to explain the formation of planets relies on concepts that were developed to reproduce the Solar System. The classical model for Solar System formation has been reviewed many times in the past \citep[see, e.g.,][]{Chambers2004, Raymond2014}. In this chapter, we outline the new, emerging paradigm for the formation of planets that is aimed at explaining the origin of both the exoplanets and the Solar System within one framework. While there are still some unanswered questions, we are certainly progressing towards a self-consistent theory that covers the stages from dust and pebbles observed in the circumstellar disks to the fully-fledged planets.

This chapter is organized as follows. We first present the constraints on the planet formation process derived from the exoplanets population in \S\ref{sub:2.1} and from the circumstellar disks observations in \S\ref{s:disks}. Next, we present the major aspects of the emerging planet formation theory, including the formation of planetesimals in \S\ref{sect:dust}, planetesimal and pebble accretion in \S\ref{sub:ppaccretion}, the radial redistribution of solids in \S\ref{sub:driftmigr}, and the accretion of gaseous envelopes in \S\ref{sect:gasacc}. In \S\ref{sect:models} we demonstrate the models for formation of planetary systems: we discuss ways to form the different types of planets in \S\ref{sub:4.1}, scenarios that reproduce the exoplanet systems and the Solar System in \S\ref{sub:4.2}, formation of planetary systems around the low mass stars in \S\ref{sub:4.3}, and we conclude by presenting simplified pebble accretion and planetesimal accretion population synthesis models in \S\ref{sub:popsynth}. We summarize the chapter and outline necessary further developments of the planet formation models in \S\ref{sect:summary}.

%%%%%%%%%%%%%%%%%%%%%%%%%%%%%%%%%%%%%%%%%%%%%%%%%%%%%%%%%%%%%%%%%%%%%

\begin{figure*}[h]
 \centering
 \includegraphics[width=0.85\linewidth]{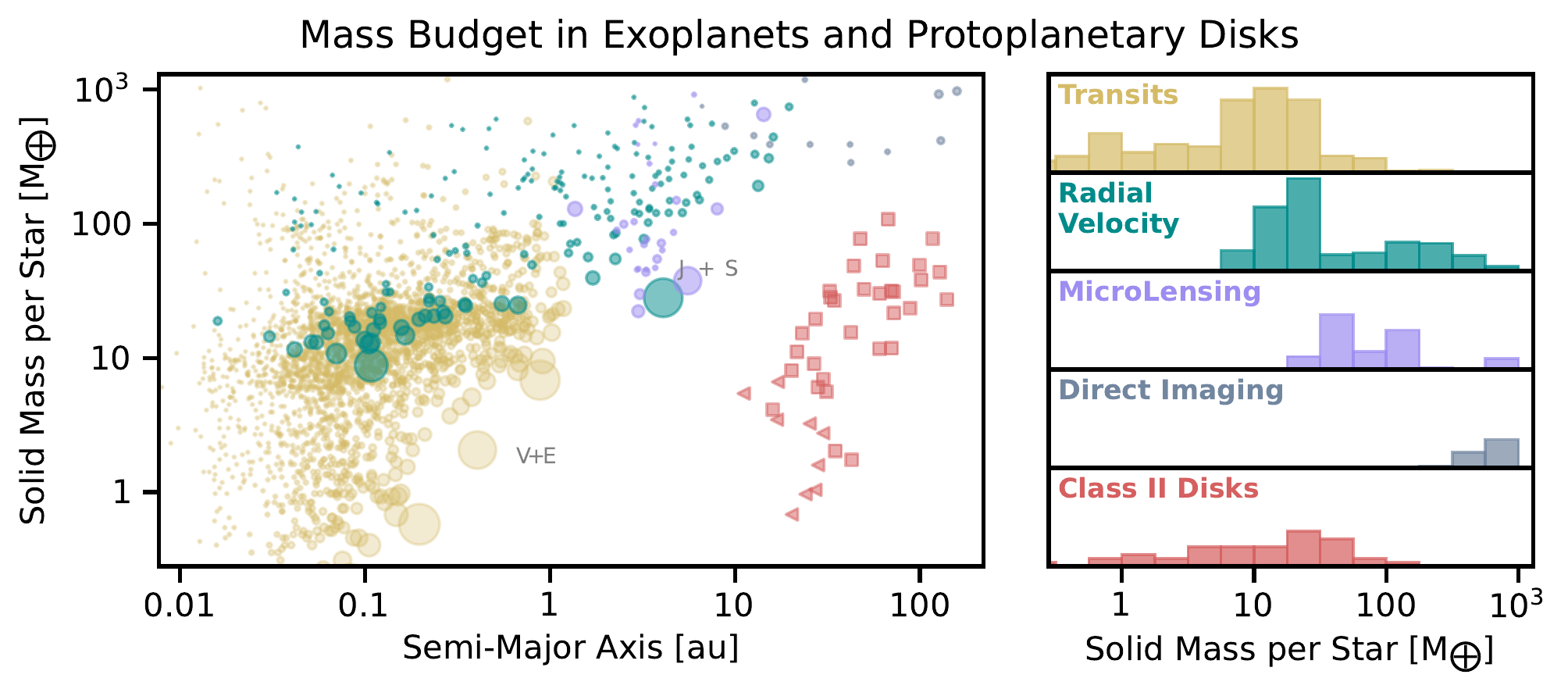}
 \caption{{The solid mass budget of exoplanets around sun-like stars from different detection methods: 
 transits ({\it Kepler}, \citealt{Thompson2018}); 
 radial velocity \citep{Fulton2021};
 microlensing \citep{Suzuki2016}; and direct imaging \citep{Vigan2021}. 
 Solid masses are estimated from observed masses and radii using the relations on \citet{thorngren16} and \citet{chenkipping17}. To correct for detection bias, each detected planet has a circle area proportional to the intrinsic occurrence of that planetary system, following \cite{Mulders2021}. For reference, Class II protoplanetary disks masses are shown in red, with symbols or upper limits placed at the disk outer radii. The right panel shows the occurrence-weighted histograms of the solid mass distributions.}
 \label{fig:massbudget}}
\end{figure*}

\section{\textbf{CONSTRAINTS ON THE PLANET FORMATION PROCESS}}\label{sect:observations}

\subsection{\textbf{Constraints from the exoplanet population}}\label{sub:2.1}

Exoplanets\index{Exoplanets|(} are a diverse population that reflect the outcome of the planet formation process in the broadest sense but our view on them is narrowed by our observational visor.
In this section, we focus on the statistical properties of exoplanets to characterize the ``typical'' outcome of the planet formation process:
super-Earths and mini-Neptunes (\S\ref{s:exo:superearths}) and cold giant planets (\S\ref{s:exo:giants}).

Our Solar System represents one possible outcome of planet formation for one particular star, one particular protoplanetary disk, and possibly other factors like galactic environment. It thus presents detailed but narrow view on the outcome of planet formation. On the other hand, the known population of exoplanets has emerged from a mix of stars with a range of masses (\S\ref{s:exo:mdwarf}) and chemical compositions (\S\ref{s:exo:metal}), from a range of protoplanetary disk properties (\S\ref{s:disks}), and formed at different times and locations throughout the Galaxy, providing a broader perspective on the possible outcomes of the planet formation process. 

Exoplanet detection methods are biased towards large planets at short distances from the star because they are easier to detect, and those planets tend to be over-represented in known exoplanet catalogues. This bias within catalogues can be corrected for by using homogeneous survey designs and quantifying the detection efficiency in exoplanet surveys. Such calculations typically yield a planet occurrence rate, or the average number of planets per star, specified for a range of planet properties such as size and orbital separation. 

However, we are fundamentally blind to certain groups of planets outside or below the detection limits of current surveys. Transit surveys can identify the smallest exoplanets, from roughly the size of Mars at 0.1 au to slightly larger than earth at 1 au. At larger separations, radial velocity surveys extend down to Neptune-mass exoplanet at 1 au and down to Jupiter mass at 10 au, while micro-lensing surveys also reach Neptune mass near ~5 au. Outside of 10 au, direct imaging surveys or mainly sensitive to super-Jupiters. 
What exoplanets lie below these detection limits
remain to be discovered. Thus the estimates of occurrence and mass of planetary systems based on detected exoplanets, as made in this chapter, always represent a lower limit to the true values. 

When different exoplanet surveys are combined, they generally paint a consistent picture of planet populations. Figure \ref{fig:massbudget} shows the amount of solid material present in mature planetary systems detected around sun-like stars. 
The most common type of exoplanet based on radial velocity and transit surveys are super-Earths and mini-Neptunes (\S\ref{s:exo:superearths}), that typically represent $\sim 10 \mearth$ in solids per planetary system, all located within 1 au.
The population of giant planets\index{Giant planets} (\S\ref{s:exo:giants}), mostly located between 1 and 10 au, represent planetary systems with large amounts of solids per planet, but that orbit a smaller fraction of stars. The giant planet population extends over multiple orders of magnitude in orbital separation and is constrained by transit, radial velocity, micro-lensing, and direct imaging surveys.

In \S\ref{s:exo:multi} we will (briefly) discuss how the observed properties of multi-planet systems constrain planet formation models.

\subsubsection{Super-Earths and sub-Neptunes}\label{s:exo:superearths}
At the time of {\it Protostars \& Planets VI}, a population of close-in super-Earths\index{Super-Earths} was known from radial velocity surveys and initial results of the {\it Kepler}  mission (see chapter by \citealt{Raymond2014}). Since then, a significant progress has been made in characterizing this population of exoplanets. In particular, the unprecedented sample size and sensitivity of the {\it Kepler} spacecraft, designed to detect transiting earth-size planets at 1 au around sun-like stars by monitoring nearly 200 000 stars, has led to a huge leap forward in statistical studies.

A key results of the \textit{Kepler} mission are that planets smaller than Neptune with orbital periods less than a year outnumber the stars they orbit. The occurrence rates, the average number of planets per star after bias corrections, are in the range $140-200\%$ for sun-like stars with spectral types FGK \citep{Mulders2018,Hsu2019,Kunimoto2020}. Because stars typically have multiple planets, this does not mean every star has such a planet, as we explain below.
The fraction of sun-like stars with planetary systems within 1 au, or the probability that a star has at least one planet, is estimated to be in the range 50-60\% \citep{Mulders2018,Zhu2018,He2019,He2021}.

Multi-planet systems are common. At least half the planets are part of nearly co-planar systems with multiple planets \citep{lissauer11}. The other half could be intrinsically single planets or highly inclined planets \citep{Johansen2012,Hansen13,Sandford2019}, or the other planets in the system are partly missed by planet detection pipelines \citep{Zink2019}. There are significant degeneracies in deriving planetary system architectures from transit surveys \citep{TremaineDong2012}, and it is possible for all planets to be part of multi-planet systems if the number of planets per system is anti-correlated with their mutual inclinations \citep{Zhu2018,He2020}.

There is a large diversity in planetary properties. The distributions of planet radii and orbital periods form smooth distributions over large ranges of detectable parameter space. The period and radius distribution of exoplanets extend up to detection limits, with little indication that these distributions end at current detection limits.

Planet occurrence rates increase rapidly with orbital period up to $\sim$10 days, but exterior to that are roughly constant in logarithm of orbital period up to orbital periods of 1 year \citep{Youdin2011,howard12,Mulders2018,Petigura2018}. 
The planet occurrence rate falls off rapidly with planet size for planets larger than $\sim{}3\,\rearth$ \citep{Youdin2011,howard12,MuldersEtal2015}. 
Between $3\,\rearth$ and the detection limit of {\it Kepler} near $\sim{}0.5\,\rearth$, the planet occurrence rate is again roughly constant in logarithm of planet radius \citep{Petigura2013}, though many studies also find slight negative dependencies on planet size \citep{Burke2015,Hsu2019}.

Follow-up radial velocity observations \citep[e.g.][]{Wolfgang2016} or transit timing variations \citep[e.g.,][]{Hadden2014} can be used to constrain the mass-radius relation and bulk density of planets. The masses of planets below $\sim{}1.6\,\rearth$ are consistent with rocky planets, while the masses above $\sim{}1.6\,\rearth$ are consistent with having substantial volatile atmospheres \citep{Rogers2015}. Using derived mass-radius relations \citep[e.g.,][]{chenkipping17}, the planet mass distribution can be estimated from the planet radius distribution (e.g. \citealt{Najita14,Manara2018,Mulders2021}, as shown in Fig.~\ref{fig:massbudget}). The typical \textit{Kepler} planetary system, with multiple planets in the range of $\sim{}0.5-3\,\rearth$, contains $5-20 \,\mearth$, mostly in solids.
These estimates are consistent with the high occurrence of planets of similar mass in radial velocity surveys \citep{howard10,Mayor2011}.

\subsubsection{Giant planets}\label{s:exo:giants}
The giant planet population is mainly constrained by long-running radial velocity surveys \citep{Mayor2011,Fulton2021}. These are complemented by transit, micro-lensing, and direct imaging surveys \citep[e.g.,][]{Santerne2016,ClantonGaudi2016}.
Radial velocity surveys have detected a population of giant planets\index{Giant planets} extending out to $10-25$ au that orbit $10-20\%$ of stars, with the intrinsic occurrence being a rising function of semi-major axis \citep{cumming08,Mayor2011}, a trend that is also seen in the {\it Kepler} data \citep{DongZhu2013,Santerne2016,Petigura2018}. Consequently, most giant planets are located exterior to $\approx$1 au \citep{Fernandes19,Fulton2021}. Although they are over-represented in exoplanet catalogues as a result of survey selection biases, the hot Jupiters only represent 0.5-1\% of giant planets \citep{Wright2012,howard12,Fressin2013}.

The main new insight into giant planets originates from the planets with the longest orbital periods. Direct imaging surveys have put limits on the occurrence of super-Jupiters in the range 10-100 au, which are on the order of $5\%$ for planets more massive than Jupiter \citep{Nielsen2019,Vigan2021}. Additionally, these surveys have shown that the planet occurrence rate in this regime decreases with semi-major axis.
This is consistent with a re-analysis of the \citet{Mayor2011} data by \citet{Fernandes19}, who find a turnover in the giant planet occurrence rate near 2-3 au. 
The California Legacy Survey presented by \citet{Fulton2021}, which extends the giant planet census out to 25~au, identifies the peak in planet occurrence rate at 2-6 au.

Overall, this leads to a consistent picture of the giant planet population across surveys: they are relatively rare at short periods where they are also detected by transit surveys, they are most common around 1-10 au where they are also seen micro-lensing surveys  \citep[e.g.][]{Suzuki2016}, and become increasingly rare at 10-100 au in direct imaging surveys.

The overall occurrence of giant planets depends sensitively on the lower mass limit adopted, because smaller planets are more common than larger planets. 
Extrapolating the radial velocity mass and semi-major axis distribution, \cite{Fernandes19} find a planet occurrence rate between 1 and 100~au of 6\% and 27\% for planets more massive than $1$ and $0.1\,M_J$, respectively. \cite{Fulton2021} estimate a planet occurrence rate between 0.3-30 au and $>0.1 M_J$ of 34\%.

\subsubsection{M dwarf exoplanets}\label{s:exo:mdwarf}

Low mass M dwarfs\index{M dwarfs} have proven to be a fertile hunting grounds for small planets due to relatively high mass and radius contrast with the host star. At the time of {\it Protostars \& Planets VI}, several earth-mass planets had been discovered in radial velocity surveys, and the first planet occurrence rates of small planets had been measured with HARPS \citep{bonfils13} and from the \textit{Kepler} mission \citep{Dressing2013}.
Since then, the planet population around M dwarfs has been characterized in more detail. Giant planets appear to be rare compared to sun-like stars both at short periods \citep{mulders2015a,HsuEtal2020} and long periods \citep{Johnson2010,Fulton2021}.

The size distribution of transiting planets around M dwarfs is dominated by planets smaller than $\approx3\,\rearth$ \citep{Dressing2013,Dressing2015,Morton2014,MuldersEtal2015}.
The occurrence of planets smaller than 4 $\rearth$ interior to $\approx150$ days is a factor 2-3 higher than that around solar-mass stars \citep{howard12,mulders2015a,Yang2020,HsuEtal2020} and the same is found in radial velocity surveys \citep{SabottaEtal2021}. 
The presence of super-Earths around M dwarfs provide a strong constraint on the efficiency of planet formation given that they have to form from lower-mass disks on average \citep[e.g.][]{MuldersEtal2015,Pascucci2016,Gaidos2017}. The average M dwarf has planets with $\approx 10$ M$_\oplus$ in solids within 0.5 au, compared to ~5 $\mearth$ for sun-like stars \citep{MuldersEtal2015}.

The architectures of planetary systems around M dwarfs are generally reminiscent of those around FGK stars. Multi-planet systems are common, both based on \textit{Kepler} data \citep[e.g.,][]{BallardJohnson2016} and on an analysis of radial velocity systems \citep{Cloutier2021}. The gradual increase in planet occurrence from F to G, K, and M stars is caused by an increase in the occurrence of planetary systems, and not by a change in planetary system architecture \citep{Yang2020,He2021}.

One key difference in the M dwarfs super-Earths\index{Super-Earths} population compared to sun-like stars is the change in the typical planet mass. Based on different models for the planetary mass-radius relation, both \cite{Wu2019} and \cite{Pascucci2018} conclude that typical planet mass scales linearly with the stellar mass.

\subsubsection{Host star metallicity}\label{s:exo:metal}
The dependence of planet properties on the host star metallicity is a key constraint for planet formation models, as stars with different metallicities sample disks with a different initial reservoir of solid building blocks. The positive scaling between giant planet occurrence rate and host star metallicity has been firmly established \citep[e.g.,][]{santos04,fischer05,Fulton2021}.

The metallicity dependence of sub-Neptune\index{Sub-Neptunes} planets is more complex. The transiting planets identified with {\it Kepler}  are found around a wider range of stellar metallicity than is typical for giant planets \citep{buchhave12,Buchhave2014,Everett2013}. A dependence of planet occurrence for super-Earths on host star metallicity is significantly weaker \citep{WangFischer2015} or perhaps absent \citep{Mulders2016,KutraWu2021}. The metallicity dependence also varies with system properties:
larger planets (2-4 $\rearth$) have a positive metallicity dependence \citep{Dawson2015,Petigura2018} though weaker than that of giant planets. 
This metallicity trend disappears for smaller planets of $1-2\, \rearth$. On top of that there is an additional dependence on orbital period: planet occurrence rates are higher around high metallicity stars within an orbital period of $\approx10$ days \citep{Mulders2016,Petigura2018}. This is also observed as an increase of the planet host star metallicity at $8.5$ days \citep{Wilson2018}. 
The origins of this trend in orbital period is still unclear.

\subsubsection{System architectures}\label{s:exo:multi}

Despite the large diversity in planet properties, the properties of planets orbiting the same stars tend to cluster together, giving rise to a common archetype of planetary system. The {\it Kepler}  multi-planet systems tend to be regularly spaced, of similar size, and with low mutual inclinations \citep{BallardJohnson2016,Millholland2017,Weiss2018,Gilbert2020}. 
While some of these correlations could arise from detection biases, forward models of planetary system populations show that observations are consistent with planet radii, orbital periods, and inclinations, that are intrinsically clustered \citep{He2019}.

A key feature of the multi-planet systems is that the majority are not in orbital resonances. The observed period ratio distribution is broad, 
implying that if planets had migrated in resonant chains,  those chains must be broken after migration  \citep{Izidoro2017} with only some remaining in near-resonant configurations. 

The architectures of giant planet systems are significantly different. Hot Jupiters are typically single, with only a few examples that have close companions. On the other hand, companions to hot Jupiters at distances larger than 50 au are common \citep{Ngo2016}.
The multiplicity of longer period giant planets is less well determined, though multi-planet systems are common in RV surveys, with an average of 1.5 detected planets per system \citep{Mayor2011}.

Whether giant planets reside in the same systems as super-Earths provide strong constraints on planet formation models, but is difficult to constrain observationally as transit and radial velocity surveys typically target different groups of stars. Nonetheless, studies that have looked at both types of planets typically find that they are correlated to some degree. \cite{Zhu2018} and \cite{Bryan2019} found that stars with super-Earths are more likely to host distant giant planets than random field stars. On the other hand, \cite{Barbato2018} find no super-Earth companions to radial velocity giants. Finally, \cite{Rosenthal2021b} quantified the presence of inner super-earths and outer giant planets and found them to be correlated in 40\% of cases.
Based on transit data alone, \citet{CHuang16} show that warm Jupiters in {\it Kepler} are often accompanied by super-Earths. Single transiting giant planets from {\it Kepler} also appear to be part of systems with known planets\index{Exoplanets|)} \citep{Herman2019}.

\subsection{\textbf{New revelations from circumstellar disks observations}}\label{s:disks}

While the exoplanet population represents the outcome of planet formation, the disks surrounding young stars constrain the initial stages of this process. Since the last {\it Protostars \& Planets} series, a major effort in characterizing the protoplanetary disks\index{Protoplanetary disks} in nearby star-forming regions has gone underway. Particularly, the Atacama Large Millimeter/submillimeter Array (ALMA)\index{Atacama Large Millimeter Array (ALMA)} has revolutionized our understanding of protoplanetary disks by providing their very first high-resolution images. In this section, we summarize the most important aspects of disks observations that influence planet formation theories. 

\subsubsection{Disk substructures}

In the standard picture, the circumstellar disks were interpreted as smooth, with the density decreasing monotonically as a function of the distance from the central star. In recent years, the high-resolution images of disks obtained in the millimeter continuum emission with ALMA, as well as in the scattered light with the Very Large Telescope (VLT)\index{Very Large Telescope (VLT)}, have identified substructures occurring at effectively all spatial scales down to the current resolution limits. 
These substructures\index{Protoplanetary disks!substructures} take various forms, such as deep dark gaps, bright rings, arcs, and spirals (see the recent review by \citealt{AndrewsReview2020} as well as the chapter by \citealt{ppvii_Benisty}). 
Ongoing planet formation is probably the favored, although not the only, explanations for these substructures seen in Class II disks \citep[see, e.g.,][]{rdong15,kezhang15,teague18,ppvii_Bae}. 

It was traditionally thought that the structure of Class II\index{Class II} disks ($> 10^6$ years, see, e.g., \citealt{ltesti14}) provide good initial conditions for planet formation theories. In the past few years, it has been increasingly clear that planet formation starts much earlier, before the disk is fully assembled. 
Importantly, a significant amount of substructure is also seen in the observable dust emission in some embedded protostellar systems. A prime example of this is the young source HL Tau\index[obj]{HL Tau} in the Taurus star-forming region, which was the first imaged with the highest resolution \citep{hltau2014}. HL Tau is still surrounded by an extended envelope, so it is categorized as an extremely young Class I source \citep[see also][]{sheehan18}. Oph~IRS 63 is also a Class I protostar in the Ophiuchus star-forming region. \citet{seguracox19} reported four annular rings in the 60--150 $M_{\earth}$ dusty disk around Oph IRS 63 surrounded by a $\sim$ 100 $M_{\earth}$ envelope. These younger stellar systems have ages of a few $10^5$ years \citep{dunham14, kristensen18}. While the embedded disks are still not well characterized \citep[see, e.g.,][]{vanthoff20}, it now seems clear that they, and not the Class II disks, provide the initial conditions of planet formation.

\subsubsection{Mass budget for planet formation}\label{sub:disk:masses}

\begin{figure}[t]
 \centering
 \includegraphics[width=\linewidth]{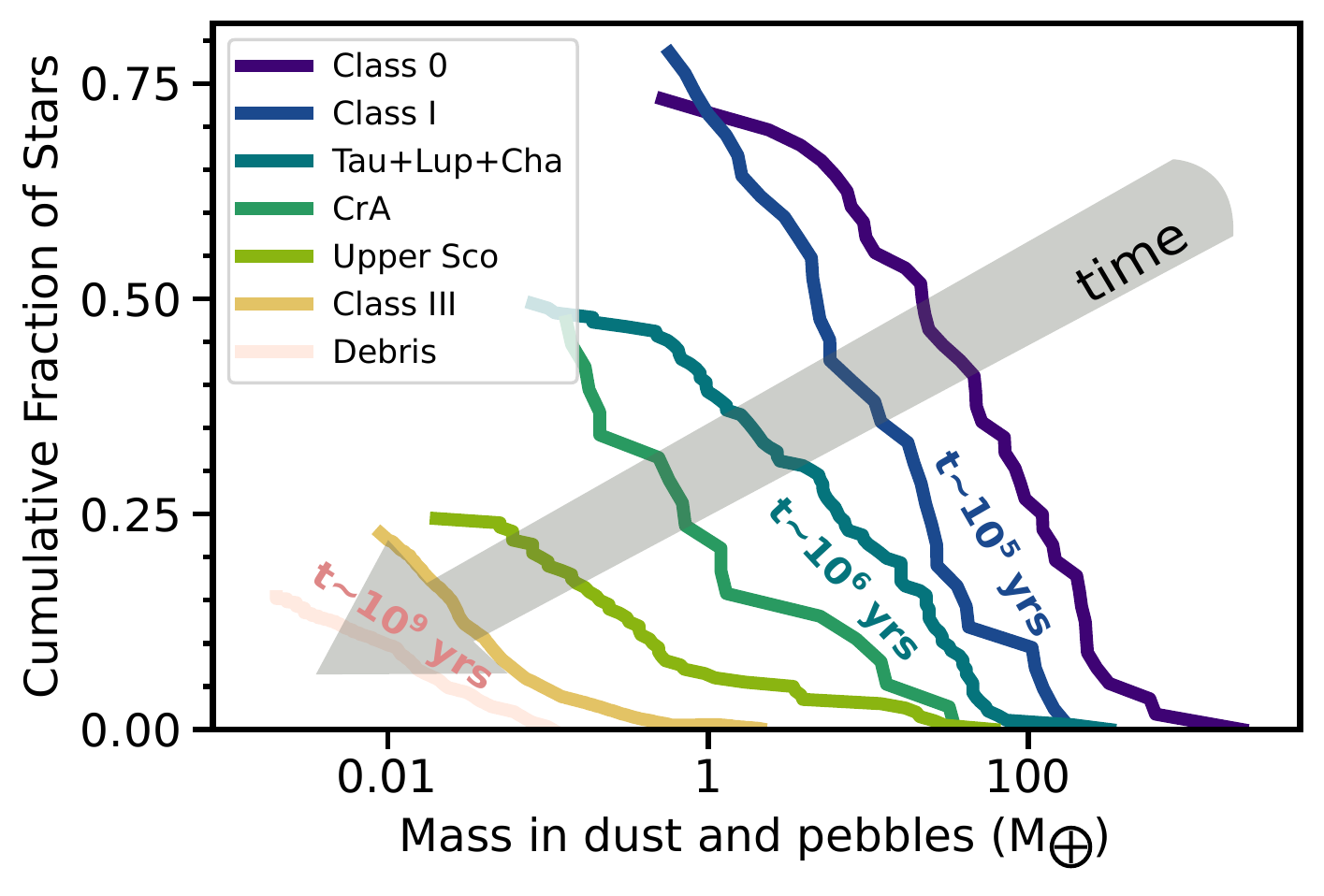}
 \caption{Cumulative distribution of dust mass around young stars at different stages of disk evolution. Stars without infrared excess are accounted for by normalizing the observed cumulative distribution of disk mass to the disk fraction. The mass distribution of Class 0/I disks is adopted from \citet{Tychoniec2020} while the Class II disks are from \citet{vdmarel21}. The older disks are taken from \citet{michel21} and \citet{holland17}. The typical ages of each sample are indicated, while debris disks have a wide range of age \citep{AMHuges18}. 
 Both the dust mass and the fraction of stars with detectable dust decrease over time.
 \label{fig:massevol}}
\end{figure}

Estimating the dust mass\index{Protoplanetary disks!mass} from dust continuum observations is not trivial (see the chapter by \citealt{ppvii_Miotello} for details). Recent surveys of protoplanetary disks in the Chameleon \citep{Pascucci2016}, Taurus \citep{flong18}, Upper Sco \citep{barenfeld16}, Perseus \citep{Tychoniec2020}, Ophiuchus \citep{ODISA}, and Orion \citep{tobin20Orion} have estimated dust masses, which we show in Fig.~\ref{fig:massevol}. The average mass of solids in Class 0\index{Class 0} disks is on the order of a few hundreds of Earth masses. The available solid mass is rapidly depleted between the young Class 0 and I\index{Class I} disks and Class II\index{Class II} disks. A significant fraction of the available solids becomes either lost or bound into larger objects within the first million years of disk evolution, the latter being an additional indication for the early start of planet formation.

While the dust continuum observations measure the solid content of disks, gas observations are necessary to estimate the dust-to-gas\index{Dust-to-gas ratio} ratio and the available gas that can be accreted to become the planetary atmosphere. Typically, a total disk mass (gas + dust) is estimated under the assumption of a gas-to-dust ratio of 100 \citep[see, e.g,][]{bergin13HD, trapman17}. However, a single-dish survey of rotational transitions of hydrogen deuteride (HD) finds that most Class II disks have low gas masses with gas-to-dust ratio around 50 and as low as 10 \citep{kama20}. The approximation of 50--100 still holds pretty well against observations toward the old TW Hya\index[obj]{TW Hya} disk \citep[3 -- 10 Myr old,][]{vacca11, bergin13HD}. Nevertheless, it is still challenging to derive the exact gas-to-dust ratio since HD observations can only be done from space. 

Rotational transitions of carbon monoxide (CO) are used to trace the bulk gas mass of a protoplanetary disk. While CO is easy to observe, finding the conversion between CO flux to a total gas mass is a challenging task \citep[see, e.g.,][]{meeus10, thi14,miotello17}. Typically, the observed integrated CO flux density is converted to CO gas mass by considering a local thermal equilibrium and an excitation temperature 15--20 K \citep[see, e.g.,][]{delaVillarmois2019,booth20}. Recently, radiative transfer models that are coupled to chemical models are used to determine the relation between the observed CO flux and gas mass \citep[see e.g.,][]{williamsbest14,miotello16}. By applying these type of models, CO based observations derive a global gas-to-dust ratio of 1--10 \citep{miotello16,flong17,gabellini19,favre19} which are lower estimates from HD. 

Overall, there is a high uncertainty in the mass determination of disks from both gas and dust observations. Dust scattering due to the 10 -- 100 microns dust grains can lead to an underestimation of solid mass \citep{zhzhu19}. However, the dust scattering is challenging to include in the radiative transfer methods (see \citealt{steinacker13} for a review) and thus this topic is still largely unexplored, and it is unclear if, and by how much, the disk masses are underestimated. Nevertheless, the consensus is that massive disks that are prone to be gravitationally unstable are rare \citep{kama20}. 

\subsubsection{Sizes of protoplanetary disks}\label{sub:disksizes}

Protoplanetary disks form in the early stages of star formation and grow with time as found in numerical models \citep{Hennebelle2016} and observed with a limited inhomogeneous sample (\citealt{Najita18}, but see \citealt[][their Sect. 2.3]{ppvii_Pascucci}). The size\index{Protoplanetary disks!Radius} evolution of disks has been typically fitted with viscously evolving disks \citep{manara19b}. The initial snapshot observations already show that the disk population is diverse \citep{Pascucci2016,ansdell16}. A handful of well-known bright disks are large with the dusty millimeter disk extending up to 200~au. In the ALMA\index{Atacama Large Millimeter Array (ALMA)} surveys, the occurrence rate of large ($\simeq 100$ au) Class II\index{Class II} dusty disks is about 15\% \citep{sanchis21}. 

The survey in the Taurus star-forming region shows that the size of Class II dusty disks with substructure can vary between 40 and 200 au \citep{flong19}. However, the unexpected result from these observations is that most protoplanetary disks are compact. This seems to be a trend in most star-forming regions which can be drawn from the dust continuum images \citep{Pascucci2016, ansdell16,ansdell17,ansdell20}. 

Similar observational techniques have been applied to the younger Class 0\index{Class 0} and I\index{Class I} sources \citep[e.g.,][]{Tychoniec2020}. Recent surveys by \citet{tobin20Orion} and \citet{sheehan20} also find that large disks are rare at these youngest stages \citep[see also,][]{maury19, ppvii_Tsukamoto}. On the other hand, it is still unclear that these compact Class 0 and I dusty disks are Keplerian disks due to the lack of high-sensitivity gas observations. Gas observations of Class 0/I disks do indicate that these disks are around 80--100 au in size \citep{murillo15,harsono18,delaVillarmois2019}. On the other hand, the emission from the more embedded protostars tend to be dominated by infalling motion (with a hint of rotational motion, see, e.g., \citealt{bjerkeli19}). High sensitivity gas observations are needed to determine the outer radii of the young embedded disks that can be compared to Class II disks. 

Disk sizes measured in the infrared through scatter light also probe the gas structure of disks. Gas observations with ALMA\index{Atacama Large Millimeter Array (ALMA)} indicate that the gaseous Class II disks are typically larger in gas than in dust \citep{sanchis21}. It indicates that there is a separation between small grains and large grains distribution in protoplanetary disks. 

\subsubsection{Observational evidence of dust growth and drift} 

The growth of dust grains can be studied through the dust continuum spectral index, as long as the dust emission is optically thin \citep[see, e.g.,][]{ltesti14}. The observed spectral index for Class II disks are commonly interpreted as an evidence for dust growth from micron to at least millimeter sizes. Dust growth in protostellar envelopes was also inferred through the dust spectral index \citep{Galametz2019}. Recently, complementary constraints on particle sizes became available from the linear polarization of self-scattered millimeter continuum emission \citep[see, e.g.,][]{Kataoka2015}, which suggest maximum grain size of $\sim 100$~$\mu$m. \citet{Sohashi20} showed that the inconsistency can be solved if the millimeter dust emission is optically thick rather than thin. The general consensus from modelling the dust emission is that the ALMA\index{Atacama Large Millimeter Array (ALMA)} results can also be explained with optically thick emission and dust grains with high scattering albedo which means in general the dust grains are between 50 -- 100 $\mu$m \citep{tazzari21}. Longer wavelength observations with the VLA at centimeters wavelengths are necessary to constrain the presence of millimeter-sized grains \citep[e.g.,][]{carrascogonzalez16}.

A prominent consequence of dust growth is the radial drift\index{Dust!radial drift} (see \S\ref{sect:theory} for theoretical perspective). As we mentioned above, the observed extent of the millimeter-sized grain is smaller than the gas. This difference can be partially explained with the optical depth effects, with gas lines being more optically thick than the continuum emission of dust. However, the inward drift is required to explain the disks in which the difference between the gas and dust radii is larger than factor of two \citep[see, e.g.,][]{facchini17, trapman20}. The disk around CX Tauri\index[obj]{CX Tau} has the gas radii extending five times further than the millimeter dust grains \citep{facchini19}. It could be that CX Tau is the norm and the current gas observations are not sensitive to detect the outer gas radii, for example, if the gas at the outer radius is too cold for CO to emit efficiently \citep[e.g.,][]{carney18} or it is obscured by the foreground cloud. 

Another evidence for radial drift comes from the observation that the dust emission at disk outer edge decreases sharply with radius, which was predicted by theoretical models including dust growth and drift \citep{Birnstiel2014, cleeves16}. However, there is increasing evidence that disk sub-structures which are decreasing the efficiency of radial drift must be common coming from comparison of the observational data of disks population and theoretical models \citep{Toci2021, Zormpas2022}.

\subsubsection{Turbulent strength}\label{sub:turb}

It is expected that the circumstellar disks are turbulent\index{Protoplanetary disks!turbulence}, either because of the magneto-rotational instability, or due to hydrodynamic instabilities. The turbulent strength is an important parameter in the planet formation models and is typically described with parameter $\alpha$ (see \S\ref{sect:theory} and \S\ref{sect:models}). The most direct way of estimating turbulence comes from measuring the non-thermal broadening of molecular emission lines. Using this method, \citet{Flaherty2015, Flaherty2018, Flaherty2020} and \citet{Teague2016, Teague2018b} found evidence for moderate turbulence levels ($\alpha \lesssim 3\cdot10^{-3}$) in the disks of TW Hya\index[obj]{TW Hya} and HD 163296\index[obj]{HD 163296}, but much higher turbulence ($\alpha \approx 10^{-1}$) for DM Tau\index[obj]{DM Tau}. Another approach to estimating turbulence comes from geometrical arguments. \citet{Pinte2016} investigated the possible height of the dust layer based on the well-known high-resolution image of HL Tau\index[obj]{HL Tau} and found $\alpha\approx3\cdot10^{-4}$. A similar conclusion was reached by \citet{Dullemond2018} who quantified the width of dust rings seen in the DSHARP sample. In line with these findings, \citet{trapman20} showed that the observed disk sizes are consistent with the viscous model with $\alpha\sim 10^{-4} - 10^{-3}$. \citet{Villenave2022} reported even lower turbulence ($\alpha\lesssim10^{-5}$) in the disk around Oph~163131\index[obj]{Oph 163131}. These values of $\alpha$ are significantly lower than predicted in the classical magneto-rotational instability picture, which gave $\alpha\approx10^{-2}$.

It is important to stress that the modern disk models predict that the turbulence strength may not be homogeneous throughout the disk, and thus it cannot be described with a single $\alpha$ value. In particular, the value of $\alpha$ may be different for different vertical layers of the disk, and the disk midplane (probed by the ALMA observations) is thought to have lower turbulence than the upper layers as it is shielded from ionizing radiation \citep[see, e.g.,][]{Lesur2014}. It was also suggested that the turbulence strength may vary radially, for example across the evaporation fronts, due to the change of dust properties \citep[see, e.g.,][]{Kretke2007}. More information on the disk structure models can be found in the chapter by \citet{ppvii_Lesur}.

\subsubsection{Comparison to exoplanet population}

We have been seeing a variety of signatures of on-going planet formation since the past {\it Protostars \& Planets} series, which help us to better understand planet formation. The timescale of planet formation may be recorded in the evolution of circumstellar disks, which is indicated by the comparison between the younger, Class 0/I\index{Class 0} objects and the older, Class II/III systems\index{Class II}. The efficiency of planet formation may be deduced from the comparison of the disk population with the exoplanets\index{Exoplanets}. Yet, we have to be careful when drawing conclusions from such comparisons because of inherent biases. In particular, such comparisons require a careful analysis of the central protostar properties.

Based on the gas kinematical information, typical protostellar masses of Class 0/I protostars are around 0.01 -- 2 solar masses \citep{oya16,vanthoff20}. From gas and dust surveys of Class II disks, the stellar masses are between 0.2 -- 2 solar masses \citep{pegues21}. Since the protostellar masses are in the same range, these two populations should be comparable. Therefore the fast mass evolution shown in Fig.~\ref{fig:massevol} may be treated an effect of planet-forming processes. A significant fraction of dust and pebbles is either turned to larger objects or lost due to radial drift on the timescale of 1 Myr. 

Most of the disks that have been observed with ALMA\index{Atacama Large Millimeter Array (ALMA)} are surrounding K and M type stars with their mass range indicated previously \citep{Pascucci2016, ansdell16, ODISA,flong18, Andrews2018}. A concern is stellar mass bias: most stars with disks are low in mass (as these are the most numerous stars in the nearby observable star forming regions), both in the Class II and 0/I phases. On the other hand, most exoplanets are detected around solar-mass stars (as a consequence of survey design). Nevertheless, it appears Class II disks are short in mass to represent the initial conditions of exoplanet formation at all stellar masses. Around sun-like stars, the known planet populations represent a solid mass reservoir of similar or larger magnitude as Class II disks \citep[][see also Fig.~\ref{fig:massbudget}]{Najita14,Manara2018,Mulders2021}, thus requiring a higher solid mass of Class 0/I disks in order to account for planet formation and the possible existence of planets outside detection limits. This is because the state-of-the-art planet formation models predict that not all of the initial mass of dust can be eventually bound in planets, as we describe in the following sections. 

%%%%%%%%%%%%%%%%%%%%%%%%%%%%%%%%%%%%%%%%%%%%%%%%%%%%%%%%%%%%%%%%%%%%%
\section{\textbf{PLANET FORMATION THEORY}}\label{sect:theory}

\begin{figure*}[ht]
 \centering
 \includegraphics[width=0.9\linewidth]{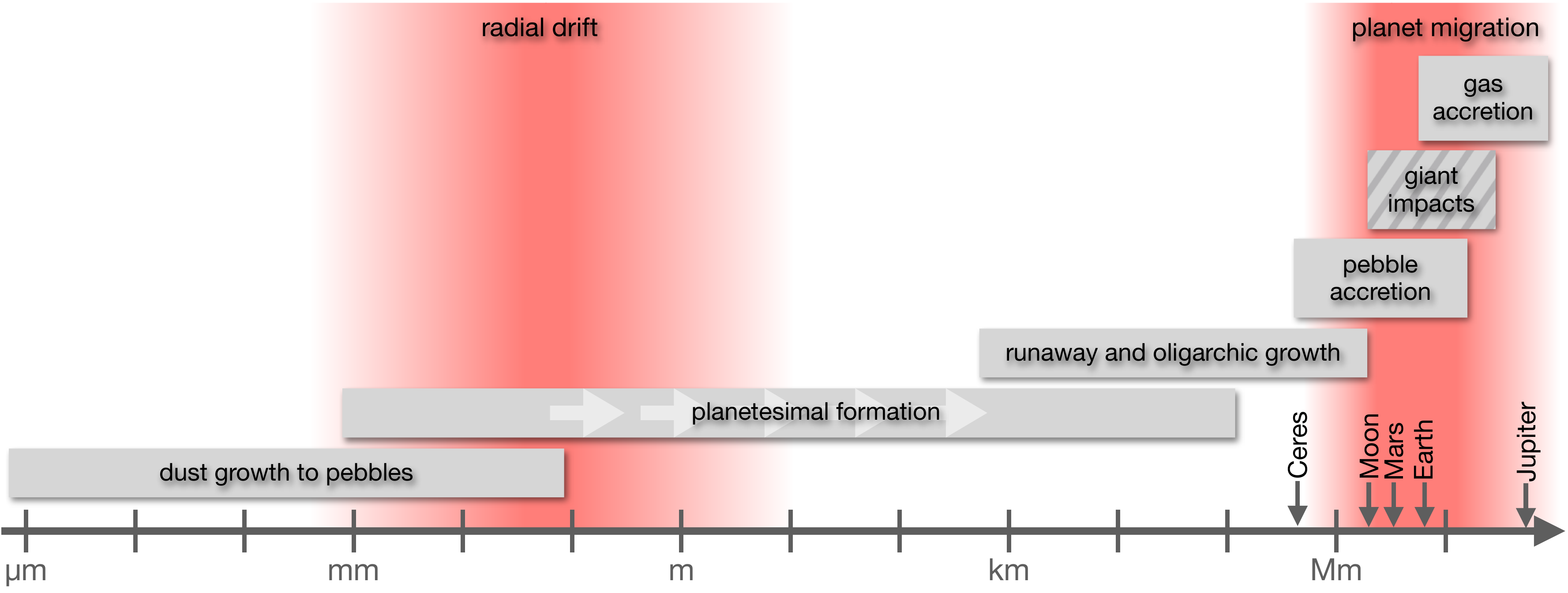}
 \caption{{Overview of processes involved in planet formation and size scales at which they operate. The grey boxes represent growth processes and the red shaded regions represent where the radial mass redistribution is effective. The white arrows in the planetesimal formation box represent the lack of the intermediate sizes during the collapse of pebble clouds. The giant impacts may happen both before and after the protoplanetary gas disk disperses, while the other processes happen in the presence of gas. }
 \label{fig:sizescales}}
\end{figure*}

In this chapter, we present the emerging paradigm of planet formation: the multistage process connecting the micron-sized protoplanetary dust and the populations of planets. Figure~\ref{fig:sizescales} introduces the size scales and processes we are discussing below.

\subsection{\textbf{Dust growth and planetesimal formation}}\label{sect:dust}

This section discusses the evolution of protoplanetary dust in the context of subsequent planet-forming processes. For a more comprehensive description of the dust evolution, we refer the readers to the review presented by \citet{Birnstiel2016}.

\subsubsection{Dust evolution}\label{sub:dustevo}

Theoretical studies of dust evolution\index{Dust!evolution|(} rely on two pillars: the understanding of protoplanetary disks structure and the understanding of collisional physics of pre-planetary dust aggregates. The former is being developed by the synergy of observations of young stellar objects, theoretical models, and radiation transfer models. The latter one is provided by laboratory experiments, recently summarized by \citet{Blum2018}, and molecular dynamics models. Already at the time of {\it Protostars \& Planets VI}, it has been clear that the dust growth to km-sizes is severely hindered by the growth barriers: bouncing, fragmentation, and radial drift\index{Dust!radial drift|(}.

\begin{figure}[h]
 \centering
 \includegraphics[width=\linewidth]{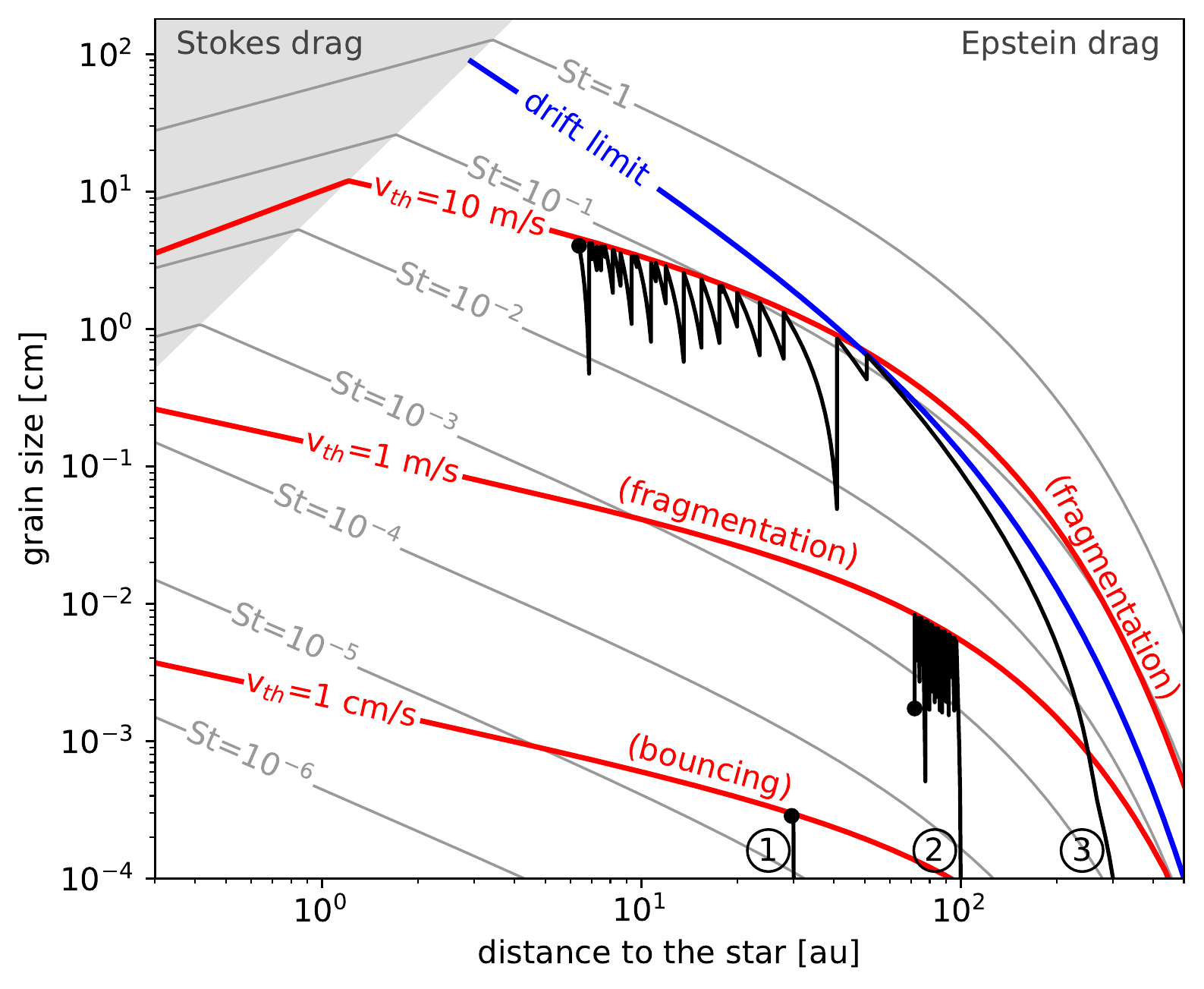}
 \caption{{Illustration of dust evolution in a standard protoplanetary disk model. The grey lines show dust aggregate size with internal density of $\rho_\bullet=1$~g/cm$^3$ corresponding to Stokes number from $10^{-6}$ to unity, depending on the orbital distance. The shaded area indicates the Stokes drag regime. The blue line shows the drift limit, above which the coagulation timescale becomes longer than the radial drift timescale, calculated using Eq.~18 in \citet{Birnstiel2012}. The drift limit does not apply in the Stokes drag regime, where coagulation is always faster than drift. The black lines show evolution tracks of three test particles that have different sticking threshold from 1~cm/s for particle 1), 1~m/s for particle 2), to 10~m/s for particle 3). The red lines show the maximum grain size that dust coagulation\index{Dust!coagulation} attains for these values of the threshold velocity. }
 \label{fig:barriers}}
\end{figure}

Dust evolution is driven by its interaction with the surrounding gaseous disk, which leads to their radial redistribution. Solid particles are embedded in a gas disk that is typically rotating with a slightly sub-Keplerian velocity due to pressure support. Dust aggregates, which in a gas-free disk would follow Keplerian orbits, interact with the gas via aerodynamic drag, lose angular momentum, and spiral towards the central star.

A particularly useful parameter describing the dynamic behavior of dust aggregates in protoplanetary disk is the Stokes number ${\rm St}$. It represents the timescale on which a dust grain couples to the flow of surrounding gas $t_{\rm stop}$ in terms of the local orbital timescale: ${\rm St} = t_{\rm stop} \Omega_{\rm K}$, where $\Omega_{\rm K}$ is the local Keplerian frequency. The smaller the Stokes number, the tighter the coupling. In the bulk of the protoplanetary disk, the Epstein drag regime is valid for dust aggregates and small grains so that the size of the aggregate ($a$), its internal density ($\rho_\bullet$), and its Stokes number ($\mathrm{St}$) are connected by  
\begin{equation}\label{eq:St}
    \mathrm{St} = \frac{\pi}{2} \frac{a \rho_\bullet}{\Sigma_{\rm{gas}}},
\end{equation}
where $\Sigma_{\rm{gas}}$ is the local vertically integrated density of gas. Figure~\ref{fig:barriers} illustrates the connection between the Stokes number and grain size as a function of orbital distance in a disk following the self-similar profile of \citet{LyndenBell1974} with total mass of 0.05~M$_{\odot}$ and critical radius of 100~au. In the astrophysical context, we define pebbles\index{Pebbles} as particles that become marginally coupled to the gas, corresponding to the Stokes numbers are in the range between $10^{-3}-10^{-2}$ and unity. This corresponds to centimeter-sized and larger particles at 1~au, but at 100~au already sub-millimeter grains are pebbles in this aerodynamical context. 

In a typical gas disk, where both the gas density and temperature decrease with the distance from the central star, the radial pressure gradient is negative throughout the disk, leading to sub-Keplerian rotation of the gas. The difference between the Keplerian rotation $v_{\rm K}$ and the gas azimuthal velocity $v_{\rm \phi,gas}$ is often described with the use of the parameter $\eta$: $v_{\rm \phi,gas} \approx v_{\rm K}(1-\eta)$. The value of $\eta$ depends on the strength of the radial pressure gradient. Here, we use the convention, in which a negative pressure gradient translates into a positive value of $\eta$. At the same time, the negative radial speed indicate inward and positive radial speed outward transport. The radial speed of dust and pebbles is the result of the loss of angular momentum driven by the interaction with the sub-Keplerian gas and the entrainment of dust in the radial gas flow, which becomes important for small dust aggregates:
\begin{equation}\label{eq:vdrift}
    v_{\rm r,solid} = \frac{v_{\rm r,gas} - 2\eta v_{\rm K} {\rm St}}{1 + {\rm St}^2},
\end{equation}
where {\rm St} is the Stokes number (see Eq.~\ref{eq:St}). The above equation is valid in the test-particle limit, when the solids-to-gas ratio is negligible. In case of enhanced abundance of solids, the radial velocity depends on the solids-to-gas ratio (and size distribution, \citealt{Nakagawa1986, Tanaka2005}). 
The radial drift velocity increases with Stokes number and reaches a maximum at $\rm{St}=1$. If the timescale for radial drift is shorter than the dust growth timescale then dust growth is effectively halted -- this barrier to growth is referred to as the the radial drift barrier, as shown in Fig.~\ref{fig:barriers}. 

The radial drift and other drag-induced sources of relative velocity (azimuthal drift, vertical settling, and differential coupling to turbulence) drive collisions between aggregates of different Stokes numbers. As particles grow and decouple further from the gas they can attain larger relative velocities compared to other particles of the same size. Thus, the collision velocities increase with dust grain size, eventually leading to disruptive collisions. The majority of dust evolution models rely on the so-called $\alpha$-disk model, in which the circumstellar disk is an accretion disk with global, isotropic turbulence, whose strength is expressed with a parameter $\alpha$. Dust evolution models show that if $\alpha\gtrsim10^{-4}$, consistent with the constraints presented in \S\ref{sub:turb}, the relative speeds driven by turbulence dominate over the other sources of collisional speeds in the disk midplane \citep[see, e.g.,][]{Birnstiel2012}. Collisions with speeds below so-called threshold velocity v$_{\rm{th}}$ lead to sticking and growth of dust aggregates. There is a significant discrepancy in the value of the threshold velocity reported by various authors. \citet{Guettler2010} reported values of around 1~m/s for silicate grains while \citet{Gundlach2015} suggested the threshold velocities are about 10 times higher for icy grains. \citet{Yamamoto2014} generally found much higher threshold velocities, from~30 m/s for silicates to~80 m/s for icy grains, which would mean that the drift barrier is more restrictive than the fragmentation barrier.

In Fig.~\ref{fig:barriers}, we assume a typical value of $\alpha=10^{-3}$ and plot evolutionary tracks of three test particles undergoing radial drift and growth corresponding to three different values of v$_{\rm{th}}$, in the range typically reported by laboratory experiments. The evolution of each of these particles is integrated for 0.8~Myr assuming a constant dust-to-gas\index{Dust-to-gas ratio} ratio of 0.5\% (in more realistic models, the dust-to-gas ratio would be a function of time and location in the disk and would depend on the evolution\index{Dust!evolution|)} of these particles). In the pessimistic case of particle 1), when the growth stalls due to bouncing, a possible outcome where particle collisions lead neither to growth or fragmentation \citep[see, e.g.,][]{zsom2010} at the velocity of 1~cm/s, grain growth is halted at a couple of~$\mu$m, when the particle is still well-coupled to the gas. This particle grows to its final size at its initial location of 30~au and stays there for the duration of the model as its radial drift is slow. In the more optimistic cases of particles 2) and 3), the growth can proceed to centimeter or even decimeter-sized pebbles. These particles reach the fragmentation barrier\index{Fragmentation barrier}, meaning that they periodically lose mass in catastrophic collisions and regrow to the maximum size indicated by the corresponding red line. The mass of the test particle after a fragmenting collision is chosen randomly from the top-heavy mass distribution as reported in experiments \citep[e.g.,][]{BukhariSyed2017}. The larger the particle grows, the faster it drifts. This is why the more sticky particle 3) overtake the less sticky particles 1) and 2). 

\subsubsection{Planetesimal formation by subsequent collisions}

The above paragraphs have introduced the problem of growth barriers that are hindering production of gravitationally bound planetesimals. In general, even if direct growth of planetesimals\index{Planetesimal formation|(} would be possible, due to the radial drift barrier it could only take place in long-lasting pressure traps (see \S\ref{sub:driftmigr}), or close to the star, where the growth timescale may be shorter than the drift timescale (see Fig.~\ref{fig:barriers}). For example, \citet{homma2019} has proposed that grains covered in organic mantles can stick efficiently if the temperature is above 200~K. This scenario could lead to direct growth of planetesimals inside of 1~au, making it possible to form close-in but not the wider orbit planets. In the previous {\it Protostars \& Planets} review, \citet{johansen2014} have discussed three scenarios in which the growth barriers can be overcome and dust growth can proceed to planetesimal sizes.

The first of these scenarios was the possibility of growth by mass transfer in high-speed collisions. It was observed in laboratory experiments that when two dust aggregates of dramatically different masses collide at a high speed ($\gtrsim10$~m/s), the smaller aggregate fragments, yet transfers part of its mass onto the large aggregates that as a result gains mass \citep{Teiser2009}. \citet{Windmark2012} and \citet{Garaud2013} have found that if collision velocity dispersion is included in the coagulation models, some aggregates may grow much larger than the majority of aggregates trapped by the bouncing and fragmentation barriers. These lucky particles would then proceed to grow and could reach planetesimal sizes. However, later works do not support this conclusion. Even if the lucky particles continue to grow past the bouncing and fragmentation barriers, their growth is too slow to prevent them from falling onto the star due to the radial drift\index{Dust!radial drift|)} and no planetesimal growth is found in global models by \citet{Estrada2016}. Furthermore, \citet{Booth2018} have revisited this scenario with an improved numerical method and found that growth beyond the bouncing barrier is unlikely.

The second scenario for planetesimal formation was porous growth. \citet{Okuzumi2012} and \citet{kataoka2013} proposed that collisions between very porous ice aggregates lead to growth past the bouncing and fragmentation barriers due to their efficient sticking even at high collision speeds and enhanced growth rate thanks to their large cross-sections. \citet{Arakawa2016} has applied the same mechanism to rocky planetesimals growth if they start with nanometer-sized silicate grains. \citet{Homma2018} have found that porous growth is unsuccessful in forming planetesimals in the early stages of protoplanetary disk evolution because the young disk is hot and the area where the dust can break through the radial drift barrier is restricted to the region inward of the snowline, where the porous icy aggregates disintegrate by evaporation. Starting with a fully-fledged disk, \citet{Kobayashi2021} found direct growth to planetary sizes inside of 10~au. However, \citet{Krijt2015} and \citet{Schraepler2018} have shown that the growth of porous aggregates is limited by erosive collisions with small projectiles. \citet{Tatsuuma2021} suggested that porous aggregates may be prone to rotational disruption due to gas flow. Also the recent global models including aggregates porosity presented by \citet{Estrada2022a} do not find growth past the fragmentation and drift barrier. It is worth noting, that even if the porous growth does not lead to direct planetesimal growth, it may still help in producing enhancements of dust density needed to trigger planetesimal formation in the hydrodynamic instabilities \citep{Krijt2016}, as discussed in the next section.

\subsubsection{Planetesimal formation by collapse of pebble clumps}

The possibilities of planetesimal formation via direct growth are limited. In the classical scenario of Solar System formation, it was proposed that planetesimals formed by fragmentation of gravitationally unstable dust sub-disk formed by efficient sedimentation \citep{Goldreich1973}. This is however prevented by the Kelvin-Helmholtz instability caused by the vertical shear between the dust sub-disk and gas, and thus a significant radial concentration is necessary \citep{youdin2002}. \citet{Ida2016} and \citet{Hyodo2019,Hyodo2021} showed that this can be possible for silicate grains close to the star. Another possibility is to concentrate solids in between the turbulent vortices which was first proposed by \citet{Cuzzi2001} and recently revived by \citet{Hartlep2020}.

The third scenario proposed by \citet{johansen2014} was planetesimal formation by collapse of self-gravitating pebble clumps triggered by the streaming instability\index{Streaming instability} originally proposed by \citet{youdin05}, which is discussed in more detail in \citet{ppvii_Lesur}. This mechanism is supported by the evidence from the Solar System\index{Solar system}, as the structure of comets and the architecture of trans-Neptunian binaries, particularly their predominantly prograde rotation, is consistent with the streaming instability models \citep{Blum2017, Nesvorny2019}. It was found in those models, that streaming instability leads to formation of mainly large planetesimals, with a characteristic size of 100~km at the main asteroid belt location \citep[see, e.g.,][]{Simon2016, Schaefer2017, Abod2019, Klahr2020}, which varies with the radial distance and the mass of the central star \citep{LiuEtal2020}. However, triggering planetesimal formation in this scenario requires pebbles (St $\gtrsim 10^{-3}$) and an enhanced dust-to-gas ratio\index{Dust-to-gas ratio} (on the order of unity in the midplane, \citealt{Carrera2015, Yang2017, Li2021}). This means that the pebbles\index{Pebbles} which are able to decouple from the gas must grow, settle to the disk midplane, and redistribute radially to increase the solid abundance before the planetesimal formation takes place.

\subsubsection{Global models of planetesimal formation}

\begin{figure}[t!]
 \centering
 \includegraphics[width=0.9\linewidth]{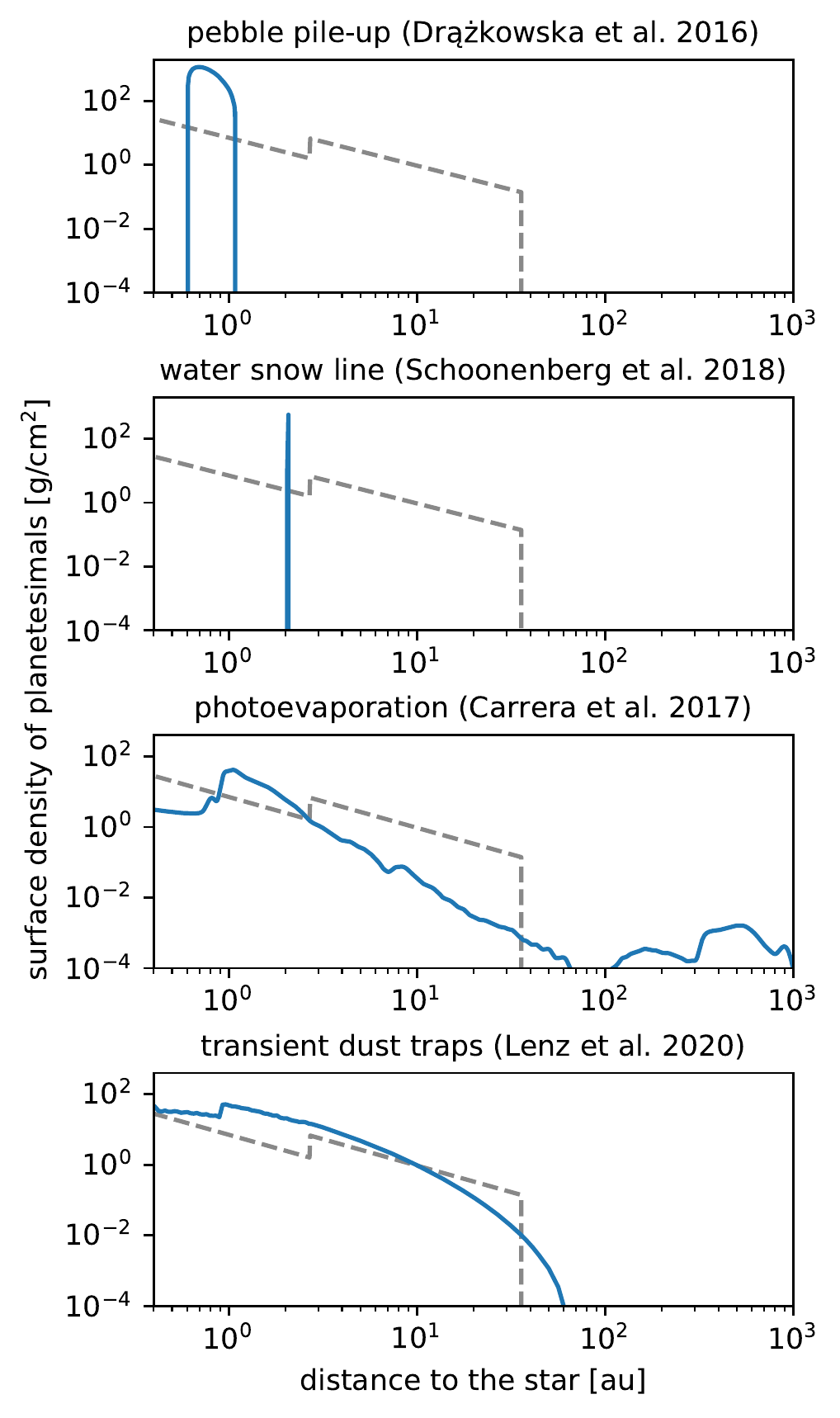}
 \caption{{Surface density of planetesimals obtained in different global protoplanetary disk evolution models. From top to bottom: \citet{Drazkowska2016}: radial drift induced pebble pile-up triggers planetesimal formation in a narrow annulus around 1~au, \citet{Schoonenberg2018}: planetesimal formation is triggered outside of the water snow line, \citet{Carrera2017}: planetesimal formation is triggered by gas removal by photoevaporation, \citet{Lenz2020}: pebble flux-regulated planetesimal formation model assuming existence of temporary dust traps across the disk. The grey dashed line shows the minimum-mass solar nebula (MMSN) distribution from \citet{Hayashi1981}.}
 \label{fig:planetesimals}}
\end{figure}

A recent development in planet formation theory was the emergence of models connecting the global evolution of gas and dust in protoplanetary disks and planetesimal formation via the streaming instability\index{Streaming instability}. We present examples of such models in Fig.~\ref{fig:planetesimals}. As an increased solid-to-gas ratio is required to trigger planetesimal formation, pebbles must be concentrated in some region of the disk. \citet{Drazkowska2016} have shown that redistribution of solids driven by dust growth and radial drift in a standard, viscous disk model, may lead to a pile-up of pebbles in the inner disk. This is because the pebbles surface density in the fragmentation-dominated close-in part of the disk (see Fig.~\ref{fig:barriers}) evolves toward a steeper profile than the typical initial condition \citep{Birnstiel2012}. This pebble pile-up locally provides conditions where streaming instability is effective and leads to planetesimal formation in a relatively narrow annulus close to 1~au. However, for this mechanism to be successful the fragmentation threshold velocity must be above 8 m/s, which is higher than many laboratory experiments predict for silicate grains. \citet{schoonenberg2017} and \citet{drazkowska2017} have shown that pebble pile-up and local burst of planetesimal formation may occur just outside of the water snow line if the sticking of ice-rich dust aggregates present outside of the snow line is increased compared to the dry aggregates inside of the snow line. This leads to a global traffic jam effect as the particles inside of the snow line are smaller and thus drift at a slower paste then the large pebbles arriving from the outer disk. Additionally, there is so-called cold finger effect which relies on the diffusion of water vapor from inside to the outside of the snow line, where it is redeposited on the existing icy aggregates, additionally enhancing the solids-to-gas ratio \citep{Cuzzi2004}. \citet{Schoonenberg2018} argued that planetesimal formation may be efficient both inside and outside of the snow line, albeit only in disks that start off with the global solids-to-gas ratio of at least 2\%. Also \citet{Charnoz2019} invoked the possibility of planetesimal formation at and inside of the water snow line if the change of turbulence level (the so-called dead zone\index{Dead zone}) is taken into account. \citet{Morbidelli2022} has that the inner ring of dry planetesimals can also form when efficient particle fragmentation at the silicate evaporation front and recondensation of silicate vapor are taken into account.

The models described in the above paragraph suggest that if the protoplanetary disk is smooth, planetesimals only form locally, in relatively narrow annuli. \citet{Lenz2019, Lenz2020} made an assumption that temporary particle traps that locally enhance the dust-to-gas\index{Dust-to-gas ratio} ratios are emerging throughout the disk. This assumption is motivated by the substructures often seen in the circumstellar disks (see \S\ref{s:disks}) and the instabilities commonly found in hydrodynamic and magnetohydrodynamic models of protoplanetary disks that lead to creation of zonal flows and vortices capable of accumulating dust. This makes the conversion of pebbles to planetesimals possible at various orbital distances, leading to planetesimal distribution closer to those assumed in classical models, such as the MMSN. A different scenario was proposed by \citet{Carrera2017}, in which the solid-to-gas ratio is enhanced thanks to the removal of gas via photoevaporation. This scenario produces a massive planetesimal belt beyond 100~au, but also a significant population of planetesimals inside of 10~au. However, \citet{Ercolano2017} have shown that their results are very sensitive to the photoevaporation profiles assumed in the models.

As visible in Fig.~\ref{fig:planetesimals}, the models consistently following the planetesimal formation process show that the resulting planetesimal surface density distribution may be very different from the initial distribution obtained from a constant solid-to-gas ratio. A significant redistribution of mass may occur before planetesimal formation takes place and the constant planetesimal-to-gas ratio assumed in traditional models is an unlikely outcome. Another interesting aspect of these models is the timescale at which the building blocks of planets form. If planetesimal formation is triggered by gas removal, most of planetesimals form very late in the disk lifetime and thus cannot participate in the formation of cores of gas-rich planets. In the models where planetesimal formation is triggered by pebble pile-ups, the planetesimal formation stage typically lasts 10$^5$~years, although it can extent throughout the whole lifetime of the disk \citep{Drazkowska2018}.

We must stress that planetesimal formation models remain unsatisfactory. State-of-the-art models of dust evolution do not necessarily lead to any planetesimal formation \citep[see, e.g.,][]{Estrada2016, Estrada2022a}. The planetesimal production found in models presented in Fig.~\ref{fig:planetesimals} greatly depends on each model's assumptions, which are not necessarily realistic. Understanding of planetesimal formation\index{Planetesimal formation|)} will be critical to models of planetary cores growth, which we discuss in the next subsection. 

\begin{figure*}[h]
 \centering
 \includegraphics[width=0.9\linewidth]{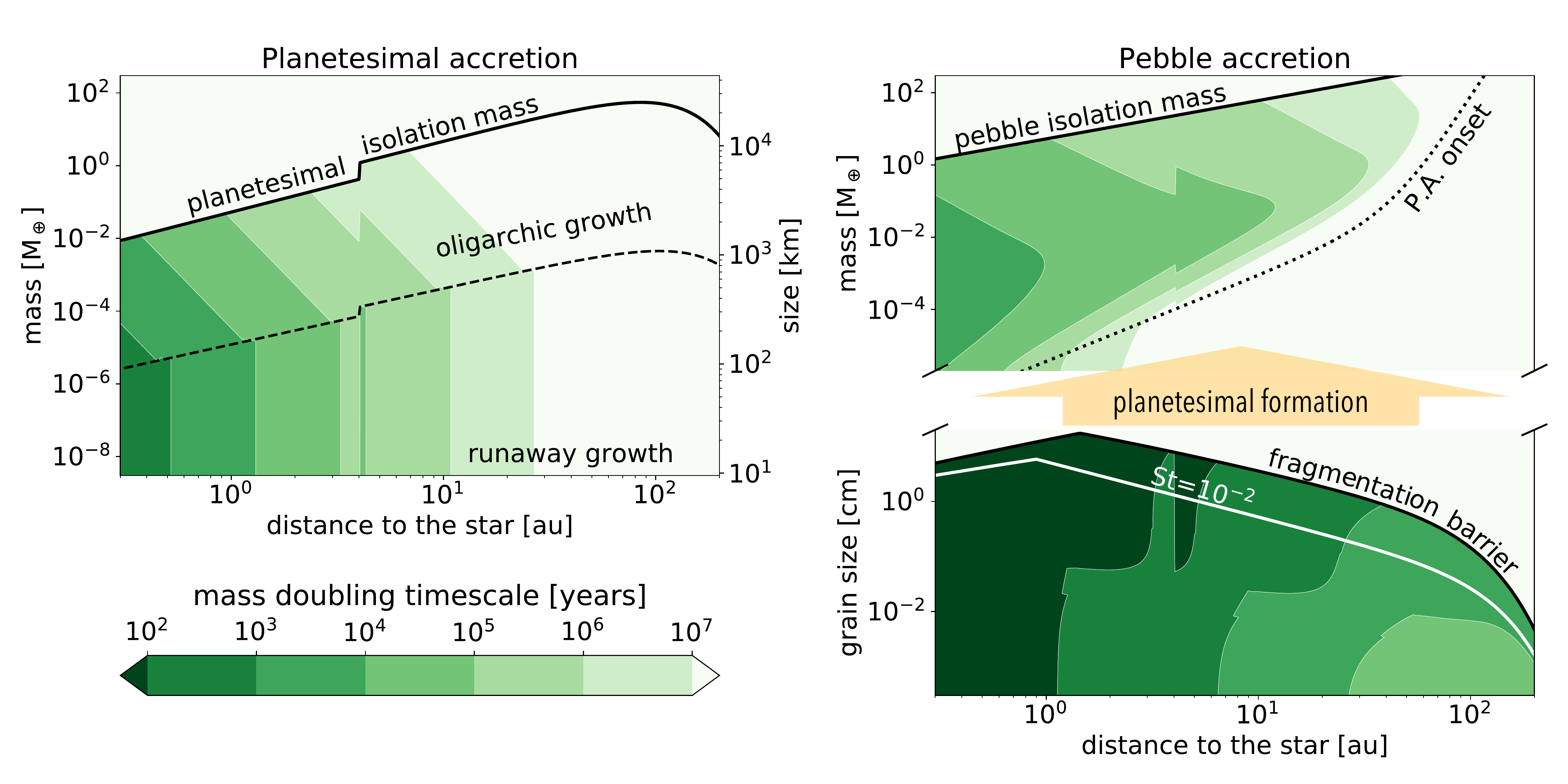}
 \caption{{Theoretical solid accretion timescales as a function of distance to the central star and the accreting body mass. The left panel corresponds to growth by planetesimal accretion and the right panel to pebble accretion (top) and dust coagulation\index{Dust!coagulation} (bottom). An identical disk model with total solids mass of 300 M$_\oplus$, surface density of both gas and solids proportional to $r^{-1}$, and the water snow line located at 4~au was used in both panels. The planetesimal accretion efficiency was calculated assuming 10~km planetesimals. The isolation mass was calculated assuming the planetary core does not migrate. The pebble accretion timescale was calculated assuming a pebble flux of 100 M$_{\oplus}$/Myr inside, and 200 M$_{\oplus}$/Myr outside of the snow line, and a pebble size corresponding to the fragmentation barrier, and agrees well with analytically predicted pebble accretion onset (dotted line).}
 \label{fig:timescales}}
\end{figure*}

\subsection{\textbf{Planetesimal and pebble accretion}}\label{sub:ppaccretion}

Here we review the two basic accretion modes for planet growth: the classical planetesimal accretion and the new paradigm of pebble accretion (see also the reviews of \citealt{Johansen2017, Ormel2017review, Liu2020}). Figure~\ref{fig:timescales} illustrates the main differences of the solid accretion timescales between these two scenarios as a function of the distance to the central star and the mass of the accreting body. Notably, the planetesimal accretion model assumes that all the available solids are in the form of planetesimals while the pebble accretion model hypothesizes that most of the solids stay in the form of pebbles, corresponding to radically different assumptions on the planetesimal formation outcome. The key difference between these two modes is whether the drag force from the disk gas plays a decisive role during their accretion.

\subsubsection{Planetesimal accretion}\label{sub:planetsimalacc}

Planetesimals larger than $10$ km in size are modestly influenced by the aerodynamic gas drag, as long as they are not on highly eccentric/inclined orbits. The gas drag is negligible and only the mutual gravitational force matters during the planetesimal-planet interaction. The accretion\index{Planetesimal accretion|(} is governed by pure gravitational dynamics, which leads to the gravitational focusing effect \citep{safronov1972,wetherill89,lissauer93,kokubo96}, with the collisional cross section  $\sigma$ being enhanced with respect to the pure geometric cross section
\begin{equation}\label{eq:crosssect}
    \sigma = \pi R_{\rm p}^2 \left(1+\frac{v_{\rm{esc}}^2}{\delta v^2}\right),
\end{equation}
where $R_{\rm p}$ and $M_{\rm p}$ are the physical radius and mass of the accreting central planetary embryo, $v_{\rm{esc}}{=}\sqrt{2GM_{\rm p}/R_{\rm p}}$ is its escape velocity, $G$ is the gravitational constant, and $\delta v$ is the relative velocity between the embryo and smaller planetesimals. For a dynamically relaxed distribution of planetesimals, in which the inclinations are similar to eccentricities, the mass accretion rate of the planetary core is given by 
\begin{equation}
    \dot{M}_{\rm{p,plts}} \simeq \Sigma_{\rm{plts}}\Omega_{\rm K}\sigma,
\end{equation}
where $\Sigma_{\rm{plts}}$ is the surface density of planetesimals, $\Omega_{\rm K}$ is Keplerian frequency.

When the random velocities among planetesimals are low such that $\delta v \ll v_{\rm{esc}}$, the gravitational focusing is dominant and the growth proceeds in the runaway regime. The planet mass accretion rate can be described as
\begin{equation}\label{eq:mdotrg}
    \dot M_{\rm p,rg} = \frac{2\pi G \Omega_{\rm K}\Sigma_{\rm{plts}} M_{\rm p} R_{\rm p}}{\delta v^2} \propto M_{\rm p}^{4/3}.
\end{equation}
The runaway growth proceeds faster when the planetesimal density is higher. To calculate the growth timescales presented in Fig.~\ref{fig:timescales}, we assumed a constant planetesimal-to-gas ratio, i.e.~$\Sigma_{\rm plts}\propto r^{-1}$ (with an exponential cut-off outside of 100~au), which is a shallower profile than the commonly used MMSN distribution with $\Sigma_{\rm MMSN}\propto r^{-3/2}$. A MMSN-like profile would have faster growth close to the star but the growth rate would decrease with radial distance even faster than in our example. \citet{ormel10b} found that the growth of the largest bodies happens on a timescale that is a fraction $C_{\rm rg}\simeq0.1$ of the collision timescale among planetesimals. They proposed that the runaway growth timescale can also be expressed as be
\begin{equation}
    \tau_{\rm{rg}} \simeq C_{\rm rg} \frac{\rho_\bullet R_{\rm 0}}{\Omega_{\rm K} \Sigma_{\rm{plts}}},
    \label{eq:Ormel_rg}
\end{equation}
where $\rho_\bullet$ and $R_0$ are the internal density and initial size of the planetesimals. As can be seen from Eq.~(\ref{eq:Ormel_rg}), $\tau_{\rm rg}$ also depends on $R_0$ where smaller planetesimals facilitate faster accretion, as we will show in \S\ref{sub:popsynth}.

The runaway growth phase ends once the growing planetary embryo becomes massive enough to dynamically stir the smaller planetesimals. The growth then transitions to the oligarchic phase, which rate becomes lower as the embryo grows (\citealt{kokubo02}, see the left panel of Fig.~\ref{fig:timescales}). 
This transition occurs at planetary core size corresponding to \citep{ormel10b}
\begin{multline}
    R_{\rm rg/oli} = 580\cdot\left(\frac{C_{\rm rg}}{0.1}\right)^{3/7}\left(\frac{R_0}{10\ {\rm km}}\right)^{3/7} \\ \left(\frac{r}{4\ {\rm au}}\right)^{5/7} \left(\frac{\Sigma_{\rm plts}}{3\ {\rm g/cm^2}}\right)^{2/7}\ {\rm km},
\label{eq:rrgoli}
\end{multline}
and the mass accretion rate in the oligarchic growth regime reads  
\begin{equation}
    \dot M_{\rm p,oli} = \pi \Omega_{\rm K} \Sigma_{\rm{plts}} R_{\rm p}^2 \propto M_{\rm p}^{2/3}.
\end{equation}
The efficiency of planetesimal accretion\index{Planetesimal accretion|)} quickly drops not only with the planet mass, but also with the orbital distance. Hence, in a standard model, forming massive planetary cores outside of the snow line takes prohibitively long compared to the gas disk lifetime \citep{Ida2004}. 
 
\subsubsection{Pebble accretion}\label{sub:pebbleacc}

Accretion of pebbles\index{Pebble accretion|(} with Stokes number between $10^{-3}$ and unity is governed jointly by the gravitational force and gas drag, the latter acting to reduce the angular momentum of a pebble during a pebble-planet interaction. This requires that the stopping time of pebbles $t_{\rm stop}$ is shorter than the encounter time between the pebble and the planet. As assisted by the gas drag, the pebble accretion radius $r_{\rm PA}$ (within which the pebbles are accreted onto the planet) can be substantially larger than the gravitational focusing radius for planetesimal accretion \citep{Ormel2010, Lambrechts2012}. 

The planet mass for which the aerodynamic effects lead to pebbles settling down the gravitational well of the planet, which is equivalent to initiating pebble accretion, can be expressed as \citep{visser2016,Liu2020} 
    \begin{multline}
M_{\rm PA\ onset}= {\rm St} \eta^{3} M_{\star}  \\ =2.5\cdot10^{-4}
\left(  \frac{{\rm St}}{0.1} \right) \left(  \frac{\eta}{0.002} \right)^{3} 
 \left(  \frac{M_{\star}}{M_\odot} \right) \mearth,
\label{eq:onset}
 \end{multline}
This mass is equivalent to a planetesimal of $500$ km in radius, similar to the transition between the runaway and oligarchic growth regime (Eq.~\ref{eq:rrgoli}). As can be seen in the right panel of Fig.~\ref{fig:timescales}, the efficiency of pebble accretion becomes very low when the planet mass is below this value. Since the planetesimals formed by streaming instability\index{Streaming instability} (see \S\ref{sect:dust}) are typically too small to accrete pebbles at a high rate, at least some planetesimal accretion contribution is needed to grow planetary cores capable of quickly growing in the pebble accretion scenario \citep{Liu2019b}.

\begin{figure}[t]
 \centering
 \includegraphics[width=0.95\linewidth]{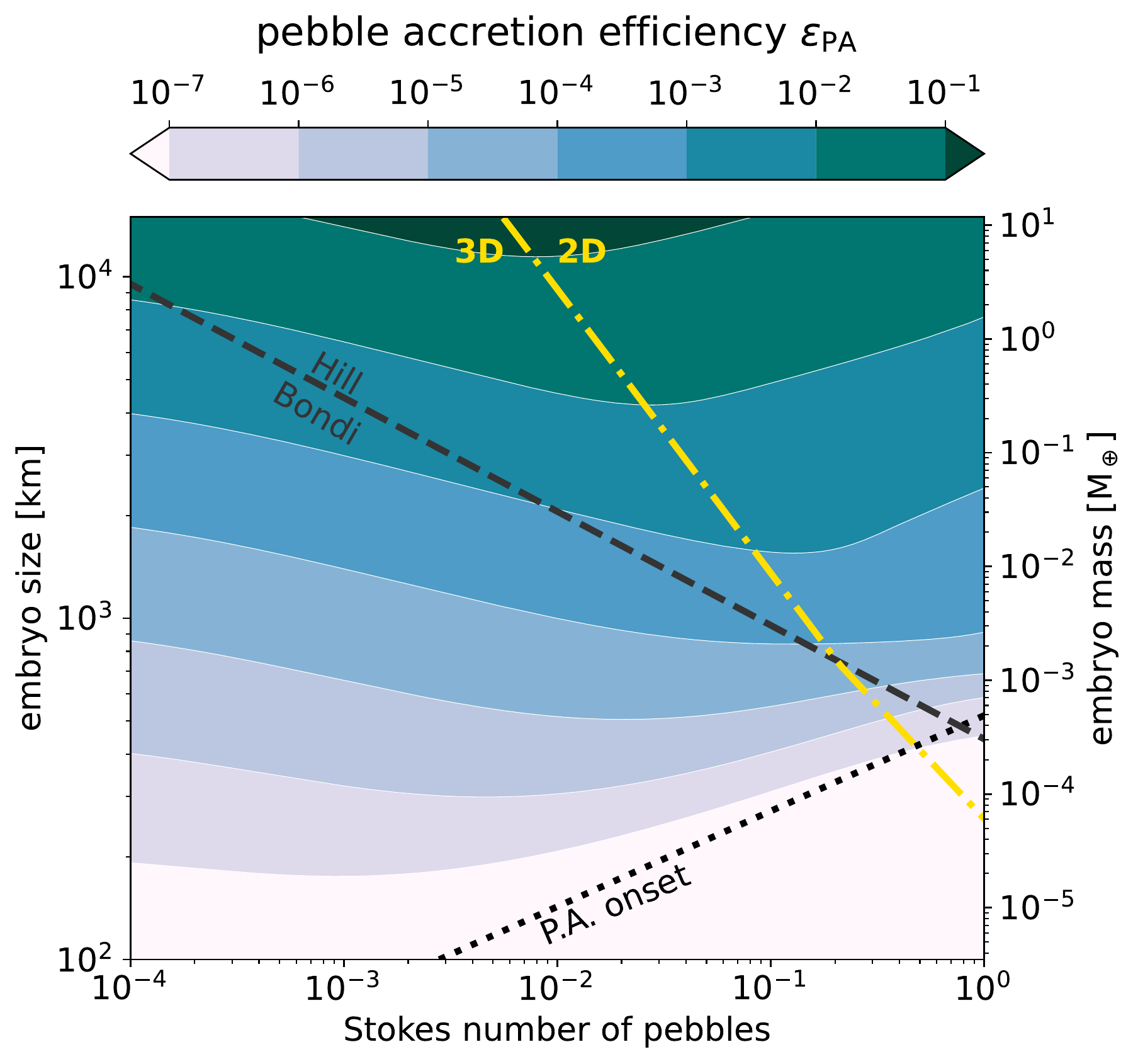}
 \caption{{The efficiency of pebble accretion $\varepsilon_{\rm{PA}}$ as a function of the planetary embryo size and the Stokes number of pebbles at 1~au in a disk with $\alpha=10^{-4}$ and the gas disk scale height of $H=0.038$~au. The dotted line corresponds to the pebble accretion onset. The dashed line shows the transition between the Bondi and Hill accretion regimes. The dashed-dotted line shows the transition between 3D and 2D accretion regimes.}
 \label{fig:PAefficiency}}
\end{figure}

By comparing the pebble accretion radius $r_{\rm PA}$ and the pebble disk scale height $H_{\rm peb}$, pebble accretion can be classified into 2D or 3D regime \citep{lambrechts2014,Morbidelli2015,Idaetal2016}. In the 2D regime ($r_{\rm PA}{>} H_{\rm peb}$), the planetary embryo is large enough to accrete from the complete layer of pebbles, while in the 3D regime ($r_{\rm PA}{<} H_{\rm peb}$) the embryo only has access to a fraction of the pebble layer. The planetary mass accretion rate in those two regimes can be approximately written as:
 \begin{equation}
\dot M_{\rm p,PA}  \approx
  \begin{cases}
  {\displaystyle 
  2r_{\rm PA} \Delta v \Sigma_{\rm peb} = C_{\rm 2D}  \sqrt{{GM_{\rm p} t_{\rm stop} \Delta v }   } \Sigma_{\rm peb} }
      \hfill \hspace{0.1cm}  \mbox{[2D]},   \vspace{ 0.2 cm}\\ 
    {\displaystyle 
  \pi r_{\rm PA}^2 \Delta v \rho_{\rm peb} =C_{\rm 3D} \frac{ 
  GM_{\rm p} t_{\rm stop} \Sigma_{\rm peb}} { H_{\rm peb} }   }   
      \hfill \hspace{0.5cm}   \mbox{[3D]},   \vspace{ 0.05 cm} 
    \end{cases}
    \label{eq:peb_accretion}
\end{equation}
 where $r_{\rm PA}{\sim} \sqrt{GM_{\rm p}t_{\rm stop}/ \Delta v}$, $ \Delta v$ is the relative velocity between the pebble and protoplanet, $\Sigma_{\rm peb}{ =} \sqrt{2\pi} H_{\rm peb} \rho_{\rm peb}$ is the surface density of pebbles, $\rho_{\rm peb}$ is the midplane density of pebbles,  $H_{\rm peb}{=}\sqrt{\alpha_{\rm t}/(\alpha_{\rm t} + {\rm St})}H$, which is set by the gas disk scale height $H$, the coefficient of turbulent gas diffusivity $\alpha_{\rm t}$ and pebbles Stokes number $St$. Depending on the planetary core and pebble sizes, the relative velocity $ \Delta v$ is determined either by gas flow ($\Delta v{=}\eta v_{\rm K}$, Bondi regime) or by the Keplerian shear ($\Delta v {=} r_{\rm PA} \Omega_{\rm K}$, Hill regime). Figure~\ref{fig:PAefficiency} takes into account all the regimes and shows the efficiency of pebble accretion $\varepsilon_{\rm PA}$ defined as the fraction of total incoming pebble flux that is accreted by the protoplanet. To create this plot, we used the pebble accretion efficiency prefactors $C_{\rm 2D}/C_{\rm 3D}$ in Eq.~\ref{eq:peb_accretion} calculated by \citet{liu2018} and \citet{ormel2018} (the Python implementation of this algorithm is publicly available at \url{https://github.com/chrisormel/astroscripts/tree/main/papers/}). This approach is valid for pebbles with $St\leq1$. For larger solids, their capture in protoplanet atmosphere muss be considered \citep{Okamura2021}. 
 
 Many pebble-driven planet formation models found in the literature focus on the $2$D Hill regime (corresponding to the top right corner of Fig.~\ref{fig:PAefficiency}: large pebbles and large planetary cores), in which the mass accretion rate can be expressed in straightforward way as \citep{lambrechts2014}
  \begin{equation}
\dot M_{\rm p,PA}  = 2 \left( \frac{\rm St}{0.1} \right)^{2/3} r_{\rm H}^2 \Omega_{\rm K} \Sigma_{\rm peb}, 
   \label{eq:peb_Hill}
\end{equation}
where 
\begin{equation}
    r_{\rm H}{=} r\left(\frac{M_{\rm p}}{3M_{\star}}\right)^{1/3}
    \label{eq:rHill}
\end{equation} is the protoplanet Hill radius. The transition between the Bondi and Hill regimes does not impact the pebble accretion efficiency in the 3D regime, while in the 2D regime it slightly changes the dependence of $\varepsilon_{\rm{PA}}$ on the Stokes number. This effect becomes more important in the outer regions of the disk, when the transition between Bondi and Hill regime shifts to higher protoplanet masses. As visible in Fig.~\ref{fig:PAefficiency}, in the 3D regime the pebble accretion efficiency increases with the Stokes number. This is because dust settling is more efficient and the protoplanet accretes from a denser midplane layer. Contrarily, in the 2D regime this effect is canceled by the increasing drift speed of pebbles and the decreasing protoplanet-pebble interaction time. We note that all the above formulas are derived for the planet on a circular orbit. When the planet is on an eccentric and inclined orbit, the epicyclic motion relative to the planet Keplerian velocity needs to be additionally accounted for \citep{Johansen2015,liu2018,ormel2018}. What is more, \citet{Kuwahara2020} pointed out that the planet-induced gas flow can suppress pebble accretion in the inner region of the disk.

\subsubsection{Pebble flux}\label{sub:pflux}

Figure~\ref{fig:PAefficiency} shows that the pebble accretion efficiency is dependent on the planetary embryo mass and on the pebble size. The final planetary growth rate is determined by the product of the pebble accretion efficiency and the incoming pebble mass flux at the planet location. Unlike planetesimal accretion that can only accrete bodies within their local regions, pebble accretion has a much larger feeding zone since these smaller bodies efficiently drift through the disk (given that there are no dust traps present, as we discuss below). As the pebble accretion efficiency is generally low, planet growth requires a large pebble reservoir \citep{Guillot2014,Ormel2017review}, so that the pebble flux is supported over long time. In our example presented in Fig.~\ref{fig:timescales}, we assumed a constant pebble flux that would only be supported for $\sim$1.5~Myrs and that the pebble size correspond to the fragmentation barrier with a constant threshold velocity of 5~m/s. In more realistic models, pebble flux is determined by dust evolution outlined in \S\ref{sub:dustevo} and in Fig.~\ref{fig:barriers} and the fragmentation threshold may depend on grain composition and thus location in the disk. 

\citet{LambrechtsJohansen2014} proposed a model for pebble flux calculation based on the idea that at a given radial distance all the dust grows to the drift limit, decouples from the gas, and drifts inwards. The dust growth timescale strongly depends on the distance so this pebble formation front moves outwards with time. Subsequent portions of solids are released maintaining the pebble flux as long as the pebble formation front has not reached the outer edge of the disk. The larger the disk, the longer it can support a large pebble flux. Thus, an extended disks of solids is often applied as initial condition in models of planetary growth by pebble accretion \citep[see, e.g.][]{Sato2016}, which may be at odds with the observed disk sizes (see \S\ref{sub:disksizes}). Using dust coagulation models, \citet{Drazkowska2021} and \citet{Andama2022} showed that the pebble flux also depends in the fragmentation threshold velocity and that a lower fragmentation threshold leads to lower but longer-lasting pebble flux due to slower drift of pebbles. Nevertheless, if pebbles are small (${\rm St} < 10^{-2}$), pebble accretion\index{Pebble accretion|)} is hindered in two ways: because its efficiency is lower for lower Stokes number and because the pebble flux is lower. 

The models discussed above do not take into account the possible disk substructures. However, in reality the evolution of dust is very sensitive to the underlying gas disk structure as the drift of pebbles depends on the pressure gradient in disk as we show in Fig.~\ref{fig:radial}. Thus, global structures in the disk may lead to complex behavior in terms of releasing pebbles into the inner disk \citep[see, e.g.,][]{Elbakyan2020,Ida2021,Andama2022b}. If disk substructures are as common and efficient in reducing the radial drift\index{Dust!radial drift} as suggested by observations, see \S\ref{sect:observations}, pebble fluxes are likely overestimated in most of the state-of-the-art models. 

\subsubsection{Isolation masses}\label{sub:miso}

A major difference between the pebble and planetesimal-driven growth is the maximum size of the planetary core that can be reached. When the core mass is low, it modestly perturbs the surrounding disk gas. The inward drifting pebbles can bypass the planetary orbit while a fraction of them get accreted. However, when the planet is massive enough to open a gap and reverse the local gas pressure gradient, the pebbles terminate their radial drift at the local pressure maximum exterior to the orbit of the planet. As such, pebbles cannot be further accreted onto the planet  and the mass growth is halted. The so called pebble isolation mass\index{Isolation mass|(} refers to such a planet mass when the pebbles start to isolate from the planet \citep{lambrechts2014,Bitsch2018,Ataiee2018}. 
The pebble isolation mass follows a similar scaling as the gap opening mass
\begin{equation}
    \label{eq:peb-iso}
    M_{\rm iso, peb} \simeq 25 M_{\oplus} \left(\frac{H/r}{0.05}\right)^3 \left(\frac{M_{\star}}{M_{\odot}}\right),
\end{equation}
where $H/r$ is the gas disk aspect ratio and the turbulent strength corresponding to $\alpha=10^{-3}$ is used. We note that in the original work by \citet{lambrechts2014} the value of $M_{\rm iso, peb} = 20 M_{\oplus}$ for $H/r=0.05$ was given, while here we applied $M_{\rm iso, peb} = 25 M_{\oplus}$ based on more accurate hydrodynamic models presented by \citet{Bitsch2018}. $M_{\rm iso, peb}$ increases with the increasing $\alpha$ and decreasing pebble size as the smaller particles are diffused more easily from the pressure maximum arising outside of the planet orbit. \citet{Zormpas2020} suggested that $M_{\rm iso, peb}$ is reduced when a realistic equation of state and radiative cooling are applied. From the pebble accretion perspective, the core mass of the planet is mainly set by the gas disk properties, such as the gas disk aspect ratio and turbulent viscosity, but independent of the local solid density.

For comparison, the well-known classical planetesimal isolation mass, valid when neither planetesimals or planets are migrating, is
\begin{multline}
    M_{\rm iso, plts} = 2\pi r \Delta r \Sigma_{\rm plts} \simeq
    \\ \simeq 0.1 M_{\oplus}\left( \frac{\Sigma_{\rm plts}}{5 \rm \ gcm^{-2}} \right)^{3/2} \left( \frac{r} {\rm AU} \right)^{3} \left( \frac{M_{\star}}{M_{\odot}} \right)^{-1/2},
\end{multline}
where we took $\Delta r {\simeq} 10 r_{\rm H}$ \citep{Kokubo1998}. This is the mass the planet reaches when it accretes all planetesimals in its feeding zone.

This isolation mass is strongly dependent on the local planetesimal density $\Sigma_{\rm plts}$ and the radial distance $r$. In a standard disk model, assuming the constant planetesimal-to-gas ratio of 1\%, the isolation mass approximates to the Mars mass in the inner terrestrial planet region and a few Earth masses beyond the water ice line, which is much lower than the pebble isolation mass (see Fig.~\ref{fig:timescales}). However, that planets can radially migrate, thus enhancing this isolation mass computed for static planets in the planetesimal-accretion case. Another qualitative difference between the pebble isolation and planetesimal isolation is that in the former case, once one planet reaches isolation mass\index{Isolation mass|)}, all other planets interior to its orbit stop accreting pebbles because the pebble flux is halted. 

\subsection{\textbf{Radial redistribution of mass}}\label{sub:driftmigr}

\begin{figure}[t]
 \centering
 \includegraphics[width=0.95\linewidth]{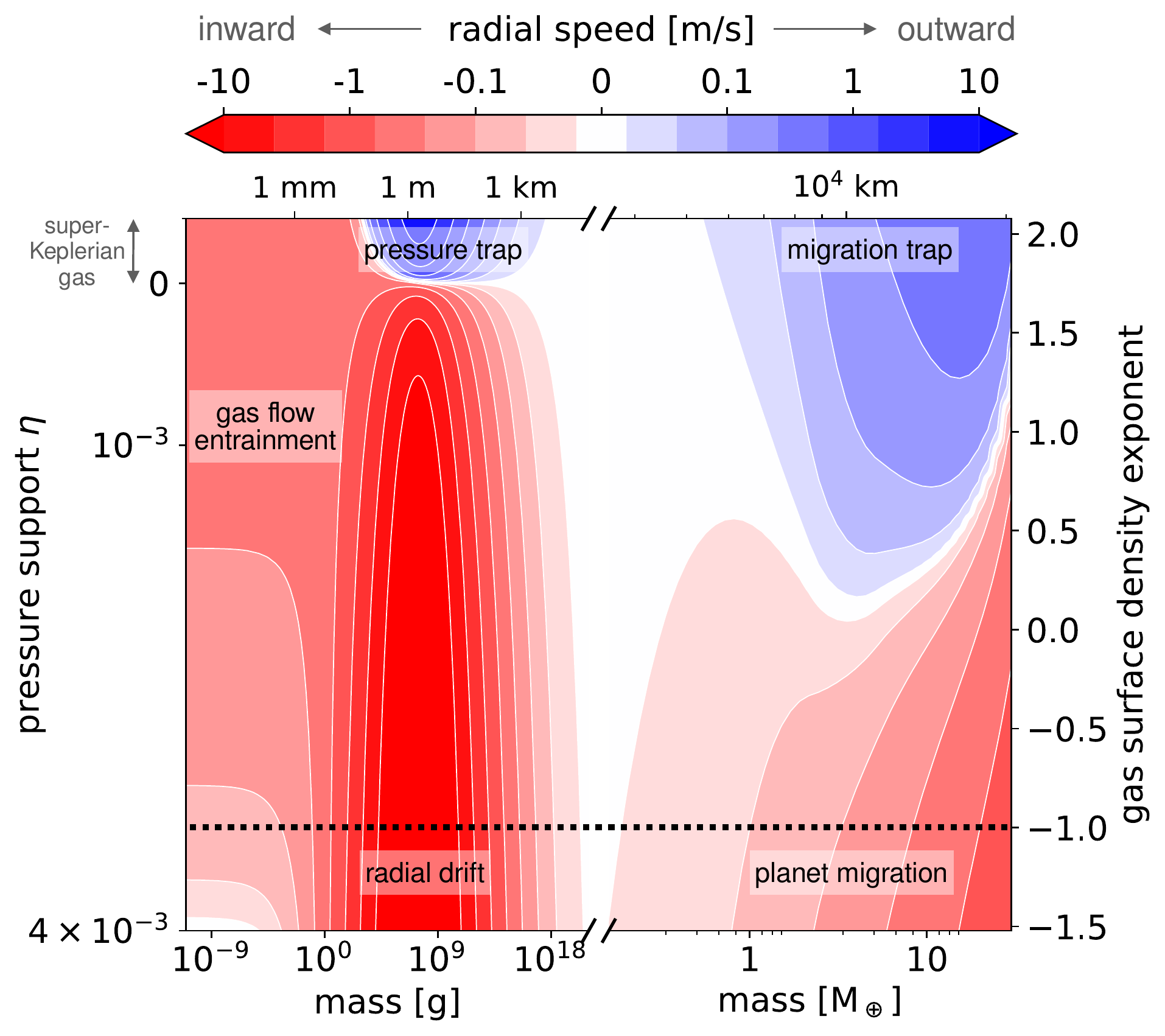}
 \caption{{The color map shows the radial speed as a function of mass (x axis) calculated assuming the surface density of gas of 300~g/cm$^2$, temperature of 200~K, a fixed exponent of the temperature distribution of -3/7, and the turbulence strength of $\alpha=10^{-3}$. We varied the gas surface density slope from its default value of -1 marked with the dotted line to modify the pressure support parameter $\eta$ (y axis). Bodies in the intermediate mass regime, between $10^{19}$~g and 0.1~M$_\oplus$, do not have significant radial velocities and thus we left them out of the plot. 
}
 \label{fig:radial}}
\end{figure}

In Fig.~\ref{fig:sizescales}, we indicated two `windows of mobility' during the planet formation process: the radial drift of pebbles and the migration of planetary cores. Both these processes were found to be possible barriers to the formation of planets as they may outperform the growth and lead to the fall of solid material onto the central star \citep{weidenschilling77, Ida2008, Misener2019}. For this reason, the concept of special locations emerged, where these radial transport processes could be slowed down or even halted, which we will discuss in this section.

\subsubsection{Radial drift}

The speed of radial velocity of dust and pebbles is described with Eq.~\ref{eq:vdrift}. In a typical disk, the radial velocity of gas is on the order of 10~cm/s but the maximum radial drift speed of pebbles is much higher, $\eta v_{\rm K}\approx 30$~m/s. Models including dust growth and radial drift show that the solids are depleted on timescales of below 1~Myr, in contrast with observations, unless some perturbations to disk structure are present \citep{Brauer2007, Pinilla2012}.

The concept of a special location relies on reducing (and eventually swapping the sign of) the $\eta$ parameter and thus lowering (or reversing the direction of) the radial drift speed. Since gas pressure is determined by its density and temperature, this can be done by modifying either of these, however to stop the radial drift this modification needs to be significant as we discuss below. The substructures observed in circumstellar disks are often interpreted as signatures of such pressure traps (\citealt{Dullemond2018}, mechanisms to produce substructures are discussed in the chapter by \citet{ppvii_Bae}. The existence of substructures poses a danger to the pebble accretion\index{Pebble accretion} process as it relies on efficient drift of pebbles leading to lasting pebble flux as we discussed in \S\ref{sub:pflux}.

\subsubsection{Planet migration}

The drift velocity described with Eq.~\ref{eq:vdrift} has its maximum at ${\rm St}=1$ and decreases for larger bodies. However, as the body mass increases, the gravitational interaction between this body and the gas disk becomes important. This interaction leads to migration\index{Planet migration} of Mars-mass and heavier planetary cores \citep{Kley2012, Baruteau_2014}. There are two migration regimes. Planetary cores that are not massive enough to open a gap in the gas fall into the type I migration regime, in which the migration speed scales linearly with the core mass. A typical timescale of the inward migration for an Earth-mass planet at 1~au is 10$^5$~years \citep{tanaka02, Ogihara2015}, comparable to its growth timescale (see Fig.~\ref{fig:timescales}). For massive cores, which start to open partial gaps, the migration speed starts to reduce compared to the pure type-I rate \citep{Kanagawa2018}, derived in the limit of planets that do not perturb the disk \citep[e.g.,][]{paardekooper10}. Planet migration is discussed in more detail in the chapter by \citet{ppvii_Paardekooper}. 

In Fig.~\ref{fig:radial}, we show the radial speed of dust, pebbles, and planetary cores calculated using the Eq.~\ref{eq:vdrift} and the type I migration prescription from \citet{paardekooper10}. In order to create the pressure and migration trap, we varied the radial pressure gradient by modifying the gas surface density slope. There are two important take-away points from this figure. First, in a typical disk with a negative surface density exponent, the pebble drift is faster than the planet migration speed. This makes pebble accretion\index{Pebble accretion} viable even if the planetary core is migrating. Second, for the radial drift to be stopped, the pressure support must be reduced to zero or even reversed, leading to a super-Keplerian gas rotation. Stopping planet migration is easier because the pressure gradient does not need to be pushed to zero but only reduced. Thus, if a pressure trap is present, there is a planet migration trap as well, because the positive barotropic torque is likely to overpower the negative Lindblad torque \citep{Morbidelli2020}. 

\subsubsection{Special locations}

In disks without other substructures, the natural place where the density and thus pressure gradient is reduced is the inner edge of the disk \citep{masset06, liu2017, Brasser2018, Miranda2018, romanova2019, Li2022}. If the inner edge of the disk is shaped by gas removal by disk wind, this effect can be even stronger, leading to outward migration of planets inside of 1~au \citep{Ogihara2015b, Ogihara2017, Kimmig2020}. 
Another location, typically also located very close to the star, is the boundary between the MRI active region and the dead zone\index{Dead zone} of the disk. The viscosity transition caused by varied level of turbulence leads to creation of a pressure trap \citep{Flock2016, Flock2019, Faure2016, Mohanty2018, Ueda2019, Jankovic2019}. In these close-in regions, planetary growth timescales are fast irrespectively of the growth mode (see Fig.~\ref{fig:timescales}). Thus, the first generation of planets could form there. Interestingly, the outer edge of a planetary gap also serves as the pressure trap, enhancing the formation of subsequent planets \citep{Kobayashi2012, Stammler2019, Eriksson2020, Shibaike2020}, potentially leading to the sequential, inside-out planet formation scenario \citep{Chatterjee2014}.

In the recent years, the water snow line was often invoked as the special location for planet formation. This has several reasons. The composition of dust and thus its opacity changes across the snow line, leading to a modification in the disk thermal structure \citep{Savvidou2020}. This can stop the drift of pebbles, favoring the accumulation of solids, planetesimal formation and, subsequently, pebble accretion \citep{Guilera2020, Charnoz2021}. In addition, the snow line can also act as a planet trap against migration \citep{Bitsch2013, Mueller2021}. 

It is worth noting that several of the models invoking the water snow line as a special location assume that water ice-rich dust aggregates are more sticky and thus grow to larger sizes than dry aggregates. This is based on laboratory experiments presented by \citet{Aumatell2014, Gundlach2015}. However, newer laboratory data do not support this conclusion, demonstrating that the icy aggregates break more easily than previously assumed when the gas temperature is significantly lower than ice evaporation point \citep{Gundlach2018, Steinpilz2019, Musiolik2019}. Thus, the fragmentation threshold velocity is in reality dependent on the local disk temperature and thus radial distance in the disk. In general, there is a large dispersion in the sticking threshold velocities reported in the literature, as we discuss in \S\ref{sect:dust}. The consequences of these new findings on planet formation models are yet to be investigated.

The special locations are not only important for retaining the solids in the disk but also for enhancing growth rates. The reduced radial transport leads to accumulation of pebbles, which speeds up the growth of planetary cores by pebble accretion\index{Pebble accretion} (although only in the Hill regime, see \S\ref{sub:pebbleacc}), which is particularly important for forming wide-orbit planets \citep{Coleman2016, Morbidelli2020, Chambers2021}. 
However, the existence of a strong pressure bump, or gap-opening planet, in the protoplanetary disk reduces the delivery of material and thus hinders the growth of the planetary cores located inside of the bump/planet \citep{Weber2018, Lambrechts2019, Brasser2020, Bitsch2021, Izidoro2021b}. This demonstrates that global models including multiple cores interacting with the disk and consistent models for dust evolution are necessary to understand the formation of planetary systems. 

%%%%%%%%%%%%%%%%%%%%%%%%%%%%%%%%%%%%%%%%%%%%%%%%%%%%%%%%%%%%%%%%%%%%%%%%

\subsection{\textbf{Gas accretion}}\label{sect:gasacc}
Gas accretion\index{Gas accretion|(} onto a planetary embryo starts when the growing embryo is massive enough to gravitationally bind the gas in its vicinity in the disk. The characteristic radius for local binding of the gas is the Bondi radius, the radius where the sound speed of the gas equals the escape velocity from the embryo:
\begin{equation}
r_B = \frac{2GM_{\rm p}}{c_s^2},
\end{equation}
where $G$ is the gravity constant, $M_{\rm p}$ the mass of the planetary embryo, and $c_s$ the gas sound speed. Note that the  term here is not the same as the "Bondi radius" used in pebble accretion nomenclature. 
In a global perspective we can define another radius where the embryo’s attraction dominates that of the parent star, the Hill radius (Eq.~\ref{eq:rHill}). 
The actual radius for accretion is usually taken to be the minimum of Bondi and Hill radii.

Inside this radius the growing embryo attracts gas from the disk, forming a planetary gaseous envelope. For low mass planets the envelope is usually hydrostatic - in pressure equilibrium with the surrounding disk gas. The envelope thermal profiles are derived by steady solution to the stellar/planetary structure equations of mass and energy conservation, energy transfer, equation of state, and hydrostatic equilibrium \citep{kippenhahn12}. 
The energy balance of the envelope has three terms: the accretion energy from solids and gas, the contraction of the gas, and the cooling by luminosity. The envelope grows as the outer envelope cools and contracts and fresh gas flows from the disk to replenish the empty space \citep{stevenson82,pollack96,rafikov2006,mordasini12}. 
The amount of nebular gas that a planet can bind is limited by the cooling and contraction of the envelope, namely its Kelvin-Helmholtz (thermal) timescale. 
The Kelvin-Helmholtz timescale is a strong function of the atmospheric opacity and of planetary mass \citep{ikoma00,horiikoma11,piso2014,Lee2014,leechiang15,lambrechts17,schulik19,bitsch21}.

As the mass of the embryo increases, its ability to accrete gas from the disk increases, and gas accretion accelerates. If the embryo becomes sufficiently massive, pressure gradients can no longer sustain the gravity, 
and the planet accretes large quantities of gas in a rapid runaway process \citep[e.g.,][]{pollack96,ayliffe12}; as a result a gas giant planet is formed. 
The minimum mass of metals to initiate runaway gas accretion is called the critical metal mass -- the mass where the total mass in metals equals that of the gas. The critical metal mass varies with disk and envelope parameters \citep{mizuno80}, among them solid accretion rate \citep{Bitsch2015b,Lambrechts2019}, and solid composition \citep{venturini17,brouwers20}. Note that the critical metal mass is a criterion that supersedes the traditional critical core mass, to include also planets with polluted envelopes\index{Polluted envelopes}, where most of the metal mass can be in the envelope.

If the embryo does not grow fast enough along the disk lifetime to reach the critical metal mass, then the planet remains metal-rich, i.e., composed mainly of metals (ices, rocks, iron), with small fraction of nebular gas.
After disk dissipation the stellar flux absorbed by the outer layers of the gas envelope can stimulate gas mass loss mechanisms \citep[e.g.,][]{lammer03,owen12}. Mass loss can be significant in low mass planets due to their low gravity and small fraction of gas, but negligible in most gas giants.

To a zero order, the process of gas accretion is similar in pebble-driven and planetesimal-driven formation scenarios. However, the interaction of the smaller size pebbles with the gas in the growing envelope may lead to differences in formation process and in the resulting planetary properties. 

\begin{figure*}[h]
 \centering
 \includegraphics[width=0.8\linewidth]{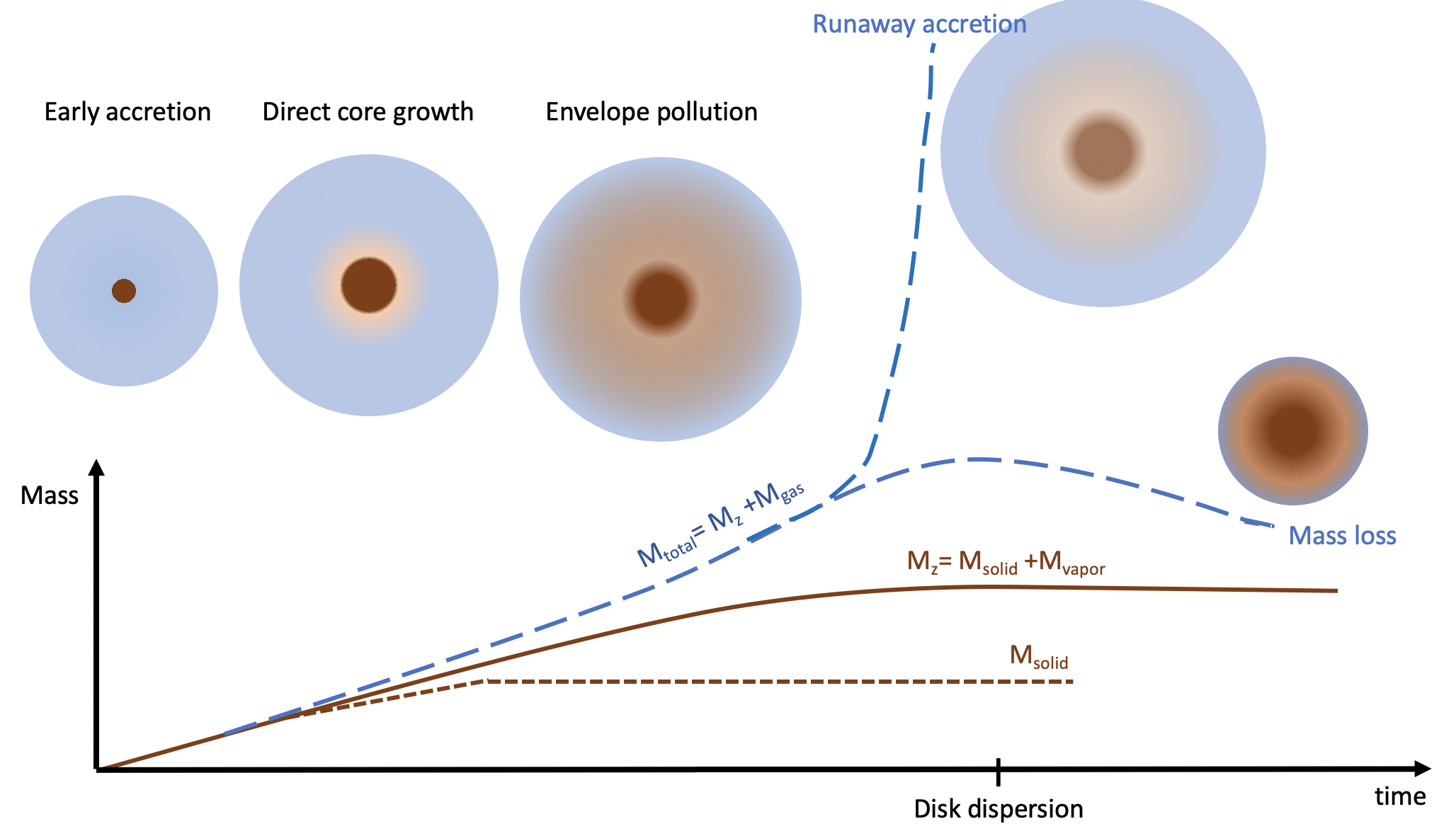}
 \caption{{Formation of polluted envelopes\index{Polluted envelopes}. Initially, solids reach the core (“early accretion”) but as the temperature and density of the envelope increase, solids become increasingly prone to breakup and sublimation, enriching the envelope in metals. At some point, temperatures have reached levels where the entire solid influx is vaporized (“envelope pollution”). The co-accretion of solids as well as nebular gas results in a gradual metallicity profile. If the planet reaches the critical mass, it will accrete gas in a runaway fashion to become a gas giant; otherwise it will end up a sub-Neptune/Neptune-like. After disk dispersal the stellar irradiation (for close-in planets) generates gas mass loss. The figure is a modified version of Fig.~1 in \citet{brouwers20}, incorporating results from \citet{ormel21}. The curves signify the mass of solids (short dashed), metals (solids and vapor; solid curve), and total (metals and gas; dashed).}
 \label{fig:pollut}}
\end{figure*}

\subsubsection{Polluted envelopes}
Accreted pebbles can sublimate and pollute the gas envelope\index{Polluted envelopes|(} with metal vapors \citep{lambrechts2014,venturini16,brouwers18}. In Fig.~\ref{fig:pollut} we present the phases of the planet growth by pebble accretion\index{Pebble accretion}. At early stage of the formation the interaction of the diluted gas with the pebbles is very weak and therefore pebbles are being accreted to the center of the planet, forming a planetary solid/liquid core of metals. As accretion of both pebbles and gas continues the atmosphere becomes hotter by the energy deposition of the accreted solids, and denser by the further accretion of gas. Consequently, the small pebble-sized particles will start to sublimate in the deep atmosphere when they hit the sublimation temperature. For refractory pebbles (e.g., rock) this temperature is about 2000\,K. For volatile pebbles (e.g., water ice) this temperature is much lower ($\sim$ 170\,K) and thus icy pebbles sublimate almost immediately when they enter the planetary envelope. At first, the envelope is saturated with the rock vapors and thus some of the pebbles still sink to the core (direct core growth). As the temperature increases (by further accretion and contraction) the envelope becomes undersaturated and the core growth terminates. Any further pebble accretion contributes solely to the pollution of the envelope.

For accretion of rocky pebbles the point where all pebbles stay in the envelope is reached when the core mass is about 1-2 M$_\oplus$ \citep{brouwers18,ormel21}. Polluted envelopes can contain large fraction of metals in vapor form, much larger than the mass in their planetary cores. Thus, sub-Neptune\index{Sub-Neptunes} planets that formed by pebble accretion can have a significant amount of their metal mass in the polluted envelope \citep{brouwers20,ormel21}.

The pollution of the envelope can have consequences on the type of the forming planet, as it is expected to lower the critical metal mass for runaway gas accretion \citep{horiikoma11,helled14,venturini16,brouwers20} and thus facilitates the formation of gas giant planets\index{Giant planets!formation}.
On the other hand, pebble fragmentation and sublimation in the envelope can act under certain conditions to limit the growth of the protoplanet \citep{Ali-Dib2020}.

It should be noted that planetesimal accretion\index{Planetesimal accretion} can also result in envelope pollution\index{Polluted envelopes|)}, but in this case pollution requires larger planetary cores (thicker envelopes) and/or fragile accreted solids \citep{podolak88,mordasini15,alibert17,brouwers18}. It is therefore limited mainly to gas giant planets \citep{lozovs17,helledsteven17,valletta20}.

\subsubsection{Composition gradients}

When pebbles sublimate before reaching the core, they deposit their energy at their sublimation location. This pushes the sublimation boundary outwards in time, resulting in insufficient available energy to mix the deeper dense vapor-rich layers upwards. Consequently, a gradual composition distribution in which metal mass fraction decreases with radius is found to be a natural outcome of planet formation by pebble accretion\index{Pebble accretion} \citep{ormel21}. Thus, the interiors of planets that formed by pebble accretion (before runaway gas accretion) are not structured in uniform layers at the end of the formation phase. The interior structure of planets formed by pebble accretion after runaway gas accretion is yet to be studied, and in particular the fate of the composition gradient.

A gradual change in the composition in the interior of the Solar System's giant planets\index{Giant planets} is the preferred scenario to explain the very precise observations made by visiting spacecrafts. The {\it JUNO}\index{JUNO} spacecraft measurements fit best models that suggest diluted core in Jupiter’s innermost region, rather than a pure distinct solid core \citep{wahl17,vazan18b,debras19}. Similarly, measurements of the Cassini spacecraft in the rings of Saturn hint towards non-uniform structure of Saturn’s interior \citep{fuller14,vazan16,mankovich21}. Interior models with gradual composition distribution explain also the (few) measurements of Uranus and Neptune \citep{marley95,helled11UN,vazanhel20,scheibe21}.

The gradual distribution of metals in planetary interiors has implications on the long term thermal evolution of the planet, and on its inferred properties \citep{stevenson85,lecontechab12,vazan16}. A stable composition gradient suppresses large scale convection \citep{ledoux47} and breaks the standard approach of adiabatic thermal evolution of planetary interiors. 
As a result, the imprints of the formation processes can remain for billions of years, to the current stage where we observe the planet. 
Non-adiabatic thermal evolution in gas giant planets affect their inferred properties \citep[see the chapter by][]{ppvii_Guillot}.
Further studies are required to explore the influence of gradual composition distribution in smaller planets, such as sub-Neptunes.

Another expected difference of pebble accretion from planetesimal accretion is the contribution of pebbles to atmospheric opacity. During the accretion phase the small pebbles can act as a dominant source of opacity and thus prolong the envelope contraction, suppressing further gas accretion.  
However, the required pebble size and accretion rate for this scenario are not very likely in the process of planet formation \citep{brouwers21}. 
Moreover, once pebble accretion subsides the atmosphere rapidly clears, followed by fast cooling and massive gas accretion \citep{ormel21}. Therefore, pebble opacity cannot (by its own) halt the gas accretion, and other mechanisms are needed to prevent sub-Neptunes from reaching runaway gas accretion\index{Gas accretion|)} and becoming gas giants.

\subsubsection{Atmospheric recycling}
In the literature, calculations of the evolution of planet envelopes thermal structure are almost exclusively conducted in 1D, assuming radial symmetry, by solving a modified version of the stellar structure equations between an inner "core" radius and an outer radius $r_\mathrm{out}$. This therefore assumes that the planet envelope is hydrodynamically isolated from its surrounding in the sense that no material leaves $r_\mathrm{out}$.  However, when a planet first acquires its envelope it does so in pressure equilibrium with the disk. The question then arises whether these planets can be characterized by an outer radius, distinguishing envelope from disk material. The standard approach has been to take $r_\mathrm{out}=\min(r_B, r_H)$ or some modification of this formula \citep{LissauerEtal2009}. Then, for growing envelopes, only radiation can leave $r_\mathrm{out}$, while disk material enters $r_\mathrm{out}$ when the envelope cools and contracts.

However \citet{OrmelEtal2015i} and \citet{FungEtal2015}, conducting isothermal hydrodynamical simulations that allowed for steep density gradients (i.e.,\, a negligible softening), found a more radical outcome: disk material penetrates deeply within the Bondi sphere and --simultaneously-- envelope material was seen to advect back to the disk. Follow-up studies \citep{CimermanEtal2017,lambrechts17,KurokawaTanigawa2018,PopovasEtal2018,BethuneRafikov2019i,KuwaharaEtal2019} have confirmed the potential for disk-planet exchange, although the flow pattern and therefore the efficacy of recycling rather sensitively depends on physical (e.g.,\, opacity or cooling rate) and computational parameters (softening radius). In addition, none of these simulations have yet
 accounted for the aforementioned sublimation of solids, which may result in a more stratified configuration.

The key implications is that an efficient disk-planet envelope recycling  interferes with the Kelvin-Helmholtz contraction of pre-planetary atmospheres. Hydrodynamically isolated atmospheres will always lose entropy by radiative cooling. With recycling, however, this entropy decrease will be arrested, because low entropy atmosphere material is replaced by high-entropy disk material. Consequently, the KH-induced contraction terminates and the atmosphere is prevented from becoming critical. Recycling is most potent near the star, because the disk-planet relative velocity and the disk gas density are the highest. For these reasons, \citet{OrmelEtal2015i} have suggested that disk-planet atmospheric recycling explains the preponderance of mini-Neptunes over hot-Jupiter exoplanets. Accounting for radiation transport, \citet{MoldenhauerEtal2021} found that super-Earth planets readily attained such a thermodynamical steady state.

%%%%%%%%%%%%%%%%%%%%%%%%%%%%%%%%%%%%%%%%%%%%%%%%%%%%%%%%%%%%%%%%%%%%%
\section{\textbf{MODELS FOR THE FORMATION OF PLANETARY SYSTEMS}}\label{sect:models}

In this section, we discuss the different models for the formation of planetary systems. In particular, we focus on the initial solid budget needed to form different types of planets (\S\ref{sub:4.1}), how planet formation theories can explain specific systems like the Solar System and interesting exoplanet systems (\S\ref{sub:4.2}), and discuss planet formation around M dwarfs (\S\ref{sub:4.3}). In the next section, we will also cover general planet population synthesis models used to understand the origins of the exoplanet population as a whole (\S\ref{sub:popsynth}). As a minimum these models contain recipes for the growth and migration\index{Planet migration} of planets. We emphasise that models of planet formation in this sense are two-fold: they are either tailored to specifically explain a given planetary system with all its constraints like the Solar System or are designed to explain the statistical exoplanet occurrence rates and properties.

In general terms, there are two models for forming a giant planet: core accretion and gravitational instability. The former assumes a bottom-up approach with a solid planetary core, or embryo, forming first and then accreting more solids and gas. The latter is a top-down model, in which a massive protoplanetary disk undergoes fragmentation into clumps. If the clump can cool quickly enough, the gravity overcomes gas pressure, and a massive, gas-rich planet forms. In this chapter, we primarily focus on the core accretion scenario, although we briefly discuss the gravitational instability in the next subsection.

\subsection{\textbf{Forming different types of planets: Earths, super-Earths, ice giants, and gas giants}}\label{sub:4.1}

It is clear that the formation of different planetary types depends crucially on the amount of available solids, as indicated by the host star metallicity giant planet relation (see \S\ref{s:exo:metal}), which shows that metal rich stars, whose disks harbor more solids and thus more planetary building blocks, host more giant planets. This trend has long been reproduced in the classical, planetesimal driven planet formation models \citep{Ida2008, ida08b, Alibert_2011, mordasini12} and more recently also in the pebble accretion scenario \citep{Bitsch2017,Ndugu2018}. 

\begin{figure}[t]
 \centering
 \includegraphics[width=\linewidth]{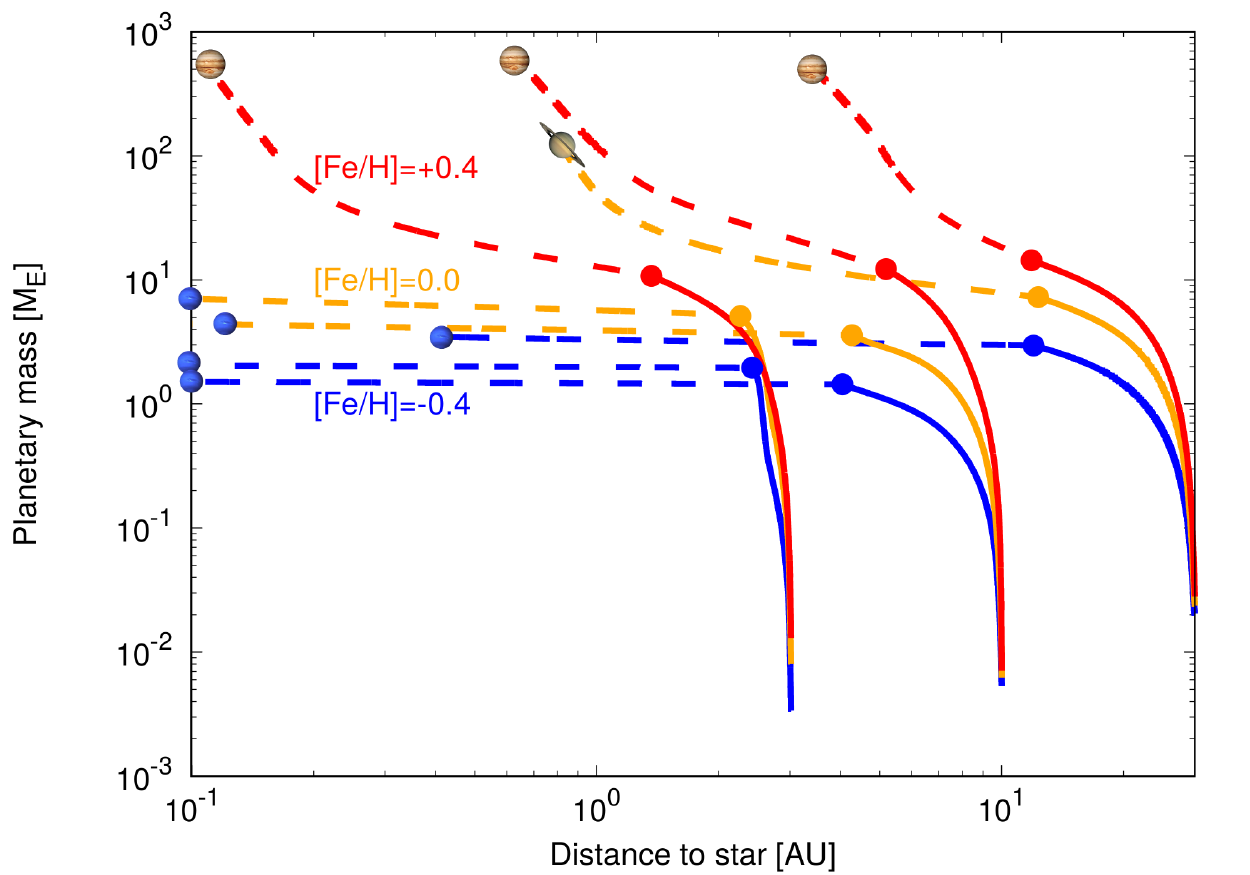}
 \caption{{Growth tracks of planets growing via pebble accretion\index{Pebble accretion} at 3~au, 10~au and 30~au in disks with different metallicities: [Fe/H]=-0.4 (blue), [Fe/H]=0.0 (orange), and [Fe/H]=+0.4 (red). The pebble flux for [Fe/H]=0.0 corresponds to 100 M$_\oplus$/Myr, with an exponential decay of 2.5\,Myr, corresponding to a total of 175 M$_\oplus$ in pebbles over 3\,Myr disk lifetime. We mark the solid accretion phase with solid lines, the pebble isolation mass with a dot and the following gas accretion phase is marked by dashed lines. The final position and planetary type is indicated at the end of the growth track, where the individual simulations form gas giants and sub-Neptunes\index{Sub-Neptunes!formation}.}
   \label{fig:tracks}
   }
\end{figure}

In Figure~\ref{fig:tracks}, we show results of integration of the migration and growth of planetary embryos starting at 3~au, 10~au and 30~au with initial masses that allow pebble accretion in the 3D/2D Hill regime (see \S\ref{sub:ppaccretion}), in disks with different metallicities carried on for 3~Myr. In this simple model, we use a constant Stokes number of $10^{-2}$ for the pebbles (roughly corresponding to a fragmentation threshold of $\sim$5~m/s in Fig.~\ref{fig:barriers}, although we must note this approach is not completely consistent with dust growth theory as the constant fragmentation threshold is not parallel with the constant Stokes number) with a pebble flux of initially 100 M$_\oplus$/Myr (for the solar metallicity [Fe/H] = 0.0) exponentially decaying in time with a decay time scale of 2.5~Myr. This results in a total amount of 175 M$_\oplus$ of pebbles over the 3~Myr integration time. Once the planet has reached the pebble isolation mass, it starts to accrete gas and can eventually turn into a giant, if the envelope contraction happens fast enough. The initial  location of planetary embryos and the mass available to them during growth regulate the types of planets that can form. We stop the integration if the planet reaches the inner edge of the disk at 0.1~au or the disk reaches it maximum lifetime of 3\,Myr.

As illustrated in Fig.~\ref{fig:tracks}, planets in the high metallicity environments where more pebbles are available, grow larger planetary cores that can eventually transition into gas giants, compared to planets growing in low metallicity environments. We want to emphasise that the outcome of the planetary growth process is a direct competition between the accretion and migration\index{Planet migration} rates. Only fast accreting and slowly migrating objects have a chance to become cold Jupiters, while slowly accreting and fast migrating objects might end up as inner sub-Neptunes close to the central star, independently if the planet grows via planetesimals or pebbles. We stress that many other processes besides the metallicity like the disk lifetime, the embryo starting position, and the disk turbulence influence the outcome of planet formation studies. We discuss this further in \S\ref{sub:popsynth}. Classically, before a growing planet can become a gas giant, it needs to reach a certain mass for the gas accretion to act efficiently \citep{pollack96}. For pebble accretion, this mass is set by the pebble isolation mass\index{Isolation mass}, depending on the disk properties (see \S\ref{sub:miso}). As disks evolve in time and cool \citep{Oka2011, Bitsch2015a, Baillie2015}, the pebble isolation mass decreases in time. The exact thermodynamical evolution of the disk also depends on the evolution of the stellar luminosity, which changes with stellar mass \citep[see, e.g.,][]{baraffe2015}. Slower growing planets thus reach a lower pebble isolation mass compared to faster growing planets, which can reach the pebble isolation mass earlier. This classical view of core and gas envelope growth has been put into question recently, where especially an envelope highly polluted\index{Polluted envelopes} with heavy metals can become critical, allowing a transition into gas giants at low core masses (see \S\ref{sect:gasacc}).

\subsubsection{Solid dominated planets} Beyond the already stated dependency of planet formation on the available solid budget, further more detailed trends have been revealed in simulations. \citet{Lambrechts2019} investigated in N-body simulations the dependency of planet growth on the pebble flux. A very low pebble flux of around 10 M$_\oplus$/Myr, results in very slow growth, where embryos of the mass of the Moon reach Mars mass in 3~Myrs. However, these planets can then further grow after the gas disk phase via collisions to form objects of a few Earth masses in a similar fashion as the classic formation scenario for the terrestrial planets\index{Terrestrial planets!formation} in the Solar System \citep[e.g.,][]{hansen09, raymond09c}.

In contrast, if the pebble flux is higher (on the order of 100-200 M$_\oplus$ over 3 Myr, as for [Fe/H]=0.0 in Fig.~\ref{fig:tracks}), planetary embryos can efficiently grow to reach the pebble isolation mass of a few Earth masses during the gas disk phase. During their growth, the planets actually migrate through the protoplanetary disk (see \S\ref{sub:driftmigr}) and reach the inner edge of the protoplanetary disk, where their migration is stopped \citep{masset06, Flock2019}. The multiple inward migrating planetary embryos then form a chain of planets in resonant configurations \citep{cossou13, Lambrechts2019, Izidoro2021}, which have to be broken in order to match the observations of period ratios of inner planets (see \S\ref{sub:4.2}).

The main difference between these two distinct modes of sub-Neptune\index{Sub-Neptunes!formation} formation is related to their formation time: at high pebble fluxes, the planets form already completely during the gas phase of the disk, while systems forming in environments with lower pebble fluxes might preferably complete their formation after the gas disk phase. 

These two modes of sub-Neptune formation could be distinguished observationally. Sub-Neptunes already forming during the gas disk phase might host a substantial hydrogen and helium envelope (on the order of a few percent of the total planet mass), while planets forming via collisions after the gas disk phase will not host such an envelope. Furthermore, systems forming in a high pebble flux environment would result in resonant systems due to efficient migration (these chains can break by dynamical interactions after the gas disk phase, but some of the resonant systems will survive, see \citealt{cossou13, Izidoro2021, Lambrechts2019}), while systems forming in low pebble flux environments need collisions to form sub-Neptunes in the first place preventing them from keeping resonant configurations at all. We discuss this issue further in \S\ref{sub:4.2}. If the pebble flux can then be related to the host star metallicity, systems in resonance should predominantly exist around metal rich stars and be devoid around metal poor stars.

Alternatively, super-Earth\index{Super-Earths!formation} systems can form via the accretion of planetesimals in the inner disk, where the accretional timescales are short enough for efficient planetesimal accretion\index{Planetesimal accretion} \citep{chiang13, Ogihara2015}, see also \S\ref{sub:ppaccretion}. The planetesimal accretion simulations by \citet{Ogihara2018} provide a good match to the {\it Kepler}  observations of inner super-Earths, opening the question if these inner systems actually form via planetesimal or pebble accretion. Distinguishing if the systems formed via migrating embryos that accrete either planetesimals or pebbles just from an architecture point of view will be extremely hard \citep{ColemanEtal2019}.

Uranus and Neptune, the ice giants\index{Ice giants} of our Solar System, consist presumably to a large fraction of icy material. However, the quality of the data of Uranus and Neptune allow interior models that do not contain any ice \citep{Guillot2005}, emphasising further the need for a space mission to explore the ice giants in our Solar System, especially because Uranus and Neptune are assumed to be the closest relatives to the very commonly observed inner sub-Neptunes around other stars.

Additionally, ice giants around a few au may be the most common type of planets in our galaxy, even more abundant that close-in sub-Neptunes \citep{Suzuki2016}. While the classic core accretion formation scenario suggests that ice giants received a constant flux of planetesimals \citep{pollack96} or pebbles \citep{lambrechts2014} preventing them to contract a massive envelope, new studies seem to indicate that envelopes highly polluted\index{Polluted envelopes} with heavy metals should become critical allowing a transition into gas giants \citep{horiikoma11, venturini16, brouwers21}. As a consequence, the exact formation pathway of these planets remains a mystery. Potential solutions to this mystery include anomalously high envelope opacities preventing cooling, late formation in gas-poor disks or atmospheric recycling (\S\ref{sect:gasacc}).

\subsubsection{Gas-dominated planets}

The classical picture of giant planet\index{Giant planets!formation} growth requires a core of several Earth masses before gas accretion can commence, but new simulations show that this simple two layer approach does not hold. Nevertheless, a planetary core needs to grow either way. Figure~\ref{fig:tracks} illustrates that efficient core growth is possible in high metallicity environments, leading to the formation of gas giants around these stars \citep{Bitsch2015b, Ndugu2018, Brugger2018, Buchhave2018}. Interestingly, the pebble accretion\index{Pebble accretion} efficiency for a single planet is quite low during core build-up \citep{lambrechts2014}, so that the high pebble flux needed to form giant planets can lead naturally to the formation of multiple giant planets within the same system \citep{Levison2015, Chambers2016, Bitsch2019b, Bitsch2020, Matsumura2021}, implying that giant planets should predominantly not be single. On the other hand, the planetesimal accretion scenario has difficulties in explaining the formation of giant planets exterior of a few au \citep{Levison10, Fortier2013, Johansen2019, Emsenhuber2020}, because of the prolonged accretion timescales at larger orbital distances (see Fig.~\ref{fig:timescales}), questioning the universal application of planetesimal accretion\index{Planetesimal accretion} to explain the exoplanet population as a whole (but see also \citealt{DAngelo2021} for a contrasting view). We discuss this further in \S\ref{sub:popsynth}.

Another formation pathway for giant planets is the gravitational instability scenario \citep{boss97, Mayer2002}. This scenario differs significantly from the core accretion scenario, because the giant plants form directly from a gravitational collapse of clumps of gas in the disk, where these clumps are generated early in very massive and cool disks \citep{Deng2017}. A gravitationally unstable disk follows the relation
\begin{equation}
 Q = \frac{c_{\rm s} \Omega_{\rm K}}{\pi G \Sigma_0} < 1 \ ,
\end{equation}
where $c_{\rm s}$ is the sound speed, $G$ the gravitational constant and $\Sigma_0$ the gas surface density. The radial dependency of $Q$ implies that mostly the outer regions of the protoplantary disk would become gravitationally unstable, resulting in planet formation at wide separations ($\sim$30 -- 300~au).
However, this formation scenario faces a few challenges. In particular, the large mass of the disk implies that the formed fragments are very massive, above a few Jupiter masses, making it very hard to explain the formation of smaller planets within this scenario. An attempt to solve this issues has been made by \citet{Nayakshin2015}. Furthermore, these fragments can be tidally disrupted \citep{Nayakshin2015a} and the planets resulting from the gravitational collapse of these fragments migrate inwards on very short timescales \citep{Baruteau2011}, making it impossible for them to stay in the outer disk. This fast inward migration is accelerated compared to the nominal type-II migration, because the fast formation of the planets does not allow them to open deep gaps needed for type-II migration before they start migrating. Nevertheless, the gravitational instability scenario could help to explain the formation of super-Jupiter planets \citep{Schlaufman2018}, particularly those found by direct imaging at wide-separations, as well as the potential planet formation signature recently observed in AB Aur \citep{Currie2022}.

\subsection{\textbf{Scenarios for the formation of exoplanet systems and the Solar System}}\label{sub:4.2}

For most exoplanetary systems we only have constraints regarding the masses and orbital distances/periods of the planets within the system. Recently, constrains regarding the atmospheric C/O content and constraints from other chemical species inside exoplanetary atmospheres have become available \citep[see, e.g.,][]{Kreidberg2014, Brewer2017, Benneke2019, Tsiaras2019, Kasper2021}, albeit there are large error bars on these constraints. The recently launched {\it James-Webb-Space-Telescope (JWST)} and the future {\it ARIEL} mission are supposed to significantly tighten these constraints and thus influence planet formation theories.

In contrast, the Solar System offers many more constraints for planet formation theories than just some basic properties of the planetary systems. In this section we will focus in particular on the dichotomy between the terrestrial and giant planets, the constraints from the asteroid belt, Jupiter composition and on the formation of Uranus and Neptune.

\subsubsection{Exoplanet systems}
One striking feature of the observations of close-in planetary systems\index{Exoplanets|(} of sub-Neptunes is that nearly all (around 95\%) of the planetary pairs in each system are not in a resonant configuration \citep{Fabrycky2014, Mills2016, Pichierri2019}, despite the fact that planet migration\index{Planet migration} theories would predict resonant configurations \citep{cossou13, Izidoro2017}, even though this has been slightly put into questions in low viscosity environments \citep{McNally2019}, where also low mass planets can open gaps, discussed further in the chapter by \citet{ppvii_Paardekooper}. Furthermore, detailed observations have revealed a radius-valley in the size distribution of these close-in planets \citep[see the chapter by][]{ppvii_Weiss}, which can be explained by a population of rocky planets without atmosphere for the peak at smaller radii (at $\approx$1.5 $\rearth$) and planets with atmosphere that could be potentially water rich for the peak at larger radii (at $\approx$2.2 $\rearth$).
Two main ways have been proposed to explain the formation of these close in planetary systems: (i) formation during the gas disk phase, where the planetary chains break apart after the gas disk dissipates \citep{Izidoro2017, Izidoro2021} or (ii) the planets form in an environment that does not support large scale migration and avoids trapping planets in resonance in the first place \citep{Lee2016}.

The ``breaking the chains'' scenario \citep{Izidoro2017, Izidoro2021} relies on the dynamical evolution of tightly packed systems after the gas disk dissipates and with it the damping forces that regulate the eccentricities and inclinations of the planets. Once these damping forces do not constrain the eccentricities of the planets any more, eccentricities rise due to dynamical interactions leading eventually to dynamical instabilities in the systems. These instabilities can result in collisions and ejections of planets, ultimately destroying the resonant chains within a few 10 Myrs after gas disk dissipation \citep{Izidoro2017, Lambrechts2019, Izidoro2021}. However, a 95\% instability rate to explain the observations is needed and not always achieved in the simulations, because the exact rate of instabilities depends crucially on the masses and distances of the planets in the initial system \citep{Izidoro2021}. The rate of instabilities could be increased by external perturbers (e.g. planetesimal rings, tidal interactions with the star, see \citealt{Morbidelli2018b}). The instabilities also increase the mutual inclinations between the planets, hiding planets from transit observations, resulting in the prediction that systems that only show one transiting planet should actually have at least 1-2 other hidden companions \citep{Izidoro2021}, which is testable by RV follow up observations for the bright nearby {\it TESS} systems.

The growth rates of planets forming via pebble accretion\index{Pebble accretion|(} are higher beyond the ice line due to the larger pebbles and higher pebble fluxes caused by the icy component. These planets then outgrow their inner companions and scatter them away as they migrate inwards, resulting in a system of water rich planets without rocky companions, which can explain the population of the second peak in the radius distribution \citep{Izidoro2021, Venturini2020a, Venturini2020b}, but fails to explain the water poor inner planets that populate the 1st peak in the radius distribution. Proposed ways to form water poor inner planets include the formation of planetary  embryos that accrete pebbles only in the inner disk \citep{Izidoro2021}, fast inward migration from icy small cores that then accrete only rocky pebbles \citep{Bitsch2019, SchoonenbergEtal2019} or the evaporation of the water component during planetary build-up (e.g. \citealt{Johansen2021}). Clearly, future work is needed to explain the composition of inner planets.

An alternative scenario relies on the formation of the planetary systems in an ``in-situ'' environment, where the planets only migrate very little and thus avoid the building of resonant chains in the first place \citep[e.g.,][]{Lee2016}. The final planetary assembly happens in a low gas density environment, where the migration rate is very slow, towards the end of the disk lifetime via collisions \citep{Dawson2016}. The local assembly of planets results in the formation of rocky planets in the inner disk regions that are devoid of water ice, but struggles to explain the formation of water rich inner planets like K2-18b \citep{Benneke2019, Tsiaras2019}. Furthermore, this scenario requires a massive solid build-up in the inner disk \citep{hansen12, Schlichting2014}, which has to happen on a short time scale towards the end of the disk lifetime, because otherwise the planets would start forming earlier \citep{Ogihara2015}. 
The structure and formation of these inner systems is discussed at length also in \citet{ppvii_Weiss}.

Further constraints that theories of exoplanet formation have to fulfill is related to the occurrence rates of sub-Neptune and cold Jupiters (defined as $M_{\rm P}>0.3 M_{\rm J}$) within the same system. Detailed observations have revealed that at least 50\% of all systems with cold Jupiters (Jupiter mass planets exterior to 1 au) should host inner sub-Neptunes with masses below 40 M$_\oplus$ \citep{Zhu2018b, Bryan2019}, see also \S\ref{s:exo:multi}.

The formation of inner sub-Neptunes\index{Sub-Neptunes!formation} and outer gas giants may be explained both in the pebble and planetesimal accretion scenarios. As the pebble flux is large enough to build the core of an outer giant planet, it can easily additionally form icy sub-Neptunes of smaller mass that migrate inwards \citep{Bitsch2019, Bitsch2020, Bitsch2021}. Similar results have been observed in simulations featuring core growth via planetesimal accretion \citep{Schlecker2020}. However, both scenarios struggle to explain the high correlation between outer cold Jupiters and inner sub-Neptunes, because the forming giant planets start to dynamically interact and mutally scatter after the gas disk phase, destroying the inner systems of sub-Neptunes in the process, possibly related to too large accretion rates of the giant planets. On the bright side, these dynamical interactions of the giant planets can explain the observed eccentricity distribution of the giant planets \citep{Marzari02, Ford08, Juric08, Raymond09b, Bitsch2020}, where eccentric giant planets are preferably found orbiting metal rich stars \citep{Dawson2013, Buchhave2018}, which can form more giants in the first place that can then mutally scatter (see also Fig.~\ref{fig:tracks}).

As discussed in \S\ref{s:disks}, the observations of protoplanetary disks with ALMA\index{Atacama Large Millimeter Array (ALMA)} have revealed detailed substructures in the dust distributions at large (several 10 to 100 au) distances, which may potentially be caused by growing planets that have reached masses close to the pebble isolation mass, where they already start to influence the drift of pebbles (see \S\ref{sub:ppaccretion}). Alternatively, giant planets could actually form directly in these rings \citep{Morbidelli2020}. If indeed planets cause, or are born in these substructures, we have to wonder about their subsequent evolution during the still ongoing gas disk phase. In models with high disk viscosity, \citet{lodato19} found that inward migration\index{Planet migration} of these objects could bring them close enough to the central star to explain the giant planet population observed in RV surveys, while \citet{Ndugu2019} concluded the opposite from their low viscosity simulations: migration is too inefficient to bring these giant planets from several 10s of au to orbits of just a few au, and instead there should be a large population of giant planets that is observable in direct imaging campaigns. However, the frequency of wide orbit giant planets from direct imaging surveys is very low (around 1\%, \citealt{Bowler2018}), putting into question that every gap/ring in these observed large disks is caused by a growing planet. Alternative explanations for the origin of these rings and gaps are discussed in chapters by \citet{ppvii_Bae} and \citet{ppvii_Pinte}.

Already now (in the pre-{\it JWST} era) constraints from planetary atmospheres to planet formation exist (e.g. \citealt{Brewer2017, Welbanks2019}). In particular, the C/O ratio of giant planet atmospheres is thought to be related to the formation location of giant planets \citep{Oberg2011, Madhusudhan2014}. This relation is caused by the change of the abundance of the main carbon and oxygen carrying species (CO, CH$_4$, CO$_2$, H$_2$O) at ice lines, where volatiles evaporate, resulting in a change in the C/O ratio in the solid and gas phase. In principle the C/O ratio of a giant planet atmosphere could then be linked to its formation location \citep{Madhusudhan2014, Madhusudhan2017, Ali-Dib2017}. This link is mostly based on the idea that the atmospheric composition is related to the bulk planetary composition, which is, at least for Jupiter, not true \citep{vazan18b, ppvii_Guillot}.

Furthermore, this picture is dramatically oversimplified, where two main processes can change the C/O ratio in the disk significantly in time: (i) the chemical evolution of the material and (ii) the radial drift and evaporation of pebbles. Simulations of the chemical evolution focus on surface reactions on the grains that can transform the carbon and oxygen bearing volatiles into more complex species (e.g. \citealt{Semenov2010, Eistrup2018}), which enrich the solid and gas C/O differently \citep[see the chapter by][]{ppvii_Krijt}. However, these processes happen on time scales of Myrs, much longer than radial drift of pebbles, making the chemical evolution potentially mostly important at late times \citep{Booth2019} or in pebble traps. On the other hand, inward drifting pebbles evaporate at ice lines \citep{Cuzzi2004, Oberg2011,Ros2013, schoonenberg2017, drazkowska2017}, which results in an enrichment of the gas phase interior to the evaporation front with the corresponding volatile species (\citealt{Garate2020, Schneider2021}, this effect is also used to explain the chemical composition of protoplanetary disks, see \citealt{Zhang2020, Banzatti2020}). This can allow forming planets to accrete gas with super solar C/H and O/H \citep{Booth2017, Schneider2021}, potentially allowing these giant planets to reach large heavy element contents of several 10 to 100 M$_\oplus$ due to the accretion of the volatile enriched gas \citep{Schneider2021}, in agreement with interior models of hot Jupiters \citep{thorngren16}. These large heavy element contents can not be explained only by the planetary core, which is typically on the order of 10-20 $\mearth$ in the pebble accretion scenario. However, both processes significantly depend on the underlying abundances of carbon and oxygen in the disk, which vary for each star, resulting in different compositions of the pebbles and gas \citep{Bitsch2020b}. Further endeavours to link the composition of atmospheric compositions could focus to include more chemical tracers, like nitrogen (e.g. \citealt{Pontoppidan2019, Turrini2021}), or focus on systems with multiple planets, giving more constraints on the formation of the individual planets\index{Exoplanets|)} in the system \citep{Bitsch2021}.

\subsubsection{Solar System}
As the Solar System\index{Solar system|(} offers many constraints ranging from planetary properties to cosmochemical constraints of meteorites that require different methods and tools to simulate, models have so far failed to explain the whole formation history of the Solar System within one simulations. Simulations thus focused to explain different aspects of the Solar System individually with the option to connect the dots between the different aspects to unveil the full formation history of the Solar System.

Recent studies have shown that meteorites\index{Meteorites} exhibit a fundamental isotopic dichotomy between non-carbonaceous (NC) and carbonaceous (CC) materials, which most likely represent materials from the inner and outer Solar System, respectively \citep{Kruijer2017}. Whereas the origin of the dichotomy is probably related to how the protoplanetary disk accreted material from the interstellar medium and redistributed it radially \citep{Nanne2019}, the persistence of this NC-CC dichotomy can be explained by the growth of Jupiter core, inhibiting substantial exchange of material from inside and outside its orbit over the planetesimal formation timescale (at least 3 My according to radioactive chronometers). This scenario requires that Jupiter produces an efficient particle barrier that only allows very little mixing of material from the outer to the inner system. The effectiveness of the barrier clearly depends on the particle size and viscosity \citep{Weber2018, Haugbolle2019}, where more detailed 2D-3D hydrodynamical simulations that include a self consistent particle growth and drift are needed to fully understand this problem (e.g., \citealt{Drazkowska2019}). 
Forming Jupiter core by sole pebble accretion with an Myr can be achieved in simulations, however, afterwards a constant bombardment of planetesimals might be needed to prevent Jupiter core to turn into a super-Jupiter planet before the end of the disk lifetime \citep{Alibert2018}.

\citet{Brasser2020} suggested that the barrier against the radial drift of outer disk material into the inner disk originated from disk effects (e.g., zonal flows, \citealt{Flock2015}) that also generate particle traps (see also chapter by \citealt{ppvii_Bae}), independently of Jupiter formation. This effect was taken recently into account in \citet{Izidoro2021b} and applied to the Solar System \citep{Izidoro2021c}, clearly showing that any kind of trap can explain the NC-CC dichotomy even without the presence of Jupiter. 
Alternatively, \citet{Lichtenberg2021} showed that even planetesimal formation in the outer disk can reduce the pebble flux delivered to the terrestrial planet region. Besides, \cite{Liu2022b} proposed that the pebble’s drift in the outer disk region could be delayed by viscous gas disk spreading,  leading the NC-CC reservoirs unmixed for several Myr.

Independently of the exact reason of cutting the pebble flux to the inner disk, the reduced pebble flux to the inner system can help to explain the dichotomy between the terrestrial\index{Terrestrial planets!formation} and giant planets in terms of their mass \citep{Morbidelli2015} and help to solve the mystery why the Earth is relatively dry despite the rapid inward motion of the water ice line during evolution of the protoplanetary disk \citep{Morbidelli2016}. A barrier, either caused by the disk effects or by the growing core of Jupiter, reduces the pebble flux to the inner system and thus automatically reduces the growth of the inner planetary embryos \citep{Lambrechts2019, Bitsch2019, Izidoro2021}, preventing efficient growth of these embryos beyond the mass of Mars, which is normally used as initial embryo size to model the subsequent formation of the terrestrial planets \citep{hansen09, raymond09c}. 

A pebble barrier does not only reduce the amount of material that can reach the inner system, but also influences the composition of the material available in the inner system, because icy material can be blocked by a barrier exterior to a few AU, where the disk is cold enough for ices to exist, even as the disk cools in time and the ice line moves interior to 1 au \citep{Davis05, Oka2011, Bitsch2015a, Baillie2015}. The water vapor, originally inside Jupiter orbit, diffuses inward faster than the water ice line moves inward \citep{Morbidelli2016}, preventing recondensation of the water vapor onto interior planetesimals and thus keeping the inner disk dry. Both effects result in the formation of a water poor Earth. Alternatively, if water rich pebbles evaporate in the planetary atmospheres during build-up of the planetary core and the vapor is transported away, the terrestrial planets could also form from initially icy pebbles \citep{Johansen2021}, however this formation procedure would result in the accretion of a lot of carbonaceous material, in conflict with our current understanding of Earth composition \citep{Kleine2020, Burkhardt2021}. If this process is really efficient, it also has important implications for the formation history of close in sub-Neptunes (see above). 

Other important clues that Solar System studies have to explain are the structure and mass of the asteroid belt. The underlying models depend crucially on the initial mass of the asteroid belt. While the Grand Tack scenario starts with an initial massive asteroid belt that is then cleared by the inward and outward migration of Jupiter and Saturn during the gas disk phase \citep{walsh11}, the low mass asteroid belt scenario initially assumes a mass depleted asteroid belt (as in \citealt{Drazkowska2016}, see Fig.~\ref{fig:planetesimals}), whose structure and composition is explained by the growth of Jupiter and Saturn scattering water rich asteroids into the belt, and of the terrestrial planets, scattering water poor asteroids into the belt \citep{hansen09, Izidoro2015, Raymond2017a, Raymond2017b}. These different scenarios clearly show the need for simulations that study the formation of planetesimals in the Solar System in detail \citep[see, e.g.,][]{Izidoro2021b}.

While the migration of planets results in resonant configurations, the giant planets of the Solar System are not in a resonant configuration. As mentioned above, these resonant configurations can be broken by instabilities after the gas disk phase. Explaining the current planetary orbits and the structure of the small body populations in the outer Solar System indeed requires that our giant planets  underwent such an instability (see \citealt{Nesvorny2018} for a detailed review). However, it is still debated when this instability occurred. In the Solar System, the timing of the instability is associated with the bombardment history on the Moon, which are caused by impacts of planetesimals \citep{gomes05}. In particular, basin-forming impacts (e.g.~Imbrium, Orientale) occured 3.9--3.7~Gyr ago, so around 600--800~Myr after the formation of the Moon itself. While clearly other basins formed before Imbrium, their exact ages are not precisely known. Two main scenarios are involved in explaining these observations: (i) the cataclysm scenario that requires a surge in the impact rate approximately at Imbrium formation, or (ii) the tail-end scenario where the lunar bombardment declined since the time of planet formation. While the cataclysm scenario is naturally explained if the giant planet instability happened 600--800~Myr after planet formation, the tail-end scenario implies an early instability of the giant planets (e.g., see \citealt{Liu2022a}). While it was originally thought that only the cataclysm scenario could reconcile the lunar crater record with the extremely low abundance of highly siderophile elements (HSE) in the lunar mantle (the latter constraining the total accreted mass), it is now proposed that the HSE were retained only after the lunar mantle crystallization and overturn, which probably occurred 150~Myr after lunar formation \citep{Elkins-Tanton2011, Morbidelli2012}. In this case the accretion tail scenario becomes compatible with both the crater record and HSE mantle and crust abundances \citep{Zhu2019}.

As for exoplanets, the composition of Jupiter and Saturn atmosphere can be used to constrain formation models. While {\it JUNO}\index{JUNO} revealed Jupiter water abundance around its equator just recently \citep{Li2020}, models have incorporated the carbon (e.g. \citealt{Madhusudhan2017, Booth2017, Schneider2021}) and nitrogen (e.g. \citealt{bosman19, Oberg19}) abundances to constrain Jupiter's formation history. Jupiter's super-solar nitrogen abundance has inspired the speculation that Jupiter formed in the far outer regions of the disk beyond 20 au (in agreement with the pebble accretion scenario, \citealt{Bitsch2015b}), where the majority of the nitrogen is frozen out and can be accreted with the solids. This inward migration of the growing Jupiter can naturally explain the asymmetry of Jupiter's Trojans, because planetesimals get trapped asymmetrically at the L4 and L5 points during migration \citep{Pirani2019a, Pirani2019b}. However, open questions about the inclination of the Trojans remain, which can be explained if the Trojans are instead captured during the instability itself \citep{Nesvorny2013}. On the other hand, as pebbles drift inwards they evaporate, enriching the gas with volatiles. As pebbles drift faster than the gas diffuses inwards, this evaporation process can lead to super-solar enrichments of volatile species \citep{Schneider2021b}, which in turn could allow Jupiter to accrete its large nitrogen content via the gas phase without the need to form Jupiter at 20--30 au, where the nitrogen is frozen out.

While the formation of ice giants\index{Ice giants} may in general be the consequence of pebble accretion, Uranus and Neptune axial tilts imply that giant impact(s) played a significant role. \citet{Izidoro2015} studied the formation of Uranus and Neptune from giant impacts of bodies of a few Earth masses that are blocked in their inward migration by the already fully formed Saturn and Jupiter, similar to the formation of sub-Neptunes at the inner edge of the disk. However, the collisions might lead to too large rotational speeds and too massive circumplanetary disks compared to Uranus and Neptune, implying that more studies are needed to explain the formation of the ice giants \citep{Ida2020, Chau2021}.

It is clear that formation scenarios specializing on different aspects of the Solar System and exoplanets require different ideas and models. However, it is clear that models of the Solar System formation and of exoplanets can benefit from each other by applying detailed knowledge of Solar System\index{Solar system|)} formation histories to exoplanets and vice versa. A valid planet formation model should thus be capable to explain the formation of both, the Solar System and exoplanets with their diversity; yet, such a model requires much more work in the future.

\subsection{\textbf{Planetary systems around low-mass stars}}\label{sub:4.3}

M-stars\index{M dwarfs} are stars of mass below 0.45\,$M_\odot$ yet above the hydrogen-burning limit of 0.08\,$M_\odot$. These low-mass stars constitute the majority of the stars in the galaxy (75\%; \citealt{Lada2006}). Being much dimmer, they and their protoplanetary disks are harder to study. Studies have indicated that total disk masses around M dwarfs and brown dwarfs are low, with a dust disk mass-to-stellar mass ratio decreasing superlinearly with mass (but with a large scatter, see \citealt{Pascucci2016,SanchisEtal2020,RilingerEspaillat2021}). In most disks, less then 1\,$\mearth$ of pebble building blocks are available to form planets. Yet the occurrence rates for close-in planets around M-stars outperforms that of solar-type stars (see \S\ref{s:exo:metal}). Perhaps this indicates that planets form early, i.e., that their pebble building blocks have already disappeared by the time we observe these disks with facilities like ALMA.

Several studies have identified a break or peak in the planet mass function, signifying perhaps a characteristic mass for planet formation. Specifically, studies find that $M_\mathrm{br} \approx2$--$3\cdot10^{-5} M_\star$ \citep{Pascucci2018,Wu2019,Liu2019} for close-in planets and $M_\mathrm{br} \approx5\cdot10^{-5} M_\star$ for the microlensing planets \citep{JungEtal2019}. The former number corresponds to 10 M$_\oplus$ planets for solar-type stars and 1 M$_\oplus$ for late M-type stars of 0.1\,$M_\odot$ -- numbers that are similar to the thermal mass scale of ${\sim}(H/r)^3 M_\star$, although the disk aspect ratio is hard to constrain. \citet{Wu2019} and \citet{Liu2019,LiuEtal2020} showed that low-mass planet-hosting stars exhibit such an upper mass limit (higher mass planets would have been readily found). Since the pebble isolation mass\index{Isolation mass} is in its essence the thermal mass (Eq.\ \ref{eq:peb-iso}), it is suggestive to argue that this peak (break) of the planet mass function is a signature of pebble accretion.

\subsubsection{Formation of TRAPPIST-1}\label{sub:trappist}
Perhaps the most iconic example of a multi-planet system is the TRAPPIST-1\index[obj]{TRAPPIST-1|(} system. TRAPPIST-1 is an M8.5 star of only 0.09\,$M_\odot$, yet it harbors seven planets, all similar in radius to Earth (within 25\%), in an ultra-compact configuration (orbits within 0.07\, au; \citealt{GillonEtal2016,GillonEtal2017}). The TRAPPIST-1 system is characterized by a multitude of two-body and three-body mean motion resonances \citep{LugerEtal2017}. Transit timing variations and photometric analysis provide further information on the dynamics, masses and, by inference, compositions of these planets \citep{GrimmEtal2018,DornEtal2018}. Recent modelling implies that the planets are consistent with a similar, rather dry composition slightly lower in density than terrestrial planets in the Solar System \citep{AgolEtal2021}.

\citet{OrmelEtal2017} realized that the key features of the TRAPPIST-1 system -- planets similar masses and compositions, and resonance -- are naturally explained by a pebble-driven growth scenario where planets form at a special location (e.g, the H$_2$O iceline), then start to migrate to the inner disk and continue to grow while migrating, whereafter the process repeats. A numerical follow-up of this scenario included the processes of pebble drift, planetesimal formation and the (stochastic) planetesimal-to-embryo coagulation phase taking place at the iceline \citep{SchoonenbergEtal2019}. The key physical properties of the TRAPPIST-1 planets were retrieved, with the caveat that the 10\% water fraction is too high in the light of the latest TTV analysis \citep{AgolEtal2021}. Yet pebble thermal ablation \citep{ColemanEtal2019} or desiccation by radionuclides as Aluminium-26 \citep{LichtenbergEtal2019} can always reduce the H$_2$O contents.

The properties of the planetary systems orbiting low-mass stars have also been modeled by planetesimal-driven population synthesis studies. \citet{MiguelEtal2020}, following \citet{IdaLin2010}, applied a semi-analytical approach with physically-motivated prescriptions for resonant capture and orbit crossing. They find that in order to assemble Earth-mass planets, disks as massive as 10\% of the star are required. In addition, the planets will be wet, having migrated from regions beyond the iceline. On the other hand, \citet{ColemanEtal2019}, focusing on TRAPPIST-1, resorted to brute force N-body integrations, seeding their simulations with 30 $0.1\,\mearth$ embryos between 1 and 5 au. This system seemed to turn immediately unstable, with mergers increasing masses to Earth values, while simultaneously migrating inwards. \citet{ColemanEtal2019} did not find a clear difference between the planetesimal- and pebble-driven simulations and also produced planets that are too wet. Similarly, \citet{OgiharaEtal2022} also conduct N-body experiments, reproduced the mass distribution of the TRAPPIST-1 planets by invoking a migration transition that is consistent with the disk wind model. Finally, \citet{BurnEtal2021} got around the ``wetness`` by invoking a hot, planetesimal-loaded inner disks (the solids-to-gas ratio lies significantly above solar value). Embryos starting at 1\,au thus accrete dry planetesimals, then migrate inwards. In this way they successfully reproduced the physical characteristics of the TRAPPIST-1 planets in about 1\% of their simulations.

Perhaps the planetesimal-driven scenario is not radically different from the pebble-driven scenario by \citet{SchoonenbergEtal2019}. Both models feature a ``birth region'' beyond the planets present location, where planets grow in a chaotic melee of impacts, before they migrate inwards to end up in resonances. However, the \citet{SchoonenbergEtal2019} model is conceptually simple, yet includes the critical pebble drift and planetesimals formation stages (by streaming instability), which population synthesis models usually ignore.  The main advantage of the population synthesis approach (see \S\ref{sub:popsynth}) lies in their statistics -- with a common ``tool set'' a synthetic exoplanet population can be generated. But the question to be asked is whether specific planet \textit{systems} can also be modelled this way.

Apart from the physical and chemical properties, multi-planet systems are often  characterized by specific dynamical properties. In the case of TRAPPIST-1, this is revealed by a multitude of two-body and three-body mean motion resonances (MMR). The dynamical structure has the potential to rule out (or confirm) certain formation scenarios. For example, \citet{LinEtal2021}, following \citet{OrmelEtal2017}'s formation-then-migration \textit{Ansatz}, constrained the time interval between the planets formation at the iceline. 
\citet{ColemanEtal2019} report that some of their simulations matches the conmensurability of the observed MMR. Yet they fail to match the fine structure seen in the observations (e.g., the values for the three-body libration angles, the higher order resonances, and the range in the planet eccentricities), which may be hard to obtain with population synthesis techniques. \citet{TeyssandierEtal2021} indeed find that the probability of planets to settle into high-order resonances with "standard" disk-driven migration is low.   \citet{Brasser2019} invoked stellar tidal damping to constrain the tidal parameters of planets b and c and \citet{PapaloizouEtal2018} argued that the TRAPPIST-1 system constituted two separate subsystems that were brought together under tidal dissipation. Finally, \citet{HuangOrmel2022} embedded the (long-term) dynamical evolution of the TRAPPIST-1\index[obj]{TRAPPIST-1|)} planets in a formation context, concluding that the two inner-most planet are most likely to have entered a gas-free cavity early in their evolution.

\subsubsection{Giant planets around low-mass stars}
While the occurrence of terrestrial planets and super-Earths around low mass stars seem to be very common, giant planets around these stars are much rarer \citep{Johnson2010}. However, giant planets around low mass stars have been detected such as GJ 3512b \citep{Morales2019}, a super-Saturn orbiting an M dwarf\index{M dwarfs} with 0.125 solar masses, corresponding to a much larger mass ratio than Jupiter in the Solar System. Even systems with multiple giants around M dwarfs have been detected, like around the two saturn mass planets around GJ 1148 \citep{Trifonov2020} or the two Jovian planets around GJ 876 \citep{Marcy2001}.

The formation of terrestrial\index{Terrestrial planets!formation} planets and super-Earths\index{Super-Earths!formation} around low mass stars is a direct consequence of the low pebble isolation mass\index{Isolation mass} and the less abundant building blocks. In this scenario the formation of giant planets, which require cores of several Earth masses, is thus very hard to explain. A large pebble isolation mass could be achieved in the outer regions of the protoplanetary disk, but there the pebble accretion rates are low, because of the low pebble surface density. Even taking the unrealistic assumption that a large amount of solids (around 150 $\mearth$) are immediately transformed to planetesimals does not solve the mystery of forming GJ 3512 b \citep{Morales2019}. On the other hand, the gravitational instability scenario could explain the formation of GJ 3512 b. The detection of these giant planets thus poses a challenge to the core accretion model around low mass stars, opening the avenue for further investigations.

%%%%%%%%%%%%%%%%%%%%%%%%%%%%%%%%%%%%%%%%%%%%%%%%%%%%%%%%%%%%%%%%%%%%%
\section{\textbf{Exoplanet population synthesis}}\label{sub:popsynth}

The purpose of synthesis\index{Population synthesis|(} models is two-fold (\citealt{Ida2004, Ida2008}, \citealt{Mordasini2009, Mordasini2018rev}, see also the {\it Protostars \& Planets VI} review by \citealt{Benz2014}). The first aim is  to create an all-encompassing self-consistent model of planet formation, from protoplanetary disks to completed planetary systems, by synthesizing, and for practical purposes simplifying, our current understanding of planet formation. This theoretical toolbox allows investigating the role and interplay of various physical processes that are often studied in isolation. 
The second component is the practical generation of synthetic exoplanet\index{Exoplanets|(} systems that can be compared to the known exoplanet census and be used to constrain theoretical uncertainties. However, historically, the outcome of synthesis models have suffered from two major sources of uncertainty. Firstly, there is the uncertainty in prior distributions of initial condition parameters that are mainly associated with our understanding of protoplanetary disks. Secondly, we face theoretical modelling choices reflecting gaps in our understanding of the physics of planetesimal and planet formation. Therefore, their predictive power, or capability to dismiss certain theoretical models, has remained limited so far (but certainly not absent, e.g.\,early synthesis efforts ingrained planetary migration as a major component of planet formation,
see \citealt{Baruteau_2014} and the chapter by \citealt{ppvii_Paardekooper}). 
However, this may change due to the ever-increasing sample of well-characterized exoplanets and the vast progress in protoplanetary disk observations, even capable of detecting planets during their formation \citep[e.g. the PDS 70 system;][]{Keppler2018} which opens a new window onto planet formation processes previously shielded from observations.

\subsection{Observed occurrence fractions} 

\begin{figure}[t]
 \centering
 \includegraphics[width=0.9\linewidth]{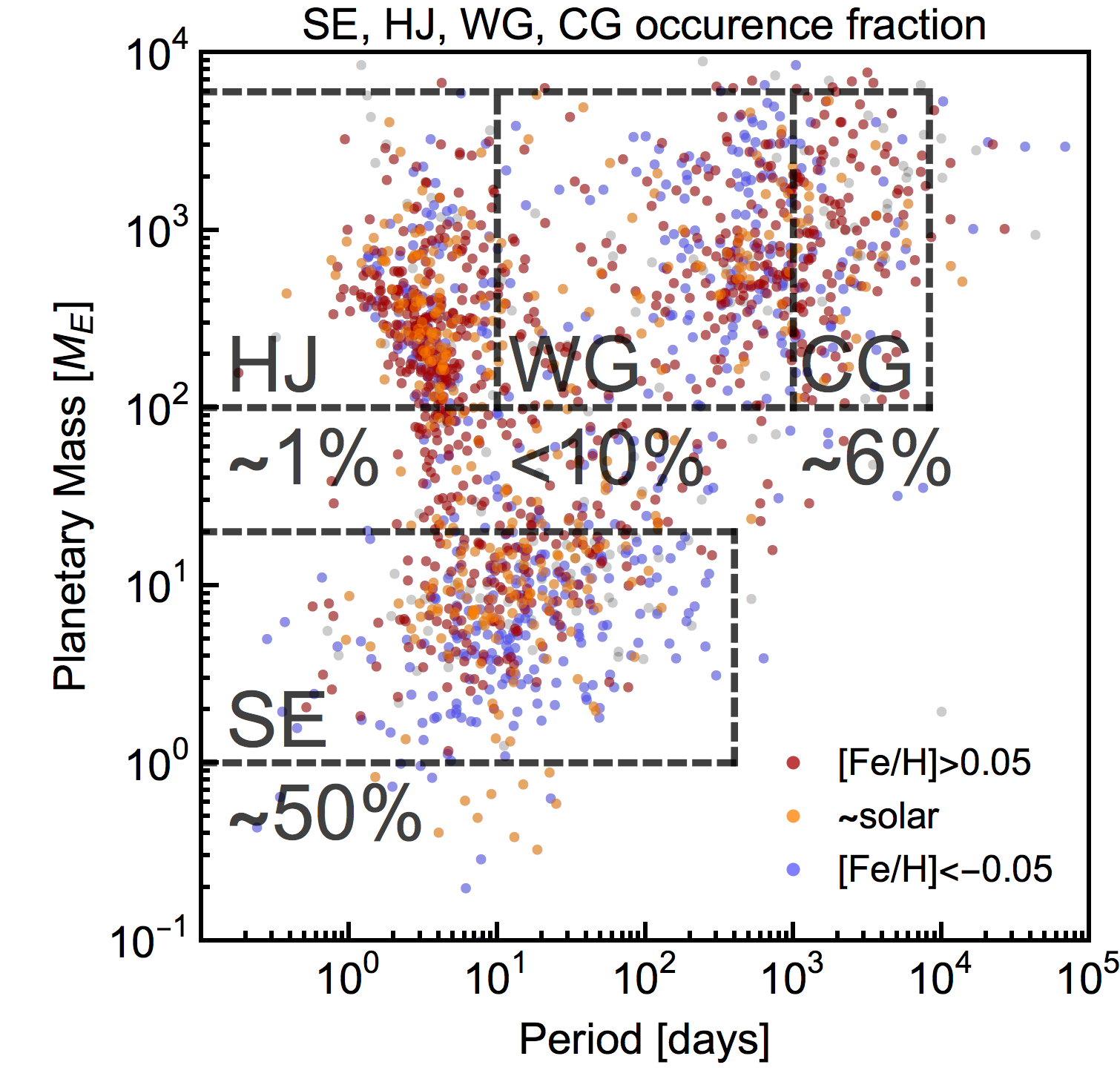}
 \caption{
Periods and masses of known exoplanets with mass determinations, mainly from RV and transit timing variations, taken from the Caltech exoplanets archive (\url{http://exoplanetarchive.ipac.caltech.edu}).
Stellar metallicities are indicated with blue, orange and red for, respectively sub-solar([Fe/H]$<-0.05$), approximately solar and super-solar ([Fe/H]$>0.05$) metallicities. Gray points are used for stars with no listed metallicity estimate. Gray dashed lines indicate the boundaries of different planetary types (hot Jupiters `HJ', warm giants `WG', cold giants `CG', and super-Earths `SE') with the associated percentages listing the observed fractions of stars with such planets.
}
\label{fig:popsynth-obs}
\end{figure}

For the sake of brevity, our discussion will be mainly focused on the comparison of synthesis models to the observed fraction of stars with different types of planets. Figure\,\ref{fig:popsynth-obs} shows the observed fractions of stars with hot Jupiters, super-Earths, warm and cold giants, given the  period and mass boundaries indicated by the gray dashed lines.

The occurrence fraction of super-Earths is about $F_{\rm SE} = 50\%$, based on {\it Kepler}-data (\citealt{He2021}, but $30\%$ according to \citealt{Zhu2018}).
The fraction of stars with cold giants is not well determined, because converting occurrence rates (the average number of planets per star) to occurrence fractions (the fraction of stars with planets) requires knowledge of the planetary multiplicity \citep{Youdin2011}. We estimate here $F_{CG} \sim 6\%$. 
This is based on the RV catalogue of \citet{Rosenthal2021a} and \citet{Fulton2021}, where we find the occurrence rate of cold giants to be $\left<o_{\rm CG}\right> \approx 12.3\%$ for planets with masses between $100$ and $6000$\,M$_{\rm E}$, and orbits within $2$--$8$\,au. We choose this outer radius edge as occurrence rates appear to decrease further at wider orbital radii where RV surveys are also less sensitive \citep{Fernandes19,Wittenmyer2020,Fulton2021}. 
The average multiplicity has not yet been determined for this catalogue. 
We therefore take a value of $\left<m_{\rm CG}\right> = 2$, which is similar to the Solar System with two giants at $5$ and $9$\,au. A value above $2$ would be unlikely: three Jupiters within this radial range approach the edge of stability \citep{chambers96,Petit2020}. Thus, we take $F_{\rm CG} =\left<o_{\rm CG}\right>/\left<m_{\rm CG}\right> \sim 6\%$.

The occurrence fraction of warm giants, $F_{\rm WG}$, is similarly not well known. Based on {\it Kepler} data, EPOS \citep{Mulders2018} achieves an acceptable fit for the warm giants with an average multiplicity of $2$ to $4$ and a resulting fraction of, respectively, $F_{\rm WG}=9$\% to $F_{\rm WG}=4$\%, assuming a mutual inclination of respectively $4$ to $6$ degree. A high multiplicity is likely given the high eccentricities of warm giants in RV surveys \citep{Rosenthal2021a}. We therefore decide on an upper limit of $F_{WG}\lesssim 10$\%. 
The fraction of stars with Hot Jupiters is about $F_{HJ} = 1\%$ \citep{HsuEtal2020}\index{Exoplanets|)}. 

\begin{figure*}[h!]
 \centering
 \includegraphics[width=0.9\linewidth]{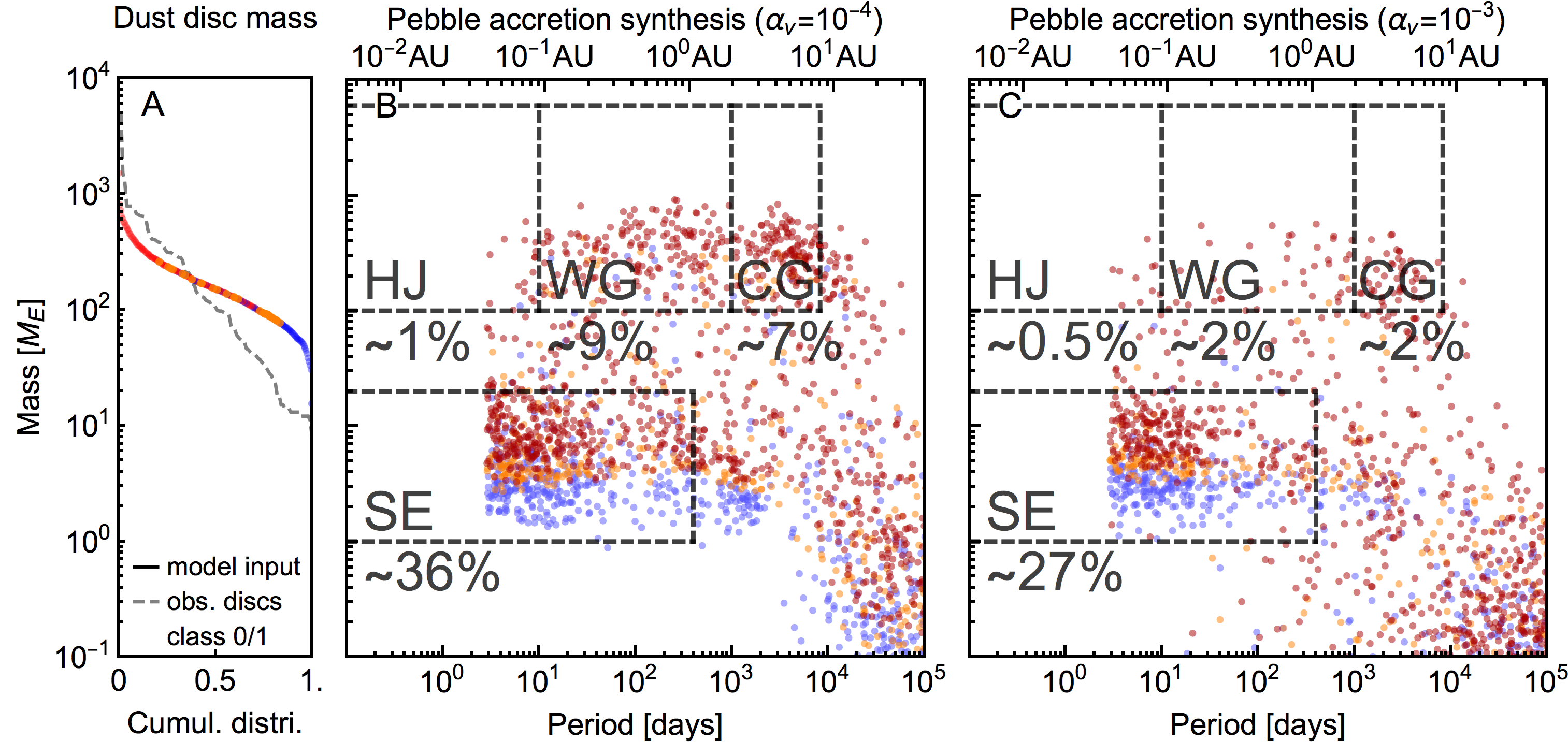}
 \caption{
 {\bf Panel A}: Cumulative distribution of the initial dust mass in each of the 2000 simulated disks with solar-mass host stars (blue, orange and red represent those disks with, respectively,  sub-solar ([Fe/H]$<-0.05$), approximately solar and super-solar ([Fe/H]$>0.05$) metallicities). For comparison, the gray dashed curve shows the dust mass distribution in observed Class 0/I disks \citep[{not limited to solar-mass stars,}][]{Tychoniec2020}.
 {\bf Panel B}: Masses and periods of planets, at the time of disk dissipation, generated through the pebble population synthesis model. Gray dashed lines indicate the boundaries of different planetary types with the associated percentages listing the synthetic fractions of stars with such planets. Color coding corresponds to the initial disk masses/metallicities of panel A.
 {\bf Panel C}: 
 Pebble population synthesis model with an increased strength of the vertical turbulent stirring parameter $\alpha_{\rm v} =10^{-3}$. The increased pebble scale height reduces the efficiency of accreting pebbles, generally reducing the final masses of the planets.
 }
\label{fig:popsynth-peb}
\end{figure*}

\subsection{A simplified pebble accretion population synthesis model} 

Following \citet{Bitsch2015b}, \citet{Ndugu2018} and \citet{Liu2019}, we construct a simplified exoplanet synthesis model, modeling the three key planet formation processes: pebble accretion\index{Pebble accretion}, planetary migration\index{Planet migration}, and gas accretion.
Similar approaches have been used in \citet{Ida2016,Ali-Dib-AJ2017,Chambers2018,Johansen2019b}. 
Two thousand protoplanetary disks are generated, following \citet{Bitsch2015a}. 
For simplicity, we set each disk to surround a star with a mass equal to the sun. 
The initial condition distributions are chosen as follows: disks have a fixed exponential decay rate of 2.5\,Myr \citep[][see also \S\ref{sect:observations}]{Mamajek2009} and an initial dust mass distribution as shown in panel A of Figure \ref{fig:popsynth-peb}, corresponding to a Gaussian stellar metallicity spread of 0.23\,dex \citep{Mayor2011}.
Thus, we choose here to model disk masses with a more narrow disk mass spread compared to the initial Class 0/I\index{Class 0} disks masses reported by \citet{Tychoniec2020}, because the large observed spread is likely mainly due to the spread in stellar masses \citep{Pascucci2018}, while we center the distribution on solar-like stars.
We assume that the pebble mass flux decays with the same exponential decay rate as the gas and we assume pebbles with ${\rm St}=10^{-2}$ (see \S\ref{sect:dust}). However, in reality, the pebble size and pebble flux are set by the outcome of dust coagulation in the outer disk \citep[different approaches can be found in][]{Bitsch2015a,Johansen2019b,Appelgren2020,Schneider2021,MuldersDraz2021}. This simplification also implies we do not explicitly determine the evolution of the corresponding outer dust disk radius. 
Finally, rapid gas disk clearing by photoevaporation is not explicitly modeled, but disk lifetimes are chosen to follow an exponential distribution with $2.5$\,Myr decay (and a minimal $1$\,Myr lifetime).

Next, we model the evolution of a single planetary embryo. Embryos are drawn uniformly in the logarithm of the orbital distance, with a mass corresponding to the onset of pebble accretion in the Hill regime, and placed randomly after $0.1$ to $0.5$\,Myr of disk evolution. The pebble scale height is set by assuming a vertical stirring parameter of $\alpha_{\rm v}=10^{-4}$ 
or $\alpha_{\rm v}=10^{-3}$  (see Sec.\ref{sub:turb} for an observational motivation).
Planets migrate in the type-I regime following \cite{paardekooper11}, while their type-II migration rate follows the prescription from \citet{Kanagawa2018}.
Gas accretion rates onto planetary cores follow \citet{Bitsch2015a} and are bound to not exceed the radial gas transport rate through the disk. 
The inner edge of the disk, where all migration is halted, is set between $0.04$ and $0.25$\,au, in line with \citet{Flock2019}.

\subsection{Comparison to observation}

\subsubsection{Planetary diversity}
The pebble synthesis model is able to reproduce the wide diversity of observed planetary types in mass and orbital period space \citep{Bitsch2015b}, as can be seen by comparing panel B  or C of Fig.\ref{fig:popsynth-peb} and  Fig.\,\ref{fig:popsynth-obs}. Of course no one-to-one correspondence can be expected. Various observational biases are present (as well as a wide spread in stellar mass  for the shown observed sample in Fig.\,\ref{fig:popsynth-obs}). Nevertheless, we can observe that a disk mass distribution that is consistent with disk masses inferred from 1 and 9\,mm emission in young Class 0/I disks (panel A of Fig.\,\ref{fig:popsynth-peb}, \citealt{Tychoniec2020}) is consistent with the diversity of the exoplanet record. This result can be contrasted to the lower dust disk mass budgets that are measured in evolved Class II disks with ages of a few Myr that are barely capable of forming super-Earths (\citealt{Manara2018}, see also \S\ref{sect:observations}). This strengthens the interpretation that planet formation is already ongoing in the youngest disk stages, well before radial drift depletes the dust reservoir of disks \citep{Appelgren2020}. 

A first apparent conclusion is thus that the observed exoplanet diversity can be linked to naturally expected variations in disk and planet formation conditions. This key result, was first seen in earlier planetesimal-based synthesis efforts \citep{Ida2004,mordasini12} is not straightforward, especially given the complex feedback dependencies between different physical processes taking place in an evolving protoplanetary disk (e.g.\,the radial migration rate depends on the planetary mass with a growth rate which, in turn, is linked to its orbital radius).

Closer inspection reveals some differences between the  observed and modeled populations of Figures \ref{fig:popsynth-obs} and \ref{fig:popsynth-peb}. 
For example, giant planets more massive than $1000$\,M$_{\rm E}$ are easy to detect but their true occurrence is in fact rare \citep[$\lesssim 1$\%][]{Bowler2018,HsuEtal2020,Fulton2021}, and even more so when excluding stars more massive than the sun \citep{Baron2019,Fulton2021}. Their lack in the synthetic model therefore does not reveal a strong discrepancy. We also note that this pebble synthesis model does not form sub-Earth-mass planets, because the high pebble accretion efficiently drives core growth up to the pebble isolation mass\index{Isolation mass}. Possibly this hints that their formation requires pebble fluxes that are reduced, by either considering less massive disks or the effects of outer planets halting the flow of pebbles \citep{Morbidelli2015,Lambrechts2019}.

\subsubsection{Occurrence fractions}
In the pebble synthesis model with a level of vertical turbulent stirring of $\alpha_{\rm v}=10^{-4}$, the fraction of stars with Hot Jupiters is $f_{\rm HJ} = 1\%$, with warm giants $f_{\rm WG} = 9\%$, with cold giants $f_{\rm CG} = 7\%$, and super-Earths\index{Super-Earths} $f_{\rm SE} = 36\%$ (Fig.\,\ref{fig:popsynth-peb}, panel B). 
This can be compared to the observed fraction of stars with these planetary classes
(Fig.\,\ref{fig:popsynth-obs}) 
and we can see a crude agreement between model and observed occurrence fractions of \emph{multiple} planetary classes.
Lower solid accretion rates, here due to a larger pebble scale height ($\alpha_{\rm v}=10^{-3}$, see also Sec.\ref{sub:pebbleacc}) lead to generally smaller planets (Fig.\,\ref{fig:popsynth-peb}, panel C). The lower occurrence fractions are then more in tension with observed values. This illustrates the sensitivity of synthesis models to chosen parameters, which are further explored in for example \citet{Ida2016,Ndugu2018,Johansen2019b,Liu2019}.
An important caveat is that these models shown here do not take multiplicity  into account. This ignores then, for example, the fraction of disks with cold giants that also could have produced super-Earths (while the inverse is less likely: a low mass disk barely capable of forming a super-Earth also producing more massive cold giants). 
Finally, a large fraction of disks, approximately half, does not form currently observable planets, but instead low-mass planets ($\lesssim 10$\,M$_{\rm E}$) on wide orbits ($\gtrsim 10^3$\,days).

\subsubsection{Observed correlations with stellar properties}
Synthesis models also recover two key observational relations between planetary occurence and  stellar metallicity and mass.
The stellar hosts of close-in giant planets commonly have super-solar metallicities 
\citep{santos04,fischer05}, while super-Earths have stellar hosts with a wider range of metallicities
\citep{Mayor2011,buchhave12}. These trends can even be seen in the observed (biased) exoplanet set of Fig.\,\ref{fig:popsynth-obs}. 
Panel B of Fig.\,\ref{fig:popsynth-peb} illustrates how the synthesis model recovers this metallicity correlation: 
in order for gas giants to complete their massive envelopes (and inwards migration) their solid cores need to emerge early and benefit from the larger mass reservoir of metal-rich disks, while the lower core and envelope masses of super-Earths do not place such strong timing constraint \citep{Bitsch2015b,Liu2019}. Another important correlation is the observed near-linear scaling of characteristic planetary mass with stellar mass \citep{Pascucci2018, Wu2019}. Synthesis efforts taking stellar mass into account show how low-mass stars (M dwarfs)\index{M dwarfs} indeed lack planets more massive than super-Earths due to the effect of a reduced pebble isolation mass around low-mass stars (\S\ref{sub:4.3}, \citealt{Liu2019},
but see also \citealt{BurnEtal2021}).

\begin{figure*}[t]
 \centering
 \includegraphics[width=0.8\linewidth]{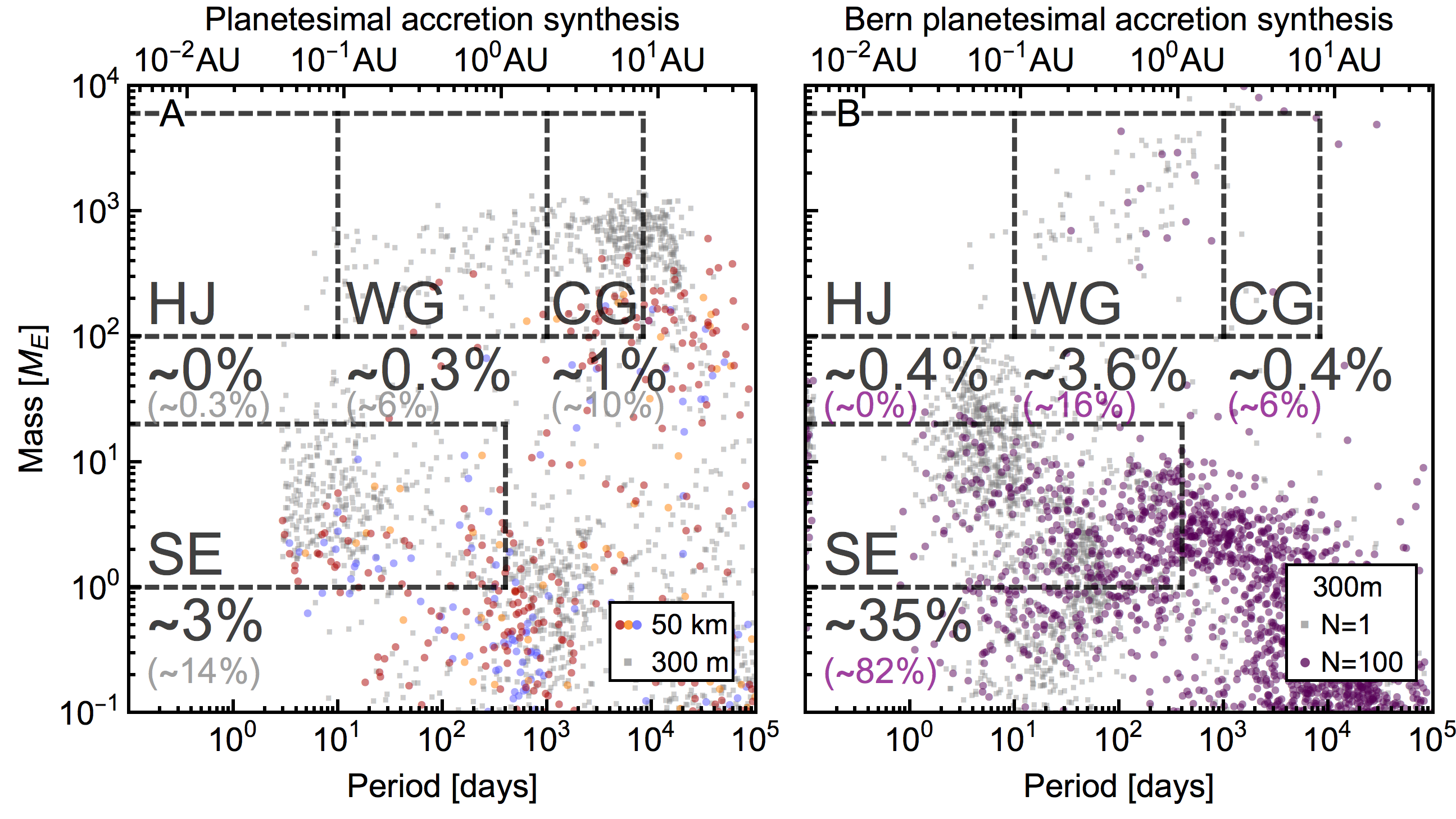}
 \caption{
 {\bf Panel A}:  Masses and periods of planets generated from the same disk distribution as in Fig.\,\ref{fig:popsynth-peb}, but now assuming planetesimal accretion instead.
 Colors follow the stellar metallicities as in Fig.\,\ref{fig:popsynth-peb} for the case of 50-km planetesimals, while the gray squares show synthesis results when choosing a smaller planetesimal size of $300$\,m 
 which promotes core growth in the outer disk.
 {\bf Panel B}: In gray, we show two thousand planets that are drawn randomly from the Bern single-planet synthesis model \citep{Emsenhuber2020a}. This model uses $300$\,m planetesimals and therefore produces an outcome similar to the $300$\,m model of panel A. However, the different initial disk distribution, combined with different model choices, result in a factor 2-3 increase in the super-Earth fraction.
 In purple we show 100 systems from the Bern 100-embryo model \citep{Emsenhuber2020a} which further increases the super-Earth fraction, while the cold giant fraction increases mainly through outwards scattering.}

\label{fig:popsynth-plan}
\end{figure*}

\subsection{A simplified planetesimal population synthesis model} 

We now repeat the above experiment, but change one aspect: we assume all pebbles are converted into planetesimals, at the start of our integrations \citep[similar to][]{Brugger2020}. This is done to highlight the effect of a different model choice that follows pioneering synthesis works \citep[e.g.][]{Ida2008, Mordasini2009}. We calculate planetesimal accretion\index{Planetesimal accretion} rates onto migrating planets following \citet{Tanaka1999,Johansen2019}, but exclude the planetesimal-shepherding effect, as also done in \citet{Emsenhuber2020a}. 
The degree of turbulent stirring of the planetesimals is chosen to be low. 
We set $\gamma = 10^{-6}$, which corresponds to low eccentricities for planetesimals with radii $R_{\rm p\epsilon}$ below $50$\,km: $e \lesssim 10^{-3}(R_{\rm p\epsilon}/50{\rm km})^{1/3}$ at 10\,au \citep[for full description see][]{ida08b,Johansen2019}. Higher values, such as $\gamma=10^{-4}$, more in line with the turbulent diffusion coefficients that are observationally inferred (see \S\ref{sub:turb}),  result in thicker planetesimal midplane scale heights that strongly suppress planetesimal accretion \citep[Sec.\,\ref{sub:planetsimalacc}, ][]{Fortier2013,Johansen2019}.

We present results for a population of planetesimals with a radius of $R_{\rm p\epsilon}=50$\,km (panel A of Fig.\,\ref{fig:popsynth-plan}) and also for smaller planetesimals 300\,m in size (gray squares in panel A of Fig.\,\ref{fig:popsynth-plan}). 
This latter choice follows the Bern model and the motivation for using small planetesimals is partly inspired by debris disk models
\citep[for more see][their Sec.\,3.3.2]{Emsenhuber2020a}. However, we do note that such planetesimal sizes are much smaller than those inferred to make up the mass distribution of the primordial asteroid belt where most mass resides in objects larger than 50\,km in size \citep{bottke05,morby09a}.

Overall, it can be seen that 50-km planetsimals lead to slow growth in wide orbits resulting in fewer warm and cold giants, in contrast to the observed fractions (Fig.\,\ref{fig:popsynth-obs}). However, the small-planetesimal model overcomes this limitation,
because smaller planetesimals are  more easily accreted with the help of the gaseous envelopes around embryos \citep[this effect is implemented following][]{inaba03,Ormel2012}. 
Even so, the fraction of stars with super-Earths\index{Super-Earths!formation} is also reduced compared to the pebble accretion case, because of less efficient growth. 
Additionally, \citet{Brugger2020} argue that the role of the pebble isolation mass is important: by halting solid growth, it can lead to smaller cores in the outer disk, resulting in inwards-migrating super-Earths, while planetesimal accretion may, under the assumption of inefficient planetesimal isolation, result in larger cores producing more cold giants instead.

Multi-embryo planetesimal synthesis models further shift the occurrence fractions \citep{Mulders2019}. 
This can be seen by comparing in panel B of Fig.\,\ref{fig:popsynth-plan} the Bern 1-planet setup (gray) with the 100-embryo setup (purple) \citet{Emsenhuber2020a}. First, we note that the Bern 1-planet setup gives results similar to the $300$\,m-planetesimal model of panel A of Fig.\,\ref{fig:popsynth-plan}. 
However, the fraction of cold giant is slightly reduced and the fraction of super-Earths increases by more than a factor two. These differences are due to the different disk distribution (not following panel A of Fig.\,\ref{fig:popsynth-peb}) and different model choices. For example, the combined effects of smaller disks, steeper solid density gradients, and inserting all embryos at $t=0$ with larger masses lead to more centrally-concentrated embryos. This, together with inefficient planetesimal isolation, promotes the formation of super-Earths.
Comparing the 1-planet and 100-embryo setup, \citet{Emsenhuber2020} highlight that the fraction of cold giants can be increased through outward planet scattering. 
These scattered cold giants would then have a different formation pathway compared to the cold giants that originate from inwards migrating planets in the pebble synthesis models \citep{lodato19, Ndugu2019}. Multi-planet synthesis models are further discussed in the chapter by \citet{ppvii_Weiss}.

\subsection{The future of exoplanet synthesis}
We identify several areas where we expect major progress in synthesis modelling to be made, with many of these efforts already underway.

The distributions for the initial condition priors of synthesis models, such a as disk masses, sizes and lifetimes, can now be drawn much closer to observed values thanks to the major progress in ALMA disk observations (see \S\ref{s:disks}, and the chapters by \citealt{ppvii_Miotello} and \citealt{ppvii_Manara}). Correspondingly, improved pebble coagulation models are finding their way in global planet formation models \citep[e.g.][]{Schneider2021,MuldersDraz2021}

A new valuable anchor point for synthesis models are direct detections of planets still forming in their protoplanetary disks \citep{lodato19,Ndugu2019}. 
This will however require uniquely linking disk gaps \citep{Zhang2018} and velocity modulations \citep{pinte20} to the presence of planets and their masses.

It is important to take into account planetary multiplicity using full N-body techniques \citep{Alibert2013,Emsenhuber2020,Izidoro2021,Matsumura2021}. 
This is necessary in order to address, for example, if cold giants enhance or suppress the occurrence of super-Earths \citep{Lambrechts2019,Bitsch2020,Schlecker2020}, which is not yet observationally known \citep{Bryan2019,Zhu2018b,Barbato2018,Rosenthal2021b}. Or, similarly to understand possible mass and period correlations between super-Earth systems 
\citep{Millholland2017,Weiss2018,ppvii_Weiss,Mishra2021}.

An additional component of synthesis models, not highlighted here, is tracing planetary composition during formation and their secular evolution \citep{Pudritz2018}. For example, this approach can explain the gap in occurrence rates around $1.8$~R$_\oplus$ \citep[the radius valley,][]{Fulton2017} as a point where envelope photo-evaporation is efficient \citep{owen2017,Jin2018}. Improved mass and radius measurements (capable of, for example, constraining the TRAPPIST-1 planets to near-Earth-like composition \citealt{AgolEtal2021}) will require more detailed compositional modeling.

Synthesis efforts need to be placed in a galactic context: different galactic populations host different exoplanet populations due to age, metallicity \citep{Bitsch2020b} and stellar environment (\citealt{Winter2020, Adibekyan2021,Mustill2021}).

The actual comparison between observations and theory is complex, and requires  observation simulators of synthetic populations\index{Population synthesis|)} that capture known survey biases \citep[e.g.][]{Mulders2019}. In a following step, the numerical process of extracting best fitting models and constraining free model parameters deserves further attention \citep[e.g.][]{Chambers2018}.

%%%%%%%%%%%%%%%%%%%%%%%%%%%%%%%%%%%%%%%%%%%%%%%%%%%%%%%%%%%%%%%%%%%%%
\section{\textbf{SUMMARY AND OUTLOOK}}\label{sect:summary}

The theory of planet formation is currently undergoing rapid evolution, and we aimed to outline the major areas of these developments in this chapter. In this summary section, we reiterate the major take away points and make suggestions for the areas where the planet formation models and observational efforts should improve.

The wealth of exoplanetary data that was collected and, more importantly, analyzed in the past years has transformed our view on many aspects of planet formation. We learned that planet formation is a robust process taking place across the range of stellar masses. The most common outcome of planet formation seem to be the super-Earths and mini-Neptunes, none of which exist in our own Solar System. What is more, these are often close-in planets and a typical exoplanet system has several Earth masses inside of 1~au, significantly more than the terrestrial planets surrounding the Sun.

At the same time, the progress in observational capabilities of the young stellar objects has shed new light on the planet-forming environments. The measured masses of solids present in the circumstellar disks turned out to be surprisingly low. At the same time, many of the disks that we have been able to image with the high resolution are showing substructures, often interpreted as the signatures of ongoing planet formation. These realizations have put new constraints on the timing of planet formation.

One of the notable emerging concepts is that planet formation needs to be starting much earlier than previously assumed, already during the Class 0/I stage when a protostellar envelope is still infalling onto the disk. These youngest circumstellar disks are still not well-characterized, and this will hopefully improve during the time to the next {\it Protostars \& Planets} series meeting. Complementary efforts should also be made in the planet formation models, that still often assume the fully-fledged Class II disks as their starting point. 

Another major conceptual change is related to the emergence of the first gravitationally bound building blocks of planets, the planetesimals. As we described in \S\ref{sect:theory}, previous planet formation models would typically assume that the planetesimals form quickly throughout the whole disk and that planet formation relies mostly on the accretion of these planetesimals (and gas). Thanks to the developments in both protoplanetary disk imaging and theoretical models, the picture is now shifting toward considering the significance of smaller solids, the pebbles, during the whole planet formation process.

The emerging paradigm of pebble accretion allows for more efficient accretion of massive planetary cores in the outer part of the protoplanetary disk (\S\ref{sub:ppaccretion}), provides better agreement with the observed structure of gas-rich planets in the Solar System (\S\ref{sect:gasacc}), makes it possible to reproduce the diverse types of exoplanets (\S\ref{sub:4.1}), and gives better agreement between the synthetic and observed population of wide-orbit giant planets (\S\ref{sub:popsynth}). However, we must note that the state-of-the-art pebble accretion models rely on the assumption that the massive planetary cores, which can quickly grow by pebble accretion, emerge relatively fast. For this, large planetesimals need to form promptly. Even if the birth sizes of planetesimal are on the order of 100 km as predicted by the streaming instability models, there still needs to be some period of accretion among planetesimals (\citealt{Liu2019b}, see also Fig.~\ref{fig:sizescales}). The pebble accretion models also required large and long-lasting fluxes of pebbles, which may be questionable given the ubiquity of sub-structures seen in the disks. Moreover, the complications related to fragmentation of pebbles which should depend on grain composition and disk temperature are usually ignored, overall resulting in too efficient pebble accretion. 

At this point, we should make another key remark, that is: the planet formation process may be spatially and temporarily fragmented. This means that the planets do not form at every location in the disk and that the planets within one disk do not necessarily form at the same time. The former point may seem obvious with the abundance of substructures observed in the protoplanetary disks (\S\ref{s:disks}) and with the knowledge about special locations where the inward drift of solids is halted and the growth processes are sustained (\S\ref{sub:driftmigr}). The later conclusion arises from analyzing the growth timescales shown in Fig.~\ref{fig:timescales}. Most of the processes involved in planet formation, from dust to core growth proceed faster closer to the star. In the core accretion paradigm, the close-in planets should essentially form at shorter timescales than wide-orbit planets. However, whether this happens or not, depends on the locations where the planetesimals and planetary cores form. One possibility of temporal fragmentation is that the planetary cores are formed at one location but over longer timescale, such as in the model presented by \citet{OrmelEtal2017} to reproduce the TRAPPIST-1 system (see \S\ref{sub:trappist}). On the other hand, \citet{Pinilla2015} presented some indirect evidence for the dependence of the planetary formation timescale on location in the disk of HD 100546, which may be fitted with two massive planets if the outer one is significantly younger than the inner one. This might be fit in with the models suggesting that formation of one massive planet facilitates the emergence of another planet outwards \citep{Kobayashi2012, Chatterjee2014}. The idea of sequential planet formation is certainly worth investigating, but the state-of-the-art planet formation models still need some developments before this could be reliably accomplished.

The major inconsistency in the present planet formation models, both the ones assuming planetesimal accretion and the ones considering pebble accretion scenario, is that they start with various assumptions about planetesimal and planetary cores formation. This could be blamed on the lack of convincing planetesimal formation models. However, now we are at the point where such global simulations connecting dust evolution and planetesimal formation start to emerge (\S\ref{sect:dust}). A notable conclusion from this work is that planetesimal formation itself is not a burst but a process which may last for a significant fraction (or even the whole duration) of the circumstellar disk lifetime, which is also consistent with the meteorite\index{Meteorites} record of the Solar System \citep{Morbidelli2020EPSL, Lichtenberg2021}. Triggering planetesimal formation may require a significant (both vertical and radial) redistribution of solids and thus the spacial distribution planetesimals may significantly vary from the that of gas (see Fig.~\ref{fig:planetesimals}). Ultimately, we need to stress that planetesimal formation is critical for determining the mass budget available for further accretion of planetary cores by planetesimal and pebble accretion.

Recently, \citet{Voelkel2021} and \citet{Coleman2021} presented the first models following pebble, planetesimal, and planetary cores formation at the same time. They included prescriptions for both pebble and planetesimal accretion to investigate the interplay of these processes. While these are preliminary studies performed for a limited set of disk parameters (and one particular planetesimal formation scenario), they shows a promising pathway toward more self-consistent planet formation models. Such an all-encompassing approach could be utilized in future versions of the planet population synthesis models, potentially allowing to constrain their free parameters. 

Planet formation is a complicated process. Connecting planet formation models to the emerging picture of disks and exoplanet demographics requires one to take a panoptic perspective, where all the planet formation concepts: pebble growth, planetesimal formation, the growth of embryos by solids and gas accretion, as well as their migration, are included in a streamlined fashion. At this moment, there is no single model that could alone explain all the aspects of the pathway from dust grains to planetary systems. Nevertheless, a significant progress has been made from the previous {\it Protostars \& Planets} series, as we described in this chapter.
\bigskip

\noindent\textbf{Acknowledgments.} We thanks the referees and the community for their feedback that helped us to improve this chapter. We thank Daniel Carrera and Christian Lenz for sharing the data needed to produce Fig.~\ref{fig:planetesimals}. We also acknowledge Łukasz Tychoniec for providing us the Class 0/I disks data points used to produce Figs~\ref{fig:massevol} and \ref{fig:popsynth-peb}. We thank Christoph Mordasini for his insightful comments on \S5, and for making available the data in panel B of Fig.~\ref{fig:popsynth-plan}. J.D., B.B., and M.L.\, acknowledge funding from the European Research Council (ERC Starting Grants 101040037-PLANETOIDS, 757448-PAMDORA, and 101041466-EXODOSS). M.L.\,acknowledges financial support by Wallenberg Academy Fellow Grant 2017.0287. G.D.M. acknowledges support from FONDECYT project 11221206, from ANID --- Millennium Science Initiative --- ICN12\_009, and the ANID BASAL project FB210003. 
A.V. acknowledges support by ISF grants 770/21 and 773/21. 
B.L. thanks the National Natural Science Foundation of China (Nos. 12222303, 12173035 and 12111530175) and the Fundamental Research Funds for the Central Universities (2022-KYY-506107-0001,226-2022-00216).
\bigskip

\bibliographystyle{pp7}
\bibliography{Chapter20.bib}

%\printindex
%\renewcommand{\indexname}{Object Index}
%\printindex[obj]

\end{document}